\documentclass[a4paper,11pt]{article}
\usepackage{jheppub} % for details on the use of the package, please see the JINST-author-manual

\usepackage{amsmath}
\usepackage{amssymb}
\usepackage{graphicx}
\usepackage{framed}
\usepackage{slashed}
\allowdisplaybreaks
\usepackage[dvipsnames]{xcolor}

\def\beq{\begin{equation}}
\def\eeq{\end{equation}}
\def\beeq{\begin{eqnarray}}
\def\eeeq{\end{eqnarray}}

\def\mathL{\mathcal{L}}

\def\PlusingN[#1]{{\left[\frac{\ln^{#1}(1-x)}{1-x}\right]_+}}

\newcommand\as{\alpha_{s}}

\def\to{\rightarrow}
\def\nn{\nonumber}

\def\q2t{{\bf q}_{\perp}^2}

\def\SIA{{\mbox{\scriptsize SIA}}}

\def\LQCD{{\Lambda_{\rm QCD}}}

\usepackage{fancyhdr}
\title{Threshold Resummation for Semi-Inclusive Single-Hadron Production with Effective Field Theory}

\author[a]{Zhen Xu,} 
\author[b,c]{Hua Xing Zhu} 

\affiliation[a]{Zhejiang Institute of Modern Physics, Department of Physics, Zhejiang University, Hangzhou 310027, China}
\affiliation[b]{School of Physics, Peking University, Beijing 100871, China}
\affiliation[c]{Center for High Energy Physics, Peking University, Beijing 100871, China}

\emailAdd{zhen.xu@zju.edu.cn}
\emailAdd{zhuhx@pku.edu.cn}

\abstract{Large double-logarithmic corrections are induced by soft gluon emissions near threshold in the semi-inclusive $e^+e^-$ annihilation (SIA) distributions, and must be resummed to all-orders in perturbation theory for reliable theoretical predictions. 
Building on strategy developed for threshold resummation for DIS structure function in momentum space using soft-collinear effective theory (SCET), we present the explicit formalism for SIA cross section. We then perform the resummation directly in momentum space for $\gamma^* \to q \bar q$, $H \to gg$ and $H \to b\bar b$ to N$^4$LL accuracy and demonstrate good convergence.
We anticipate that these results will benefit the extraction of the light-quark, the heavy-quark as well as the gluon fragmentation functions. \\

}

\begin{document}
\maketitle
\flushbottom

\section{Introduction}

Semi-inclusive $e^+e^-$ annihilation (SIA), where a specific final-state hadron is detected, 
is an important process for studying the dynamics of quark and gluon fragmentations in Quantum Chromodynamics (QCD). 
In particular, it provides one of the cleanest ways to extract the fragmentation functions (FFs) describing the non-perturbative fragmentation of a parton into the specified hadron detected in the final state via collinear factorization. It is well-known that the perturbative matching coefficients in the factorization formula contain large double logarithmic corrections in the production threshold \cite{Sterman:1986aj,Appell:1988ie,Catani:1989ne,Korchemsky:1992xv,Catani:1996yz,Contopanagos:1996nh,Vogt:1999xa,Cacciari:2001cw,Moch:2005ba,Moch:2009my,AH:2020xll}, which is specified in the SIA process by the limit $x = 2 p \cdot q/q^2 \to 1$, where $x$ is the momentum fraction, $p$ is the hadron momentum, and $q$ is the center-of-mass frame total momentum. To achieve reliable theoretical prediction, such large logarithms must be resummed to all orders to preserve perturbative convergence.

The all-order resummation of such double logarithmic corrections for SIA distribution for $e^+e^- \to \gamma^*/Z \to $ \emph{hadrons} has been achieved to the next-to-next-to-next-to-leading logarithms (N$^3$LL) \cite{Cacciari:2001cw,Moch:2009my} 
using the conventional method in Mellin moment $N$-space, where the large $\ln^k(1-x)/(1-x)$ corrections as $x \to 1$ are transformed to $\ln^{k+1}N$ as $N \to \infty$. 
In this work we consider the resummation of the large logarithms directly in momentum space \cite{Becher:2006nr,Becher:2006mr,Manohar:2003vb,Pecjak:2005uh,Chay:2005rz,Manohar:2005az,Idilbi:2006dg,Chen:2006vd,Idilbi:2005ky,Becher:2007ty}
based on the Soft-Collinear Effective Theory (SCET) \cite{Bauer:2000yr,Bauer:2000ew,Bauer:2001yt,Bauer:2001ct,Beneke:2002ph}. 
Building on the factorization theorem for DIS structure function, we discuss the resummation formalism for SIA distribution near threshold. We then resum the large logarithms to N$^4$LL accuracy, improving the previous results in the literature.

The main motivation for the resummation is to improve the accuracy of extracting the fragmentation functions. 
The SIA distribution for $e^+e^- \to \gamma^*/Z \to $ \emph{hadrons} primarily constrains the combination $D_q^h + D_{\bar q}^h$ of the quark and anti-quark fragmentation functions \cite{Bender:1984fp,Derrick:1986jx,TASSO:1990cdg,L3:1991nwl,TOPAZ:1994voc,ALEPH:1996sio,DELPHI:1996sen,DELPHI:1996oqw,OPAL:1996fae,OPAL:1997asf,OPAL:1999ldr,L3:2004cdh}. 
While the semi-inclusive electron-proton DIS provides the complementary measurements sensitive to the asymmetric contribution of the quark and anti-quark contributions \cite{H1:1997mpq,ZEUS:1999ies}, the  $e^+e^- $ process has the cleanest environment due to the absence of beam remnants. 
On the other hand, 
the gluon fragmentation contributes in $e^+e^- \to \gamma^*/Z \to $ \emph{hadrons} only at higher order in perturbation theory or by scaling violations \cite{Braaten:1993rw,Bourhis:1997yu,DELPHI:1997oih,DELPHI:1999alp,ParticleDataGroup:2022pth}, leading to reduced constraining power from such process. 
The heavy-quark fragmentation function has been discussed and measured with the vector boson as the intermediate state 
\cite{Mele:1990yq,Mele:1990cw,Braaten:1994bz,Nason:1999zj,Melnikov:2004bm,Mitov:2004du,Neubert:2007je,Biello:2024zti,ALEPH:1995hae,ALEPH:1995zdi,SLD:1999cuj,SLD:2002poq,OPAL:2002plk,DELPHI:2011aa}. 
With the enriched physics prospects in the promising future $e^+e^-$ Higgs factories such as the CEPC \cite{CEPCStudyGroup:2018rmc,CEPCStudyGroup:2018ghi,CEPCStudyGroup:2023quu}, the FCC-ee \cite{FCC:2018byv, FCC:2018evy} and the ILC \cite{Behnke:2013xla,Bambade:2019fyw}, we then discuss the resummation for  $H \to $ \emph{hadrons}, which is anticipated to significantly benefit the extraction of the gluon fragmentation from the $H \to gg$ process as well as the heavy-quark fragmentation from the $H \to b \bar b$ process.

Furthermore, the fragmentation functions are important for understanding the dynamics of jets in the final states of collider events. We anticipate that the soft gluon resummation to be also important for processes at the Large Hadron Collider (LHC), where the calculation of the higher order single-hadron distributions is much more challenging. Progress has been made in this direction in the moment-space approach in e.g. \cite{Laenen:1998qw,Jager:2004jh,deFlorian:2005yj,deFlorian:2007tye,deFlorian:2013qia,Catani:2013vaa,Kumar:2013hia,Almeida:2009jt,Hinderer:2014qta,Hinderer:2018nkb,Forslund:2020lnu} and in the momentum-space approach in e.g. \cite{Ahrens:2011mw,Ferroglia:2013awa,Liu:2017pbb,Liu:2018ktv}.  
The fragmentation functions are also important for understanding the substructure of jets 
\cite{Procura:2009vm, Jain:2011xz, Kang:2016mcy, Lee:2024icn, Lee:2024tzc} and parton showers
\cite{vanBeekveld:2023lsa,vanBeekveld:2024jnx,vanBeekveld:2024qxs}. 
Very recently, a global analysis of the fragmentation functions to charged hadrons has been performed with including the high precision LHC data with a wide kinematic coverage \cite{Gao:2024nkz,Gao:2024dbv}, which have led to significant constraints on FFs especially of gluons.  
The prospects of extracting the fragmentation functions at future $e^+e^-$ colliders have also been studied in \cite{Zhou:2024cyk}, which is based on the fixed-order program FMNLO \cite{Liu:2023fsq}. 
The impact of the small-$x$ resummation on the extraction of the fragmentation functions has been considered in \cite{Anderle:2016czy}, while an analogous analysis on the large-$x$ side has not beed addressed in the literature, to the best of our knowledge.

Another motivation for resumming the large threshold logarithms in SIA comes from the recent interests in the projected $N$-point energy correlators in the collinear limit~\cite{Chen:2020vvp}, which has been used to extract precisely the strong coupling constant $\alpha_s$ from jet substructure by the CMS collaboration~\cite{CMS:2024mlf}. In the collinear limit, the projected $N$-point energy correlators factorizes into the so-called hard function extracted from the SIA coefficient function, 
and a jet function which describes the measurement encoding the direct angular dependence. The large threshold logarithms in the SIA coefficient function can potentially lead to large corrections in the hard function for the energy correlators.
Therefore, it is well-motivated to discuss the large-$x$ resummation to higher orders for the SIA process.
We leave the more detailed phenomenological analyses of the large-$x$ resummation effects in the extraction of fragmentation functions as well as in the event-shape observables to the future study.

The structure of the rest of this paper is the following. We present the resummation formalism in Section \ref{sec:res_formalism}, with a brief recap of the moment-space resummation and the discussion of momentum-space resummation from effective field theory approach. The resummation results to N$^4$LL for $\gamma^* \to $ \emph{hadrons} are presented in Section \ref{sec:gamma2Hadrons}, where we provide the numerical results with the scale-uncertainty and the large-$x$ terms in the fixed-order result to N$^3$LO. We check explicitly the momentum-space results in this section with the resummation coefficients from the moment-space approach after extracting the four-loop anomalous dimension $B_q$, and find perfect agreement order by order in perturbation theory. 
Section \ref{sec:Higgs2Hadrons} contains the resummation results to N$^4$LL for $H \to $ \emph{hadrons} via $H \to gg$ and $H \to b \bar b$ decays, including numerical results with scale-uncertainty and expansion to N$^3$LO, checked successfully with existing results in the literature. 
We briefly summarize and conclude in Section \ref{sec:summary}. 
The ingredients required for performing the resummation to N$^4$LL are collected in  Appendix~\ref{sec:ingredients}, and the predictions of the large-$x$ terms at N$^4$LO are given in Appendix~\ref{sec:PredictionN4LO}. The large-$x$ coefficients expanded to N$^4$LO based on the N$^4$LL resummation are provided in the ancillary file with the arXiv submission of this paper.

\section{Resummation formalism}
\label{sec:res_formalism}

The SIA cross section has been referred to as single-particle inclusive distribution in the $e^+e^-$ annihilation via the production of a neutral gauge boson
\beq
e^+ (l_1) + e^- (l_2)  \to V  (q)  / H (q) \to h(p) + X (p_X) \,  ,
\eeq
where $h (p)$ represents the detected hadron $h$ with momentum $p$, and $Q^2 = q^2$ is the squared center-of-mass collision energy. 
The unpolarized differential SIA cross section can be written as \cite{Nason:1993xx}
\begin{align}
\label{unpolarizedSIA}
  \frac{d^2\sigma}{dx\,d\cos\theta} = \frac{3}{8}(1+\cos^2\theta)\frac{d\sigma_T}{dx}
  + \frac{3}{4}\sin^2\theta\frac{d\sigma_L}{dx} + \frac{3}{4}\cos\theta
  \frac{d\sigma_A}{dx}, 
\end{align}
where $x = {2p\cdot q}/{Q^2} \to 1$ is the fraction of beam energy carried by the detected hadron, and $\theta$ is the relative angle between the detected hadron and the beam direction in the center of mass frame. 
The three terms $\sigma_L$, $\sigma_T$ and $\sigma_A$ represent the longitudinal, transverse and asymmetric contributions to the cross section, respectively. 
The asymmetric $\sigma_A$ is parity-violating and comes from the interference between the virtual photon $\gamma^{*}$ and the $Z$-boson contributions. 
After integrating the angles, only the longitudinal $\sigma_L$ and transverse $\sigma_T$ components are present, with the latter providing the dominant contribution to the total cross section. 
Near the threshold region, defined with the energy fraction $x  \to 1$, the phase space for extra emissions is severely restricted and only soft gluon radiations are permitted. The associated infrared (IR) divergences are cancelled with the virtual contributions, while large double logarithmic corrections remain in the form of the plus distributions 
\beq
\mathcal{L}_{n}(1-x) \,  = \, \left[\frac{\ln^{n}(1-x)}{1-x}\right]_+ \, ,
\eeq
with $n \le 2k-1$ in the $k$-th order contribution in the strong coupling constant $\alpha_s$. We use the parameter $a_s \equiv \alpha_s/(4\pi)$ for the perturbative expansions in the strong coupling constant through out this paper.

\begin{figure}[ht]
	\centering
	\includegraphics[width=6.5cm]{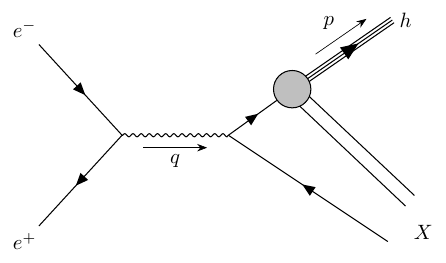} \qquad\qquad
	\includegraphics[width=5.5cm]{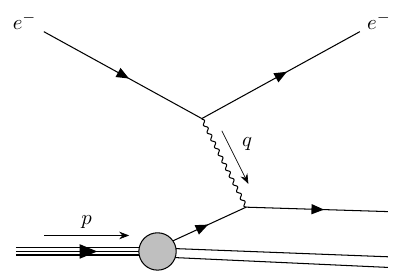} \quad
	\caption{Born level diagrams describing the SIA (left) and DIS (right) processes.}
	\label{fig:SIA-DIS}
\end{figure}

We first focus on the $\gamma^* \to $ \emph{hadrons} case, where we have the quark $q$ and anti-quark $\bar q$ in the final state of the Born-level kinematics. 
The longitudinal cross section or the pure  flavour-singlet contributions  are suppressed
by a relative power of $(1-x)$ in the $x \to 1$ limit. 
The leading power threshold corrections come from the flavour non-singlet contributions, which can be written as \cite{Cacciari:2001cw}
\begin{align}
\label{Def_SIA}
\frac{1}{\sigma_{0}}\frac{d\sigma_{T}(x,Q^2)}{dx} =  \sum_q\int_x^1 \frac{d\eta}{\eta}   
\;C_q^{(\SIA)}(\frac{x}{\eta}, Q^2,\mu^2,\mu_F^2)
\,D_q(\eta,\mu_F^2)  + \mathcal{O}\bigg(\frac{\LQCD}{Q}\bigg) \,, 
\end{align}
where the convolution is over the momentum fraction $x$ and between the coefficient function $C_q^{(\SIA)}$ and the (partonic) fragmentation function $D_q$ for a parton of flavour $q$. The anti-quark contribution is not distinguished from the quark contribution due to charge conjugate invariance. 
The normalization is over the Born-level total cross section $\sigma_0 = N_c \cdot \sum _q e_q^2\,  4 \pi \alpha^2 / (3 Q^2 )$, where $N_c=3$ is the number of colors in QCD,  $e_q$ is the electric charge of the quark $q$ and $\alpha$ is the electromagnetic coupling.

\subsection{Factorization Theorem in Moment Space}

In the conventional method, threshold resummation is performed in moment space \cite{Sterman:1986aj,Appell:1988ie,Catani:1989ne,Korchemsky:1992xv,Catani:1996yz,Contopanagos:1996nh,Vogt:1999xa,Cacciari:2001cw,Moch:2005ba,Moch:2009my}. 
In terms of $N$-moments, defined for a plus-distribution $f(x)_+$ at large-$x$ as 
\beq
f_N = \int_0^1 dx \left(x^{N-1} -1 \right) f(x)_+  \, ,
\eeq
 the large-$x$ logarithm $[\ln^k(1-x)/(1-x)]_+$ becomes $\ln^{k+1}N$ at large-$N$. 
 The coefficient function in terms of $N$-moments has the exponentiated form 
 \cite{Cacciari:2001cw,Contopanagos:1996nh,Moch:2009my}
\beq
\label{coefficient_moment}
C_N(Q^2,\mu_F^2) = g_0 (Q^2,\mu_F^2) \exp{[G_N(Q^2,\mu_F^2)]} + \mathcal{O}(N^{-1} \ln^k N) \, .
\eeq
The resummation exponent $G_N$ is described by the radiative factors
\beeq
\label{exponent_moment}
 \exp{[G_N(Q^2,\mu_F^2)]} = \Delta_N (Q^2,\mu_F^2)  \cdot J_N (Q^2) \, ,
\eeeq
which are given by integrals over two functions $A$ and $B$ of the running coupling, 
\beeq 
\Delta_N (Q^2,\mu_F^2) &=& \exp \left\{
\int_0^1 d\eta \,\frac{\eta^{N-1} -1}{1-\eta} \;
\int_{\mu_F^2}^{(1-\eta)^2Q^2} 
\frac{dq^2}{q^2} \,A[\as(q^2)] \right\} 
\eeeq
and
\beeq 
J_N(Q^2) =
\exp \Bigg\{
 \int_0^1  d\eta   \, \frac{\eta^{N-1} -1}{1-\eta}  
 \, \Bigg[ \int_{(1-\eta)^2Q^2}^{(1-\eta)Q^2} \frac{dq^2}{q^2} A[\as(q^2)]  
+  B[\as((1-\eta)Q^2)] \Bigg] \Bigg\} \,. \quad
\eeeq
The function $A$ is the universal cusp anomalous dimension for a quark field, and $B$ function entering the jet function $J_N$ in the moment-space resummation describes the collinear emissions. 
The expansion coefficients in the coupling constant of the functions $g_0$, $A$ and $B$ will be required for implementing the resummation to given accuracy.  
It is not very surprising that the leading power threshold logarithms share great similarity in structure for the coefficient functions for SIA cross section and DIS structure function, because, at the Born level, they are kinematically related by crossing, relating the final-state detected hadron with the initial-state target nucleon, depicted in Figure~\ref{fig:SIA-DIS}. The difference should therefore come from that the space-like quantities are now changed to their time-like counter-parts, and that extra soft emissions are from the initial-state or final-state.

The relation of the large-$x$ contributions can also be understood from the reciprocity relation \cite{Gribov:1972rt,Chen:2020uvt}. The fact that the physical cross sections should be renormalization group (RG) invariant tells us that the evolution of the coefficient functions for the SIA and DIS can be determined as the counter-part to that of the FFs and the PDFs respectively. 
Moreover, the large-$x$ limit / large-$N$ limit of time-like and space-like splitting functions, which govern the RG evolution of FFs and PDFs \cite{Gribov:1972ri,Lipatov:1974qm,Dokshitzer:1977sg,Altarelli:1977zs}, is the same, with the difference starts at the subleading powers. 
This crucially leads to the fact that the resummation exponent is identical in the SIA and DIS cases \cite{Cacciari:2001cw,Moch:2009my}, to be discussed in more detail below.

We also note that the threshold resummation effects have been found to be larger in the time-like case than in the space-like case \cite{Moch:2009my}, and the subleading power corrections, suppressed by $(1-x)$ or by $1/Q$, should be interesting for further investigation.

\subsection{Factorization Theorem in SCET}

We first focus on the $\gamma^* \to q \bar q$ case, which is closely related to the DIS case at the leading power of large-$x$ limit \cite{Becher:2006nr,Becher:2006mr} due to the crossing symmetry and the reciprocity relation. The factorization formula can be readily generalized to the $H \to g g $ and $H \to b \bar b$ cases, discussed in Section~\ref{sec:Higgs2Hadrons}. 
An important feature of the factorization theorem in the effective theory is that the coefficient function in the large-$x$ limit can be factorized into the hard function $H_{\rm SIA}$ and the jet function $J_{\rm SIA}$. 
The factorization theorem of SIA cross section at the leading power of large-$x$ limit reads, 
\beq
\label{SIAfact}
   F^{(\SIA)}(x,Q^2) = H_{\rm SIA}(Q^2,\mu) \, Q^2   \, x
    \int_x^1\!d\xi\,
   J_{\rm SIA}\big( Q^2(1-x/ \xi),\mu \big)\,D_{q}^{}(\xi,\mu) +\mathcal{O}\left(\frac{\Lambda_{\rm QCD}}{Q}\right)\,. 
\eeq
The structure of the factorization theorem is as similar as possible to that of the DIS structure function, with the change of space-like components to their time-like counter-parts. 
We note another difference that the leading power corrections from non-perturbative effects are of order $\Lambda_{\rm QCD}/Q$ in the time-like case for both the longitudinal and transverse components \cite{Nason:1993xx}. 
The following statements are in order in the comparison of the time-like SIA case and the space-like DIS case.

\begin{itemize}
\item Hard Wilson coefficient can be extracted from the time-like on-shell form factor 
\beq
H_{\rm SIA}(Q^2,\mu) = |C_V(Q^2,\mu)|^2 \,. 
\eeq
It can be shown that the ratio of resummed coefficient functions for SIA via $\gamma^* \to q \bar q$ and DIS is given by the ratio of the squared time-like form factor and squared space-like form factor. 
\item Jet function for SIA $J_{\rm SIA}$ is the same as the jet function for the DIS structure function, from the fact that they describe the same degree of freedom.  
\end{itemize}

The following two points are important for the factorization theorem near threshold in SCET. 

\begin{itemize}
\item Matching of the QCD current, yielding the hard function. 
\item Matching of the hadronic tensor, yielding the jet function.
\end{itemize}

Diagrammatically, this amounts to crossing the photon line from space-like to time-like, and the initial-state target nucleon to the final-state identified hadron. As for the final-state jet, the only difference is that the soft radiation comes from the initial- or final-state, which however does not change its kinematics at the leading power in the soft limit.

\subsection{Renormalization Group Equations}

The hard function $C_V(Q^2,\mu)$ can be extracted from the time-like on-shell form factors, which are fundamental ingredients for many QCD calculations. 
The RG evolution equation for the hard function $C_V(Q^2,\mu)$ takes the following form
\begin{equation}\label{HardSIA_RGE}
   \frac{d}{d\ln\mu}\,C_V(Q^2,\mu)
   = \left[ \Gamma_{\rm cusp}(\alpha_s)\,\ln\frac{Q^2}{\mu^2}
   + \gamma^V(\alpha_s) \right] C_V(Q^2,\mu) \,,
\end{equation}
where both $ \Gamma_{\rm cusp}(\alpha_s)$ and $ \gamma^V(\alpha_s) $ can be perturbatively expanded and determined from the fixed-order calculation for $C_V(Q^2,\mu)$. 
The quantity $ \Gamma_{\rm cusp}(\alpha_s)$ is called the cusp anomalous dimension and plays an important role in the study of the infrared and collinear behaviour of QCD. 
The solution to the hard function RGE has relatively simple structure
\begin{equation}\label{HardSIA_RGE_Sol}
   C_V(Q^2,\mu) = \exp\left[ 2S(\mu_h,\mu) - a_{\gamma^V}(\mu_h,\mu) \right]
   \left( \frac{Q^2}{\mu_h^2} \right)^{-a_\Gamma(\mu_h,\mu)}\,
   C_V(Q^2,\mu_h) \,,
\end{equation}
with the exponents given by
\begin{equation}\label{RGE_Exponent_S}
S(\nu,\mu) = - \int\limits_{\alpha_s(\nu)}^{\alpha_s(\mu)}\!
d\alpha\,\frac{\Gamma_{\rm cusp}(\alpha)}{\beta(\alpha)}
\int\limits_{\alpha_s(\nu)}^\alpha \frac{d\alpha'}{\beta(\alpha')} \,, 
\end{equation}
and
\begin{equation}\label{RGE_Exponent_a}
a_{\gamma^V}(\nu,\mu) = - \int\limits_{\alpha_s(\nu)}^{\alpha_s(\mu)}\!
d\alpha\,\frac{\gamma^V(\alpha)}{\beta(\alpha)} \,, \qquad
a_\Gamma(\nu,\mu) = - \int\limits_{\alpha_s(\nu)}^{\alpha_s(\mu)}\! d\alpha\,\frac{\Gamma_{\rm cusp}(\alpha)}{\beta(\alpha)} \,. 
\end{equation}
The RG equation for the jet function $J(p^2,\mu)$ is an integro-differential equation \cite{Becher:2006qw}, 
\beeq
\label{JetSIA_RGE}
   \frac{dJ(p^2,\mu)}{d\ln\mu}
   = &-& \left[ 2\Gamma_{\rm cusp}(\alpha_s)\,\ln\frac{p^2}{\mu^2}
   + 2\gamma^J(\alpha_s) \right] J(p^2,\mu) 
   \nn\\
   &-& 2 \Gamma_{\rm cusp}(\alpha_s) \int_0^{p^2}\!dp^{\prime 2}\,
   \frac{J(p^{\prime 2},\mu)-J(p^2,\mu)}{p^2-p^{\prime 2}} \,.
\eeeq
The exact solution can be obtained using the associate jet function defined by the Laplace transform \cite{Becher:2006qw,Becher:2006mr,Becher:2006nr}
\begin{equation}\label{Laplace}
   \widetilde J\Big( \ln\frac{Q^2}{\mu^2},\mu \Big)
   = \int_0^\infty\!dp^2\,e^{-s p^2}\,J(p^2,\mu) \,, \qquad
   s = \frac{1}{e^{\gamma_E} Q^2} \,.
\end{equation}
The solution of the RGE for the original jet function is given by
\begin{equation}\label{JetSolution}
   J(p^2,\mu) = \exp\left[ - 4S(\mu_i,\mu) + 2 a_{\gamma^J}(\mu_i,\mu) \right]
   \widetilde J(\partial_\eta,\mu_i) \left[ \frac{1}{p^2}
   \left( \frac{p^2}{\mu_i^2} \right)^\eta \right]_{\!*}\,
   \frac{e^{-\gamma_E\eta}}{\Gamma(\eta)} \,,
\end{equation}
with $\eta=2a_\Gamma(\mu_i,\mu)$ and the star-distribution defined as 
\beq
\int_0^{Q^2} d p^2 \left[ \frac{1}{p^2} \left(\frac{p^2}{\mu_i^2}\right)^\eta \right]_* f(p^2) = \int_0^{Q^2} d p^2 \frac{f(p^2)-f(0)}{p^2}  \left(\frac{p^2}{\mu_i^2}\right)^\eta  + \frac{f(0)}{\eta}  \left(\frac{Q^2}{\mu_i^2}\right)^\eta  \, ,
\eeq
for a smooth function $f(p^2)$. 
The large-$x$ resummed SIA cross section, which generalizes the DIS result with  the space-like ingredients substituted by the time-like ones, including the change of the PDF $\phi_q$ to the FF $D_q$, then reads
\begin{eqnarray}\label{SIAresummed}
  F^{(\SIA)}(x,Q^{2})
   &=& |C_V(Q^2,\mu_h)|^2
    \left( \frac{Q^2}{\mu_h^2} \right)^{-2a_\Gamma(\mu_h,\mu_i)}
    \exp\left[ 4S(\mu_h,\mu_i) - 2a_{\gamma^V}(\mu_h,\mu_i) \right] 
    \nonumber\\
   &&\times \exp\left[ 2a_{\gamma^\phi}(\mu_i,\mu_f) \right]
    \widetilde J\Big( \ln\frac{Q^2}{\mu_i^2}+\partial_\eta,\mu_i \Big)\,
    \frac{e^{-\gamma_E\eta}}{\Gamma(\eta)}\,\int_x^1\!d\xi\,
    \frac{D_q^{}(\xi,\mu_f)}{\left[ \left( \xi-x \right)^{1-\eta}  \right]_{\!*}} \,,   \qquad\;
\end{eqnarray}
where $a_{\gamma^\phi} = a_{\gamma^J} - a_{\gamma^V}$, and $\gamma^\phi = \gamma_J - \gamma_V$ can be determined from the $\delta(1-z)$ term of the Altarelli-Parisi splitting function $P_{i \leftarrow i} (z)$ for both quarks and gluons \cite{Becher:2006nr,Becher:2006mr}. 
We have also approximated the prefactor $x$ at the leading power in the numerical evaluation below. 
The resummed coefficient function can be identified by comparing Eq.(\ref{Def_SIA}) with Eq.(\ref{SIAresummed}), 
\begin{eqnarray}\label{SIAcoe}
  C^{(\SIA)}(x,Q^{2})
   &=& |C_V(Q^2,\mu_h)|^2
    \left( \frac{Q^2}{\mu_h^2} \right)^{-2a_\Gamma(\mu_h,\mu_i)}
    \exp\left[ 4S(\mu_h,\mu_i) - 2a_{\gamma^V}(\mu_h,\mu_i) \right] 
    \nonumber\\
   &&\times \exp\left[ 2a_{\gamma^\phi}(\mu_i,\mu_f) \right]
    \widetilde J\Big( \ln\frac{Q^2}{\mu_i^2}+\partial_\eta,\mu_i \Big)\,
    \frac{e^{-\gamma_E\eta}}{\Gamma(\eta)}\,  \left[\frac{1}{ \left( 1-x \right)^{1-\eta}  }\right]_{\!*} \,. \qquad
\end{eqnarray}
Achieving different resummation order requires the ingredients calculated to corresponding orders, summarized in Table \ref{tab:ords}. The logarithmic order for the resummed coefficient function in the momentum-space approach is lowered by one compared to the moment-space approach, as they are connected through convolution \cite{Becher:2006nr,Becher:2006mr,Becher:2007ty,Bonvini:2012az}.

\begin{table}
\begin{center}
\begingroup
\renewcommand{\arraystretch}{1.2}
  \begin{tabular}{|c|c|c|c|c|c|}  \hline
   Resummation Order &  Log Accuracy $ \sim a_s^n L^k$ &   $J$, $C_V$  &  $ \gamma_J$, $ \gamma_V$  &  $\beta[\alpha_s]$  &  $\Gamma_{\rm cusp}$ \\  
    \hline
    LL  & $k=2n$ & tree & tree & 1-loop & 1-loop  \\
    \hline
    NLL  & $2n-1 \le k \le 2n$  & tree & 1-loop & 2-loop & 2-loop \\
    \hline
    N$^2$LL & $2n-3 \le k \le 2n$ & 1-loop & 2-loop & 3-loop & 3-loop \\
    \hline
    N$^3$LL & $2n-5 \le k \le 2n$ & 2-loop & 3-loop & 4-loop & 4-loop \\
    \hline
    N$^4$LL & $2n-7 \le k \le 2n$ & 3-loop & 4-loop & 5-loop & 5-loop \\
    \hline
\end{tabular}
\endgroup
\end{center}
\vspace{-0.2cm}
\caption{Definition of the resummation order, the corresponding logarithmic accuracy for the resummed coefficient function and the required inputs. RG consistency for the anomalous dimensions are checked with the large-$x$ end-point limit of the splitting function to three-loop. And the large-$x$ end-point limit of the splitting function at four-loop and $\gamma_V^{(3)}$ has been used to extract $\gamma_J^{(3)}$. 
}
\label{tab:ords}
\end{table}

\section{Resummation for $\gamma^{*} \to $ hadrons}
\label{sec:gamma2Hadrons}

In this section, we present the numerical result for $\gamma^{*} \to hadrons $  to N$^4$LL prediction in momentum space with its scale uncertainty. We observe very good perturbative convergence from NLL to N$^4$LL as well as the reduction of the scale uncertainty obtained from varying the hard scale of the process. We then perform the resummation also in moment space to N$^4$LL, and cross check  their agreement order by order in perturbation theory.

\subsection{Resummation Result to N$^4$LL}

\begin{figure}[ht]
	\centering
	\setlength{\abovecaptionskip}{-0.0cm}
	\setlength{\belowcaptionskip}{-0.0cm} 
	\includegraphics[width=12.cm]{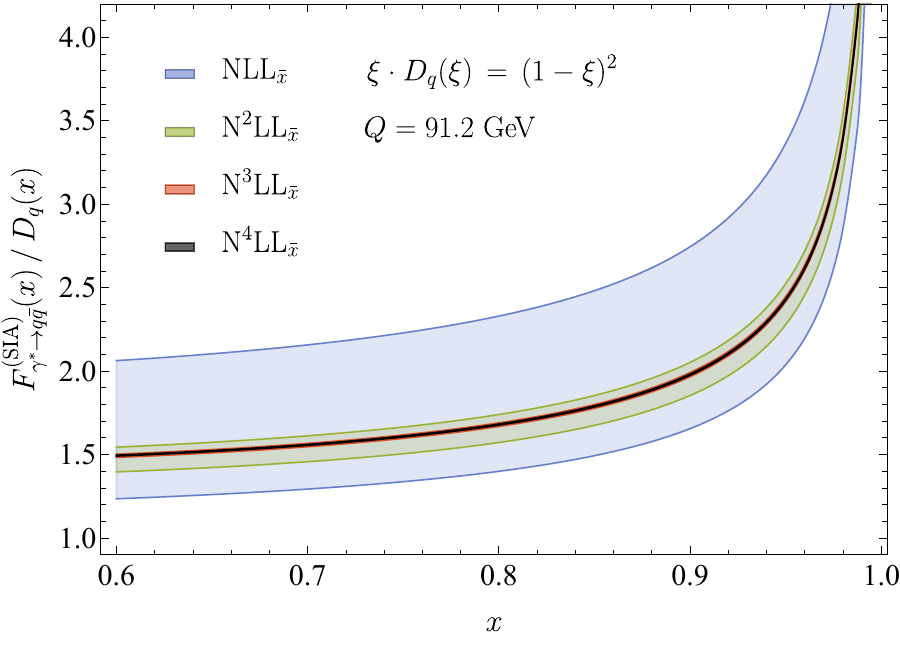}\\
	\caption{SIA cross section for $\gamma^{*} \to q \bar q$ to N$^4$LL in the large-$x$ limit, with a typical form for the fragmentation function $D_q^{}(\xi) = (1-\xi)^2/\xi$ near the end-point, as considered in \cite{Moch:2009my} in the conventional approach in moment space. Bands represent variation of the hard scale $\mu_H \in [Q/5,5Q]$. We set the number of light flavour quarks $n_f=5$. }	\label{fig:SIAN4LL_Quark}
\end{figure}
\begin{figure}[ht]
	\centering
	\setlength{\abovecaptionskip}{-0.0cm}
	\setlength{\belowcaptionskip}{-0.0cm} 
	\includegraphics[width=12.2cm]{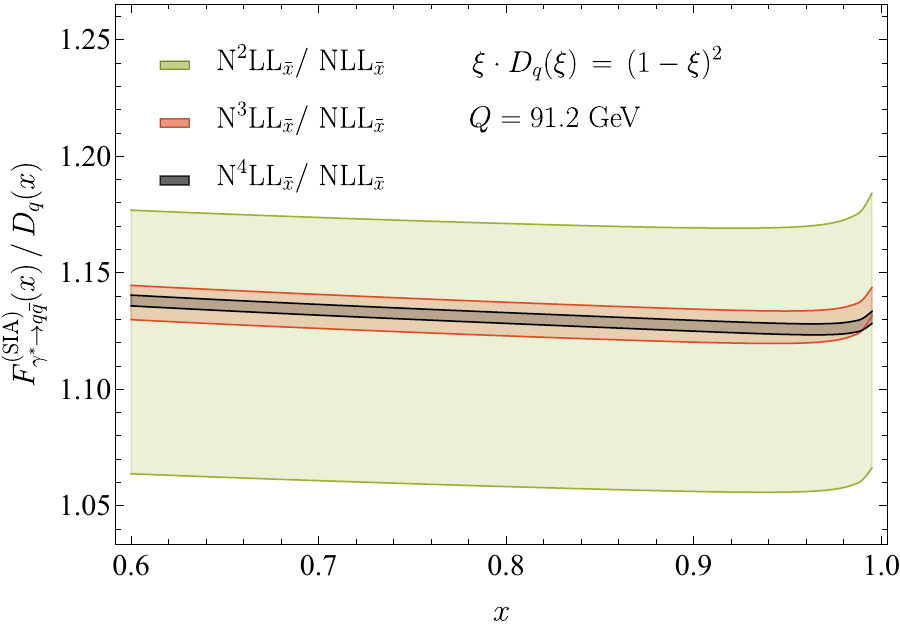}\\
	\caption{The ratio of N$^n$LL result and NLL result for $\gamma^{*} \to q \bar q$, with bands representing the variation of the hard scale $\mu_H \in [Q/5,5Q]$ for the numerators. }
	\label{fig:SIAN4LLvsNLL}
\end{figure}

The perturbative ingredients needed for the N$^4$LL resummation can be extracted from the literature, with the 5-loop cusp estimated with uncertainty and all others known in exact, which are presented in Appendix~\ref{sec:ingredients}. 
To illustrate the impact of the large-$x$ resummation in phenomenologically relevant situations, we use a typical form for the fragmentation function near the end-point. For simplicity, we exploit the parametric form $D_q^{}(\xi) = (1-\xi)^2/\xi$ as an example and omit here its factorization scale $\mu_f$ dependence.

Figure~\ref{fig:SIAN4LL_Quark} shows the prediction from NLL to N$^4$LL at the center of mass energy of 91.2 GeV, where good perturbative convergence is observed through to N$^4$LL. 
The hard scale is chosen as the c.o.m. energy $\mu_H = Q$, and the intermediate matching scale is set to be the invariant mass of the final-state jet $\mu_i = M_X = Q\sqrt{1-x}$.  We then vary the hard scale around the canonical scale by a factor of 5 as an estimate of the theory uncertainty. 
Figure~\ref{fig:SIAN4LLvsNLL} shows more clearly the comparison of the higher order resummations to the NLL order resummation. 
{For $x=0.9$ at the central scale, the N$^2$LL result modifies NLL result by $9.536\%$, the N$^3$LL result modifies N$^2$LL result by $2.354\%$, and the N$^4$LL result modifies N$^3$LL result by $0.504\%$}.

\subsection{Fixed-Order Expansion to N$^3$LO}

One-loop and two-loop results for the fragmentation coefficient function have been calculated in \cite{Rijken:1996vr,Rijken:1996ns,Rijken:1996npa,Mitov:2006wy,Mitov:2006ic}. The large-$x$ limit has the following form 
\beeq
C_{\rm SIA}[q] &=& \mathL_{-1} + a_s C_F \left[4 \mathL_1  -3 \mathL_0  +  \left(8 \zeta_2-9\right) \mathL_{-1}\right]   
\nn\\
%%%%
 && + a_s^2 \bigg( C_A C_F \bigg[-\frac{22}{3} \mathL_2 +\left(\frac{367}{9}-8\zeta_{2} \right) \mathL_1 + \bigg( \frac{44\zeta_{2}}{3}+40\zeta_{3}-\frac{3155}{54} \bigg)  \mathL_0  
 \nn\\
%%%%
 && \qquad
+
 \left(-\frac{49 \zeta_2^2}{5}+\frac{215 \zeta_2}{3}+\frac{140
   \zeta_3}{3}-\frac{5465}{72}\right) \mathL_{-1}
 \bigg] 
  \nn\\
%%%%
  &&  +C_F n_f \bigg[ \frac{4}{3} \mathL_2 -\frac{58}{9} \mathL_1  + \left( -\frac{8\zeta_{2}}{3}+\frac{247}{27}  \right) \mathL_0
 +\left(-\frac{38 \zeta_2}{3}+\frac{4
   \zeta_3}{3}+\frac{457}{36}\right) \mathL_{-1}
  \bigg] \bigg) 
 \nn\\
%%%%
  &&  +C_F^2 \bigg[8 \mathL_3 -18 \mathL_2 +\left( 16\zeta_{2}-27 \right)\mathL_1 + \left(-8 \zeta_3 +\frac{51}{2} \right) \mathL_0 
   \nn\\
%%%%
 && \qquad
+ \left(30 \zeta_2^2-39 \zeta_2-78   \zeta_3+\frac{331}{8}\right) \mathL_{-1}
\bigg]  
\, + \, \mathcal{O}(a_s^3) \, ,
\eeeq
where we have used the short-hand notations
\beq
\mathL_{n} \,  = \, \left[\frac{\ln^{n}(1-x)}{1-x}\right]_+ \, , \qquad \mathL_{-1} \, = \, \delta(1-x)  \, .
\eeq
All the terms at the large-$x$ limit agree with our resummed result. 
Furthermore, we present the three-loop result at the leading power of large-$x$ limit, including the $\delta(1-x)$ terms, predicted from the resummation formula. 
\beeq
C_{\rm SIA}^{(3)}[q] &=& 
C_F^3 \bigg[	8 \mathL_5-30 \mathL_4-36 \mathL_3 +	
   \bigg(72 \zeta_2+16 \zeta_3+\frac{279}{2}\bigg) \mathL_2
\nn\\
%%%%
&&
+ \bigg(-\frac{104 \zeta_2^2}{5}-84 \zeta_2-360 \zeta_3+\frac{187}{2}\bigg) \mathL_1
\nn\\
%%%%
&&
+  \bigg( 6 \zeta_2^2-64 \zeta_2 \zeta_3-123 \zeta_2+274 \zeta_3+432
   \zeta_5-\frac{1001}{8} \bigg) \mathL_0
   \nn\\
%%%%
&&
+
\bigg(
\frac{26864 \zeta_2^3}{315}-\frac{1248 \zeta_2^2}{5}-128 \zeta_2 \zeta_3-\frac{200 \zeta_2}{3}-\frac{304 \zeta_3^2}{3}-318 \zeta_3+1240 \zeta_5-\frac{7255}{24}
 \bigg) \mathL_{-1}
    \bigg]
\nn\\
%%%%
&+& 
C_F^2 C_A \bigg[-\frac{220}{9} \mathL_4 +\bigg(\frac{1732}{9}-32 \zeta_2\bigg) \mathL_3 
+   \bigg(\frac{460 \zeta_2}{3}+240 \zeta_3-\frac{8425}{18}\bigg)\mathL_2
\nn\\
%%%%
&&
+ \bigg(-\frac{196  \zeta_2^2}{5}+\frac{416 \zeta_2}{3}-\frac{160 \zeta_3}{3}-\frac{5563}{18}\bigg)  \mathL_1 
\nn\\
%%%%
&&
+ \bigg(39 \zeta_2^2+80 \zeta_2 \zeta_3-\frac{4627 \zeta_2}{27}-\frac{3304
   \zeta_3}{9}-120 \zeta_5+\frac{16981}{24}\bigg) \mathL_0 
   \nn\\
%%%%
&&
+
\bigg(
    -\frac{8072 \zeta_2^3}{63}+\frac{77732 \zeta_2^2}{135}+\frac{7900 \zeta_2 \zeta_3}{9}-\frac{8498 \zeta_2}{27}+\frac{1016 \zeta_3^2}{3}
       \nn\\
%%%%
&& \quad
-\frac{6419 \zeta_3}{3} -\frac{4952 \zeta_5}{9}+\frac{9161}{12}
\bigg) \mathL_{-1}
    \bigg]
\nn\\
%%%%
&+&
C_F^2 n_f \bigg[ \frac{40}{9} \mathL_4 -\frac{280}{9}  \mathL_3 +\bigg(\frac{683}{9}-\frac{64 \zeta_2}{3}\bigg) \mathL_2 
+  \bigg(-\frac{32 \zeta_2}{3}+\frac{112 \zeta_3}{3}+\frac{83}{9}\bigg) \mathL_1
\nn\\
%%%%
&&
+ \bigg( -16 \zeta_2^2+\frac{790 \zeta_2}{27}-60 \zeta_3-\frac{2003}{108} \bigg) \mathL_0	
   \nn\\
%%%%
&&
+
\bigg(
    -\frac{10352 \zeta_2^2}{135}-\frac{1072 \zeta_2 \zeta_3}{9}+\frac{512 \zeta_2}{27}+\frac{1348 \zeta_3}{3}-\frac{592 \zeta_5}{9}-\frac{341}{36}
\bigg) \mathL_{-1}
    \bigg]
\nn\\
%%%%
&+&
 C_A^2 C_F \bigg[ \frac{484}{27}  \mathL_3  +\bigg(\frac{88 \zeta_2}{3}-\frac{4649}{27}\bigg) \mathL_2 + \bigg(\frac{176 \zeta_2^2}{5}-\frac{680 \zeta_2}{3}-264   \zeta_3+\frac{50689}{81}\bigg) \mathL_1
\nn\\
%%%%
&& 
+ \bigg(  -\frac{652 \zeta_2^2}{15}-\frac{176 \zeta_2 \zeta_3}{3}+\frac{32126
   \zeta_2}{81}+\frac{21032 \zeta_3}{27}-232 \zeta_5-\frac{599375}{729}  \bigg) \mathL_0 
   \nn\\
%%%%
&&
+
\bigg( 
\frac{15704 \zeta_2^3}{315}-\frac{1276 \zeta_2^2}{135}-516 \zeta_2 \zeta_3+\frac{80929 \zeta_2}{81}-\frac{248 \zeta_3^2}{3}+\frac{105712 \zeta_3}{81}
   \nn\\
%%%%
&& \quad
-\frac{416 \zeta_5}{3}-\frac{1909753}{1944}   
   \bigg) \mathL_{-1}
    \bigg]
\nn\\
%%%%
&+&
C_F n_f C_A \bigg[ -\frac{176}{27} \mathL_3 +\bigg(\frac{1552}{27}-\frac{16 \zeta_2}{3}\bigg) \mathL_2 +  \bigg(\frac{512 \zeta_2}{9}+16 \zeta_3-\frac{15062}{81}\bigg) \mathL_1 
\nn\\
%%%%1
&&
+ \bigg( \frac{208 \zeta_2^2}{15}-\frac{9920 \zeta_2}{81}-\frac{776
   \zeta_3}{9}+\frac{160906}{729} \bigg) \mathL_0
   \nn\\
%%%%
&&
+
\bigg( 
-\frac{3208 \zeta_2^2}{135}+\frac{232 \zeta_2 \zeta_3}{3}-\frac{29650 \zeta_2}{81}-\frac{18314 \zeta_3}{81}+\frac{8 \zeta_5}{3}+\frac{142883}{486}
   \bigg) \mathL_{-1}
    \bigg]
\nn\\
%%%%
&+&
C_F n_f^2 \bigg[ \frac{16 \mathL_3}{27}-\frac{116 \mathL_2}{27} + \bigg(\frac{940}{81}-\frac{32 \zeta_2}{9}\bigg) \mathL_1 
+ \bigg( \frac{232 \zeta_2}{27}-\frac{32 \zeta_3}{27}-\frac{8714}{729}  \bigg) \mathL_0
   \nn\\
%%%%
&&
+
\bigg( 
\frac{112 \zeta_2^2}{27}+\frac{764 \zeta_2}{27}-\frac{152 \zeta_3}{81}-\frac{9517}{486}
\bigg) \mathL_{-1}
    \bigg]
   \nn\\
%%%%%
&+&
{N_{F,V}} \frac{d_F^{abc}d_F^{abc} }{N_F} \left(
-\frac{32 \zeta_2^2}{5}+160 \zeta_2+\frac{224 \zeta_3}{3}-\frac{1280 \zeta_5}{3}+64
\right) \mathL_{-1} \, ,
\eeeq
where the color structures are collected in Eq.~(\ref{eq:casimirs}) in Appendix. The $\delta(1-x)$ terms at three-loop contain the contribution from the three-loop jet function constant term calculated in \cite{Bruser:2018rad} and given in  Eq.~(\ref{three-loop-jet-constant-quark}), as well as the three-loop hard function constant term obtained from the form factor \cite{Gehrmann:2006wg,Moch:2005tm,Baikov:2009bg,Lee:2010cga,Gehrmann:2010ue} given in Eq.~(\ref{hard_fn_three_loop}). The term proportional to $N_{F,V}$ comes from a new flavour structure in the so-called singlet diagrams where the virtual gauge boson does not couple to the final state quarks, appearing in the calculations of the quark form factor from the three-loop order \cite{Gehrmann:2006wg,Moch:2005tm,Baikov:2009bg,Lee:2010cga,Gehrmann:2010ue}. 
In the current case with purely electromagnetic interaction via a virtual photon, it is defined as the weighted sum of the quark charge, commonly denoted as
\beq
\label{def:NFV}
N_{F,V} = N_{F,\gamma} \equiv \frac{1}{e_q} \sum _{q^{\prime}} e_{q^{\prime}}  \,,
\eeq
where the sum is over the light-flavour quarks running in a quark loop, and $e_q$ is the electric charge of the external quarks present in the Born level diagram. 
This factor should be considered symbolic, as its value cannot be given in the quark fragmentation coefficient function in general due to the fact that the fragmentation functions for different quark flavours to a given hadron are not the same, thus giving different contributions. However, this does not cause any practical issues since such a contribution can be taken into account unambiguously once the set of fragmentation functions is specified. 
Such a term from the new flavour structure only leads to a very small contribution in the $\gamma^* \to q \bar q$ case from N$^4$LL order. In the results presented in Figure~\ref{fig:SIAN4LL_Quark}, its effect is below $0.1\%$ on the N$^4$LL correction, with N$^3$LL subtracted, and below $0.001\%$ on the full N$^4$LL result at $x = 0.9$. 
Such a new flavour structure does not appear in the $H\to gg$ and $H \to b \bar b$ cases discussed in the next section.

\subsection{Comparison with Moment-Space Result}

In this subsection, we perform the resummation in moment space and cross check with the momentum-space result. 
The functions $A_q$ and $B_q$ determine the resummation exponent, where $A_q$ is the universal cusp anomalous dimension and  $B_q$ describes the collinear emissions. 
Following the discussion relating the momentum-space resummation with the moment-space resummation from \cite{Becher:2006mr,Becher:2006nr}, $B_q$ can be related to the SCET jet function, its anomalous dimension and the cusp anomalous dimension as  
\begin{equation}
\label{Bq_SCET_Jet}
e^{\gamma_E\nabla}\,\Gamma(1+\nabla)\,B_q(\alpha_s)
= \gamma^J(\alpha_s) + \nabla\,\ln\widetilde J(0,\mu)
- \left[ e^{\gamma_E\nabla}\,\Gamma(\nabla) 
- \frac{1}{\nabla} \right] \Gamma_{\rm cusp}(\alpha_s) \,,
\end{equation}
where the differential operator $\nabla$ is defined as
\begin{equation}
   \nabla = \frac{d}{d\ln\mu^2}
   = \frac{\beta(\alpha_s)}{2}\,\frac{\partial}{\partial\alpha_s} \,.
\end{equation}
The fact that the collinear degrees of freedom are described by the same jet function and that the cusp anomalous dimension is universal implies that the anomalous dimension $B_{q}$ is the same in both space-like DIS and time-like SIA cases.  
Defining the expansion coefficients in the coupling constant as
\beq
A_q =  \sum_{n=1} a_s^n  A_{q,n} \qquad \text{and} \qquad B_q = \sum_{n=1}  a_s^{n} \,B_{q,n} \, ,
\eeq
the N$^4$LL resummation requires the expansion coefficients of $B_{q}$ to the 4-th order in the coupling constant. 
Using the formula in Eq.~(\ref{Bq_SCET_Jet}), we then find the analytic expressions for the required coefficients of $B_{q}$ 
\begin{eqnarray}
   B_{q,1} &=& 
   -3C_F  \,, \\
   %%%%
    \nonumber\\
   %%%%
   B_{q,2}
    &=&  C_F^2 \left(12 \zeta_2-24 \zeta_3-\frac{3}{2}\right) 
    + C_F C_A \left(\frac{44 \zeta_2}{3}+40 \zeta_3-\frac{3155}{54}\right)  
    + C_F n_f  \left(\frac{247}{27}-\frac{8 \zeta_2}{3}\right)  , \qquad \\
   %%%%
    \nonumber\\
   %%%%
   B_{q,3} &=&  C_F^3   \left(-\frac{288 \zeta_2^2}{5}+32 \zeta_3 \zeta_2-18 \zeta_2-68 \zeta_3+240 \zeta_5-\frac{29}{2}\right) 
   \nonumber\\
   %%%%
   &+&C_F^2 C_A  \left(	-\frac{272 \zeta_2^2}{5}-16 \zeta_3 \zeta_2+287 \zeta_2-\frac{712 \zeta_3}{3}-120 \zeta_5-46	\right)
   \nonumber\\
   %%%%
   &+&
     C_F^2 n_f  \left(	-50 \zeta_2+\frac{32 \zeta_3}{9}+\frac{5501}{54}	\right)
   + C_F n_f^2   \left(	\frac{232 \zeta_2}{27}-\frac{32 \zeta_3}{27}-\frac{8714}{729}	\right)
\nonumber\\
   %%%%
   &+&
   C_F n_f  C_A   \left(	\frac{208 \zeta_2^2}{15}-\frac{9920 \zeta_2}{81}-\frac{776 \zeta_3}{9}+\frac{160906}{729}	\right)
   \nonumber\\
   %%%%
   &+&C_F C_A^2    \left(	-\frac{652 \zeta_2^2}{15}-\frac{176 \zeta_3 \zeta_2}{3}+\frac{32126 \zeta_2}{81}+\frac{21032 \zeta_3}{27}-232 \zeta_5-\frac{599375}{729}	\right)  \,,	\\
   %%%%
    \nonumber\\
   %%%%
       B_{q,4}  &=&
{\gamma_J^{(3)}[q]}   + 
 C_F n_f^3 \bigg(\frac{16 \zeta_2^2}{5}-\frac{1864 \zeta_2}{81}+\frac{752 \zeta_3}{243}+\frac{124903}{6561}\bigg)                                                                                                                                                                                                                                                                                                       \nn \\   &+& C_A^2 C_F^2 \bigg(-\frac{22352 \zeta_2^3}{45}-\frac{481637 \zeta_2^2}{135}-\frac{8932 \zeta_2 \zeta_3}{9}+\frac{307033 \zeta_2}{54}+\frac{616 \zeta_3^2}{3}+\frac{160994 \zeta_3}{27}
   \nn\\
%%%%
&&
-\frac{17776 \zeta_5}{9}-\frac{51325}{72}\bigg)                                                                                                                                                                                                                                                                                                       + C_A C_F^3 \bigg(\frac{77792 \zeta_2^3}{63}-880 \zeta_2^2+1496 \zeta_2 \zeta_3-\frac{2717 \zeta_2}{6}+\frac{2992 \zeta_3^2}{3}
   \nn\\
%%%%
&&
-4565 \zeta_3-616 \zeta_5-\frac{10769}{12}\bigg)                                                                                                                                                                                                                                                                                                       + C_A^2 C_F n_f \bigg(\frac{3536 \zeta_2^3}{189}+\frac{3398 \zeta_2^2}{15}+760 \zeta_2 \zeta_3-\frac{2896561 \zeta_2}{729}
   \nn\\
%%%%
&&
+\frac{3056 \zeta_3^2}{9}-\frac{662540 \zeta_3}{243}-\frac{1288 \zeta_5}{9}+\frac{15772433}{2916}\bigg)                                                                                                                                                                                                                                                                                                       \nn \\   &+& C_A C_F n_f^2 \bigg(-\frac{2464 \zeta_2^2}{45}-\frac{128 \zeta_2 \zeta_3}{3}+\frac{408683 \zeta_2}{729}+\frac{41788 \zeta_3}{243}+\frac{32 \zeta_5}{3}-\frac{1414898}{2187}\bigg)                                                                                                                                                                                                                                                                                                       \nn \\   &+& C_F^2 n_f^2 \bigg(-\frac{10664 \zeta_2^2}{135}-\frac{64 \zeta_2 \zeta_3}{9}+\frac{7178 \zeta_2}{27}+\frac{12652 \zeta_3}{81}+\frac{32 \zeta_5}{3}-\frac{102445}{243}\bigg)                                                                                                                                                                                                                                                                                                       \nn \\   &+& C_F^3 n_f \bigg(-\frac{14144 \zeta_2^3}{63}+\frac{1128 \zeta_2^2}{5}-272 \zeta_2 \zeta_3+\frac{19 \zeta_2}{3}-\frac{544 \zeta_3^2}{3}+806 \zeta_3+112 \zeta_5+\frac{482}{3}\bigg)                                                                                                                                                                                                                                                                                                       \nn \\   &+& C_A C_F^2 n_f \bigg(\frac{4064 \zeta_2^3}{45}+\frac{150022 \zeta_2^2}{135}+\frac{1976 \zeta_2 \zeta_3}{9}-\frac{70060 \zeta_2}{27}-\frac{112 \zeta_3^2}{3}-\frac{164470 \zeta_3}{81}
   \nn\\
%%%%
&&
+\frac{2704 \zeta_5}{9}+\frac{2465317}{972}\bigg)                                                                                                                                                                                                                                                                                                       + C_A^3 C_F \bigg(-\frac{19448 \zeta_2^3}{189}-\frac{7631 \zeta_2^2}{45}-\frac{8668 \zeta_2 \zeta_3}{3}+\frac{6106508 \zeta_2}{729}
   \nn\\
%%%%
&&
-\frac{16808 \zeta_3^2}{9}+\frac{2193533 \zeta_3}{243}+\frac{4180 \zeta_5}{9}-\frac{673606301}{52488}\bigg) \,,                                                                                                                                                                                                                                                                                                      
\end{eqnarray}
where the four-loop jet function anomalous dimension $ \gamma_J^{(3)}[q]$ contains a small uncertainty and is given in Eq.~(\ref{gamma_quark_jet_4_loop}) in Appendix~\ref{sec:ingredients}. 
These results agree with the calculation presented in \cite{Das:2019btv}, 
where the four-loop coefficient $B_{q,4}$ has now been determined more accurately above due to the calculations in \cite{Moult:2022xzt,Duhr:2022yyp,Duhr:2022cob}. 

The resummation exponent in Eq.~(\ref{coefficient_moment}) can be cast into \cite{Moch:2005ba,Moch:2009my,Das:2019btv}
\beq
G_N = \ln{\tilde N} g_1(\lambda) + g_2(\lambda) + a_s g_3(\lambda) + a_s^2 g_4(\lambda) + a_s^3 g_5(\lambda) + \cdots \, ,
\eeq
where $\lambda = \beta_0 a_s \ln{\tilde N}$ and ${\tilde N} = N \exp(\gamma_E)$. The treatment of absorbing the Euler-Mascheroni constant $\gamma_E$ into the resummation exponent is referred to as the $\tilde N$-exponential scheme. The resummation exponent is the same for SIA via $\gamma^* \to q \bar q $ and the DIS process as noted above. The terms $g_1$, $g_2$, $\dots$, $g_5$ will be needed for the LL, NLL, $\dots$, N$^4$LL resummation, respectively, and can be found in \cite{Das:2019btv}.

The $N$-independent $g_0$ multiplying the resummation exponent can also be expanded in terms of the coupling constant as
\beq
g_0 =  1+  \sum_{n=1}  a_s^{n} \,g_{0,n} \, .
\eeq
%%%%
The term $g_0$ can be determined in the SCET framework 
following \cite{Becher:2006nr,Becher:2006mr} and the above discussion of the factorization, 
\beeq
\label{g0_SCET}
g_0^{\rm SCET}(Q^2,\mu_f) &=& |C_V(Q^2,Q)|^2\,\,\widetilde J(0,Q)\, \exp\left[ \int_{\mu_f^2}^{Q^2}\!\frac{dk^2}{k^2}\,  \gamma^\phi(\alpha_s(k)) \right]
\, , 
\eeeq
where $C_V$ here is the time-like quark form factor. 
The four-loop coefficient $g_{0,4}$ contributes from the ninth column at N$^4$LL in the tower expansion \cite{Vogt:1999xa,Moch:2005ba,Moch:2009my,Das:2019btv} of the moment-space approach, and it can be extracted analytically up to the scale-independent constant term of $\widetilde J^{(4)}[q]$ at the four-loop order.
We note again that the resummation orders in the momentum-space and moment-space approaches are not fully equivalent. The N$^4$LL in the momentum-space approach does not require $\widetilde J^{(4)}[q]$ and yields the first eight columns when converted to the tower expansion. 
We then determine the analytic expression for $g_{0}$ at $\mu_f = Q$ to three-loop from the SCET framework as
\beeq
g_0^{\rm SCET}  &=& 1+ a_s
C_F (10 \zeta_2-9 )  
+ a_s^2 \bigg[
C_F^2 \bigg(\frac{244 \zeta_2^2}{5}-\frac{105 \zeta_2}{2}-66 \zeta_3+\frac{331}{8}\bigg)                                                                                                                                                                                                                                                                                                       
\\
   &+& C_F n_f \bigg(-\frac{143 \zeta_2}{9}+\frac{4 \zeta_3}{9}+\frac{457}{36}\bigg)                                                                                                                                                                                                                                                                                                       + C_A C_F \bigg(-\frac{69 \zeta_2^2}{5}+\frac{1657 \zeta_2}{18}+\frac{464 \zeta_3}{9}-\frac{5465}{72}\bigg)                                                                                                                                                                                                                                                                                                        
\bigg]    \nn \\       
&+& a_s^3 \bigg[
C_F n_f^2 \bigg(\frac{428 \zeta_2^2}{135}+\frac{2762 \zeta_2}{81}+\frac{80 \zeta_3}{81}-\frac{9517}{486}\bigg)                                                                                                                                                                                                                                                                                                       + C_A C_F^2 \bigg(-\frac{60142 \zeta_2^3}{315}+\frac{24433 \zeta_2^2}{27}
   \nn\\
%%%%
&&
+\frac{2540 \zeta_2 \zeta_3}{3}-\frac{50681 \zeta_2}{108}+\frac{536 \zeta_3^2}{3}-\frac{49346 \zeta_3}{27}-\frac{3896 \zeta_5}{9}+\frac{9161}{12}\bigg)                                                                                                                                                                                                                                                                                                       
+ C_A^2 C_F \bigg(\frac{21248 \zeta_2^3}{315}
   \nn\\
%%%%
&&
-\frac{13309 \zeta_2^2}{135}-\frac{6008 \zeta_2 \zeta_3}{9}+\frac{70849 \zeta_2}{54}-\frac{248 \zeta_3^2}{3}+\frac{115010 \zeta_3}{81}-\frac{416 \zeta_5}{3}-\frac{1909753}{1944}\bigg)                                                                                                                                                                                                                                                                                                       \nn \\   &+& C_F^2 n_f \bigg(-\frac{16742 \zeta_2^2}{135}-104 \zeta_2 \zeta_3+\frac{1273 \zeta_2}{54}+\frac{10766 \zeta_3}{27}-\frac{784 \zeta_5}{9}-\frac{341}{36}\bigg)                                                                                                                                                                                                                                                                                                       \nn \\   &+& C_A C_F n_f \bigg(-\frac{556 \zeta_2^2}{135}+\frac{800 \zeta_2 \zeta_3}{9}-\frac{37181 \zeta_2}{81}-\frac{21418 \zeta_3}{81}+\frac{8 \zeta_5}{3}+\frac{142883}{486}\bigg)                                                                                                                                                                                                                                                                                                       \nn \\   &+& C_F^3 \bigg(\frac{56528 \zeta_2^3}{315}-\frac{1701 \zeta_2^2}{5}-236 \zeta_2 \zeta_3-\frac{239 \zeta_2}{12}-\frac{176 \zeta_3^2}{3}-411 \zeta_3+1384 \zeta_5-\frac{7255}{24}\bigg)
 \nn \\   &+& 
 {N_{F,V}} \frac{d_F^{abc}d_F^{abc} }{N_F} \bigg(-\frac{32 \zeta_2^2}{5}+160 \zeta_2+\frac{224 \zeta_3}{3}-\frac{1280 \zeta_5}{3}+64\bigg)                                                                                                                                                                                                                                                                                                                                                                                                                                                                                                                                                                                                           
\bigg] \,.
\eeeq
%%%%
The SCET result $g_0^{\rm SCET}$ should reproduce the $g_0$ term considered in moment space \cite{Moch:2005ba,Moch:2009my,Das:2019btv} without the $\gamma_E$ terms in the perturbative expansions \cite{Becher:2006nr,Becher:2006mr}. This corresponds to the coefficient $g_0$ in the $\tilde N$-exponential scheme for SIA via $\gamma^* \to q \bar q $, which are known to three-loop from \cite{Moch:2005ba,Moch:2009my,Das:2019btv},
where the new flavour structure proportional to $N_{F,V}$ above is denoted in another common notation by $fl_{11}^{\rm}$ in, for example, \cite{Larin:1996wd,Vermaseren:2005qc}. 
The $g_0$ term extracted from the two approaches agrees with each other after identifying $N_{F,V} = n_f \cdot fl_{11}$, which in turn confirms the expectation from \cite{Becher:2006nr,Becher:2006mr}.

\begin{table}[ht]
\begin{center}
\begin{tabular}{||c || c | c | c | c | c | c | c | c | c||} 
 \hline
 $k$ & $c_{k1}$ & $c_{k2}$ & $c_{k3}$ & $c_{k4}$ & $c_{k5}$ & $c_{k6}$ & $c_{k7}/10$ & $c_{k8}/10^2$   \\ [0.3ex] 
 \hline\hline 
1 & 2.66667 & 7.07848 & 13.1298 & -- & -- & -- & -- & -- \\ \hline

2 & 3.55556 & 25.6908 & 105.621 & 104.338 & 357.883 & -- & -- & -- \\ \hline

3 & 3.16049 & 43.3408 & 309.335 & 1016.50 & 2305.97 & 2090.11 & 933.218 & -- \\ \hline

4 & 2.10700 & 46.6020 & 514.068 & 3125.96 & 11774.1 & 23741.1 & 4593.23 & 433.659 \\ \hline

5 & 1.12373 & 36.4525 & 577.143 & 5393.82 & 32365.2 & 121093. & 28821.6 & 3936.75 \\ \hline

6 & 0.499436 & 22.3131 & 481.110 & 6314.54 & 55037.7 & 322832. & 125737. & 30122.9 \\ \hline

7 & 0.190261 & 11.1933 & 315.972 & 5515.83 & 65426.2 & 544828. & 318836. & 126533. \\ \hline

8 & 0.063420 & 4.75033 & 170.252 & 3808.07 & 58765.0 & 653202. & 530087. & 310424. \\ \hline

9 & 0.018791 & 1.74552 & 77.4998 & 2160.26 & 41980.1 & 597519. & 636593. & 508626. \\ \hline

10 & 0.005011 & 0.565212 & 30.4703 & 1035.67 & 24725.4 & 437349. & 588886. & 609666. \\ \hline

11 & 0.001215 & 0.163529 & 10.5267 & 428.658 & 12329.1 & 264924. & 438509. & 567621. \\ \hline

12 & 0.000270 & 0.042748 & 3.23980 & 155.750 & 5311.74 & 136185. & 271308. & 427808. \\ \hline

\end{tabular}
\vspace{-0.5cm}
\end{center}
\caption{Coefficient $c_{ka}$ of $a_s^k \ln^{2k-a+1}$$N$ to N$^4$LL from the moment-space resummation, with the number of light-flavour quarks $n_f=5$. 
The uncertainty from the anomalous dimension $B_{q, 4}$ starts from the eight column but is negligible. 
This table does not include the tiny contribution from the $N_{F,V}$ term as explained in the text below Eq.~(\ref{def:NFV}). 
The ninth column requires the 5-loop cusp anomalous dimension as well as the four-loop $N$-independent coefficient $g_{0,4}$, which is not yet available for the SIA process. 
\label{table:SIA-Moment-N4LL}
}
\end{table}

These known results are sufficient for determining the coefficients to the eighth column in the tower expansion, which are sufficient for cross check of the momentum-space result. 
More specifically, we expand our momentum-space results in the strong coupling constant $a_s$ and transform them into Mellin moment space. 
The large logarithmic terms in momentum space are transformed to the Mellin $N$-moment space as the following 
\beq
\left[\frac{\ln^{2k-1}(1-x)}{1-x}\right]_+	\quad \longrightarrow  \quad\; 
\sum_{a=0}^{2k} \tilde c_{ka} \ln^{a} N + \mathcal{O}\bigg(\frac{1}{N}\bigg) \, .
\eeq
For each $k$, its moment can be calculated by restoring the regulator and observing that 
\beq
\frac{1}{(1-x)^{1+\epsilon}} = - \frac{1}{\epsilon} \delta(1-x) + \sum_{m=0}^{\infty} (-1)^m \frac{\epsilon^m}{m!} \left[\frac{\ln^m(1-x)}{1-x} \right]_+ \, ,
\eeq
and the exponent of the logarithmic terms is the same as the exponent of the regulator $\epsilon$. Once the divergence has been regulated with $\epsilon$, we can change the order of the integration and the expansion, and calculate first the moment of the LHS directly as 
\beq
 \int_0^1 dx \frac{x^N}{(1-x)^{1+\epsilon}} = \frac{\Gamma[-\epsilon] \, \Gamma[1+N]}{\Gamma[1+N-\epsilon]} \, ,
\eeq
which can be expanded in $\epsilon\to 0 $ and $N\to \infty$. Matching the coefficient of $\epsilon$ at the given power yields the corresponding order of the moment needed.

We find small deviation from the result in \cite{Moch:2009my} at the sixth column from $a_s^5$ order. 
However, as claimed in \cite{Moch:2009my}, the first six columns of their result are exact up to numerical truncation. 
In particular, we find that the values of higher-order coefficients are sensitive to numerical accuracy.
Using the results for the ingredients determined more accurately, we cross-check our momentum-space result with the moment-space result for SIA coefficients in $e^+e^- \to \gamma^* \to q \bar q$. 

After performing moment-space resummation to high numerical accuracy, we find perfect agreement between the results from the momentum-space approach and the moment-space approach through to N$^4$LL, shown in Table~\ref{table:SIA-Moment-N4LL}.

\section{Resummation for  $H \to $ hadrons}
\label{sec:Higgs2Hadrons}

We now investigate the threshold contributions to the semi-inclusive distribution for $ H \to $ \emph{hadrons}. 
The loop-induced Higgs decay $H \to gg $ reaches a branching ratio of about $8\%$ \cite{Inami:1983,Spira:1995rr,Chetyrkin:1997iv,Baikov:2006ch}.
The Higgs decay $H \to b \bar b$ is the dominant Higgs boson decay with a branching ratio of about $58\%$ \cite{Chetyrkin:1996sr,Mondini:2019gid,ATLAS:2021tbi,CMS:2023tfj}. 
These processes constitute the main channels for $H$ decays to hadrons, given the situation that the masses of the first two generations of quarks are much smaller than the $b$-quark mass. 
The investigation of threshold resummation for $ H \to $ \emph{hadrons} has been relatively rare in the literature, see e.g. \cite{Idilbi:2005ni,AH:2019phz,Das:2024pac} for a relevant study of Higgs production in hadron-hadron collisions and  \cite{Corcella:2004xv,Blumlein:2006pj} for  $H \to b \bar b$ decays at N$^3$LL. This is mainly due to the reduced interest in phenomenology as the Higgs boson has not yet been produced at Leptonic colliders. 
%%%%%%
On the other hand, since photons and $Z$ bosons do not distinguish between quarks and antiquarks, $e^+e^- \to \gamma^* / Z \to $ \emph{hadrons} primarily constrains the combinations $D_q^h + D_{\bar q}^h$. Gluon fragmentation contributes only at higher order in perturbation theory or by scaling violations \cite{DELPHI:1999alp,ParticleDataGroup:2022pth}. 
Upon the call for precision Higgs study at leptonic colliders such as the CEPC \cite{CEPCStudyGroup:2018rmc,CEPCStudyGroup:2018ghi,CEPCStudyGroup:2023quu}, the FCC-ee \cite{FCC:2018byv, FCC:2018evy} and the ILC \cite{Behnke:2013xla,Bambade:2019fyw}, we perform the study of threshold resummation for the $ H \to gg$ and $ H \to b \bar b$ processes and emphasize that they serve as a promising candidate for extracting the gluon and heavy-quark fragmentation functions as well as for performing electroweak precision measurements. 

The factorization formula at the leading power shares the same structure as in the $\gamma^* \to $ \emph{hadrons} case discussed above. 
After presenting the factorization formula, we then perform the resummation to N$^4$LL for both $ H \to gg$ and $H \to b \bar b$ processes. 
We check the $ H \to gg$ results with fixed-order coefficient function calculated to NNLO in  \cite{Moch:2007tx,Almasy:2011eq}, 
and check $H \to b \bar b$ results with the N$^3$LL coefficients from \cite{Blumlein:2006pj}, finding perfect agreement in both cases.

\subsection{Resummation Formula}

The Higgs decays into gluon-gluon mainly through a top-quark loop. The process is conjugate to the Higgs hadroproduction via gluon-gluon fusion, which dominates the production cross section at high energies. 
Therefore, the discussion follows closely with the threshold resummation for Higgs hadroproduction in \cite{Idilbi:2005ni}, which in turn shares great similarities to the Drell-Yan production of $W/Z$ bosons. 
The factorization formula is however distinct for the SIA cross section in $H \to gg$ from the above cases due to different kinematics. The factorization formula in fact has the same structure as the $\gamma^* \to q \bar q$ discussed above, 
\begin{align}
\label{Def_SIA_Higgs}
F^{(\SIA)}_{H\to {hadrons}}(x,Q^{2})  &= \frac{1}{\sigma_{0}}\frac{d\sigma_{H\to {hadrons}}^{(\SIA)}(x,Q^2)}{dx}  \\
&= \int_x^1 \frac{d\eta}{\eta}   
\;C^{(\SIA)}_{H \to \Psi}(\frac{x}{\eta}, Q^2,\mu^2,\mu_F^2)
\;D_{\Psi}(\eta,\mu_F^2)  + \mathcal{O}\bigg(\frac{\LQCD}{Q}\bigg) \,, \qquad \nn
\end{align}
where $\Psi = gg $ or $b \bar b$, with $D_{\Psi}$ now the gluon fragmentation or the $b$-quark fragmentation function, respectively. 
The coefficient function $C^{(\SIA)}_{H \to \Psi}$ for $\Psi = gg $ or $b \bar b$ in the large $x$-limit can be factorized as Eq.~(\ref{SIAfact}) into the hard function extracted from the $H \to gg $ or $H \to b \bar b $ form factor and the gluon or quark jet function, respectively. 
The perturbative ingredients available are sufficient for the resummation to N$^4$LL, with the 5-loop cusp estimated with uncertainty, given in Appendix~\ref{sec:ingredients}.

The $H \to gg$ process is described by the Higgs Effective Field Theory (HEFT) in the infinite top-quark-mass limit.  The HEFT Lagrangian is given by 
\beeq
\mathL_{\rm HEFT} = -\frac{\lambda_0}{4} H F^{\mu \nu}_{a} F_{a, \mu \nu}  \, .
\eeeq
The bare coupling $\lambda_0$ is renormalized with the renormalization constant $Z_{\lambda}$ as 
\beeq
\lambda_0 = Z_{\lambda} \,  \lambda \, ,
\eeeq
and the $H \to gg$ form factor then receives the renormalization scale dependence. 

\subsection{Resummation Result to N$^4$LL}

\subsubsection{$H \to gg$}

We present the numerical result for the threshold resummation to N$^4$LL for $H \to gg$ process. The perturbative convergence with decreasing scale uncertainty only takes place after we include the scale-dependence from the HEFT coupling renormalization. 
We therefore normalize the distribution $H \to gg$ with the decay rate, which has been calculated to N$^4$LO \cite{Inami:1982xt,Chetyrkin:1997iv,Baikov:2006ch,Moch:2007tx,Davies:2017rle,Herzog:2017dtz,
Spiridonov:1984br, Kramer:1996iq, Chetyrkin:1997un, Schroder:2005hy, Chetyrkin:2005ia,Liu:2015fxa,Chen:2023fba}.
For the N$^4$LL resummation, we use the N$^3$LO result with its scale-dependence, 
\beeq
K_{H \to gg} &=&1 +a_s \left[C_A \left(\frac{22 L_{\mu}}{3}+\frac{73}{3}\right)+ n_f \left(-\frac{4 L_{\mu}}{3}-\frac{14}{3}\right) \right]
    \\
%%%%
&+&
a_s^2 \bigg[C_A^2 \left(\frac{121
   L_{\mu}^2}{3}+313 L_{\mu}-\frac{242 \zeta_2}{3}-110 \zeta_3+\frac{37631}{54}\right)
   \nn \\
%%%%
&+&
   C_A n_f
   \left(-\frac{44 L_{\mu}^2}{3}-\frac{340 L_{\mu}}{3}+\frac{88 \zeta_2}{3}-4
   \zeta_3-\frac{6665}{27}\right)+C_F n_f \left(-8 L_{\mu}+24
   \zeta_3-\frac{131}{3}\right)
   \nn \\
%%%%
&+&
   n_f^2 \left(\frac{4 L_{\mu}^2}{3}+\frac{28 L_{\mu}}{3}-\frac{8
   \zeta_2}{3}+\frac{508}{27}\right)\bigg]
   \nn \\
%%%%
&+&
a_s^3 \bigg[C_A^3 \bigg(\frac{5324 L_{\mu}^3}{27}+\frac{22528 L_{\mu}^2}{9}+L_{\mu} \left(-\frac{10648
   \zeta_2}{9}-\frac{4840 \zeta_3}{3}+\frac{965285}{81}\right)
   \nn \\
%%%%
&&
\quad\quad\quad -\frac{45056
   \zeta_2}{9}-\frac{178156 \zeta_3}{27}+\frac{3080
   \zeta_5}{3}+\frac{15420961}{729}\bigg)
   \nn \\
%%%%
&+&
   C_A^2 n_f \bigg(-\frac{968 L_{\mu}^3}{9}-\frac{4042
   L_{\mu}^2}{3}+L_{\mu} \left(\frac{1936 \zeta_2}{3}+\frac{704   \zeta_3}{3}-\frac{170263}{27}\right)
   \nn \\
%%%%
&&
\quad\quad\quad   + \frac{8084 \zeta_2}{3}+\frac{9772 \zeta_3}{9}-\frac{80   \zeta_5}{3}-\frac{2670508}{243}\bigg)
   \nn \\
%%%%
&+&
   C_A C_F n_f \left(-\frac{286 L_{\mu}^2}{3}+L_{\mu}
   \left(352 \zeta_3-\frac{8569}{9}\right)+\frac{572 \zeta_2}{3}+1364 \zeta_3+160
   \zeta_5-\frac{23221}{9}\right)
   \nn \\
%%%%
&+&
   C_A n_f^2 \bigg(\frac{176 L_{\mu}^3}{9}+\frac{692 L_{\mu}^2}{3}+L_{\mu}
   \left(-\frac{352 \zeta_2}{3}+\frac{32 \zeta_3}{3}+\frac{9187}{9}\right)
   \nn \\
%%%%
&&
\quad\quad\quad -\frac{1384 \zeta_2}{3}+\frac{56 \zeta_3}{9}+\frac{413308}{243}\bigg)
%%%%
   + C_F^2 n_f \left(6 L_{\mu}+192
   \zeta_3-320 \zeta_5+\frac{221}{3}\right)
   \nn \\
%%%%
&+&
   C_F n_f^2 \left(\frac{52 L_{\mu}^2}{3}+L_{\mu}
   \left(-64 \zeta_3+ \frac{1534}{9}\right)-\frac{104 \zeta_2}{3}-240
   \zeta_3+440\right) 
   \nn \\
%%%%
&&
   +n_f^3 \left(-\frac{32 L_{\mu}^3}{27}-\frac{112 L_{\mu}^2}{9}+L_{\mu} \left(\frac{64
   \zeta_2}{9}-\frac{4064}{81}\right)+\frac{224 \zeta_2}{9}+\frac{64
   \zeta_3}{27}-\frac{57016}{729}\right)\bigg] \, , \nn \quad 
\eeeq
where $L_{\mu} = \ln{(\mu^2/Q^2)}$. The order of the $K$ factor used for the normalization is in accordance with the order of the hard function used in the resummation result. 

\begin{figure}[ht]
	\centering
	\setlength{\abovecaptionskip}{-0.0cm}
	\setlength{\belowcaptionskip}{-0.0cm} 
	\includegraphics[width=12.2cm]{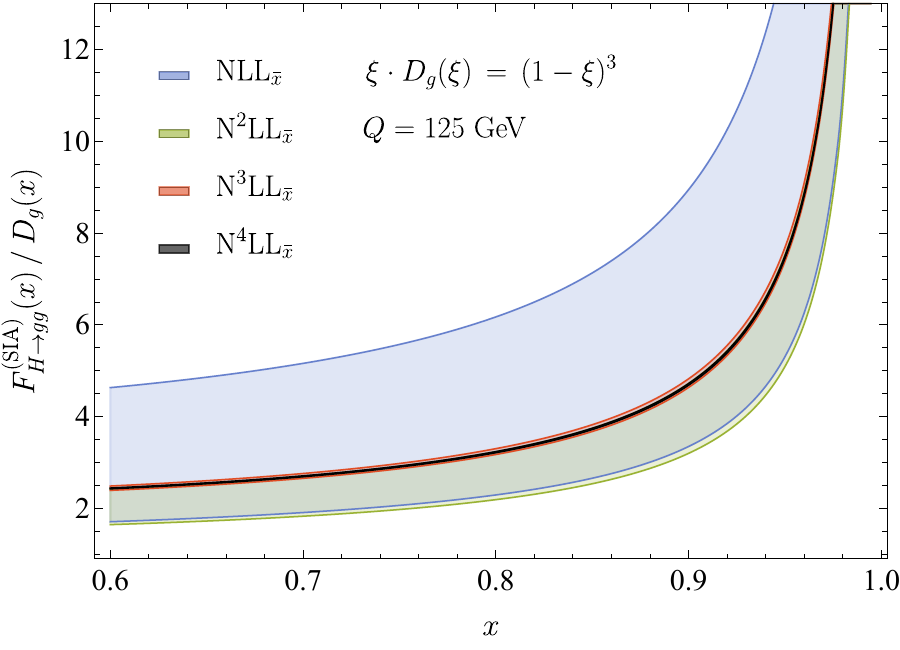}\\
	\caption{SIA cross section for $H \to gg$ to N$^4$LL in the large-$x$ limit, with a typical form for the fragmentation function $D_g^{}(\xi) = (1-\xi)^3/\xi$ near the end-point. Bands represent variation of the hard scale $\mu_H \in [Q/5,5Q]$. We set the number of light flavour quarks $n_f=5$.}
	\label{fig:SIAN4LLg}
\end{figure}

\begin{figure}[ht]
	\centering
	\setlength{\abovecaptionskip}{-0.0cm}
	\setlength{\belowcaptionskip}{-0.0cm} 
	\includegraphics[width=12.2cm]{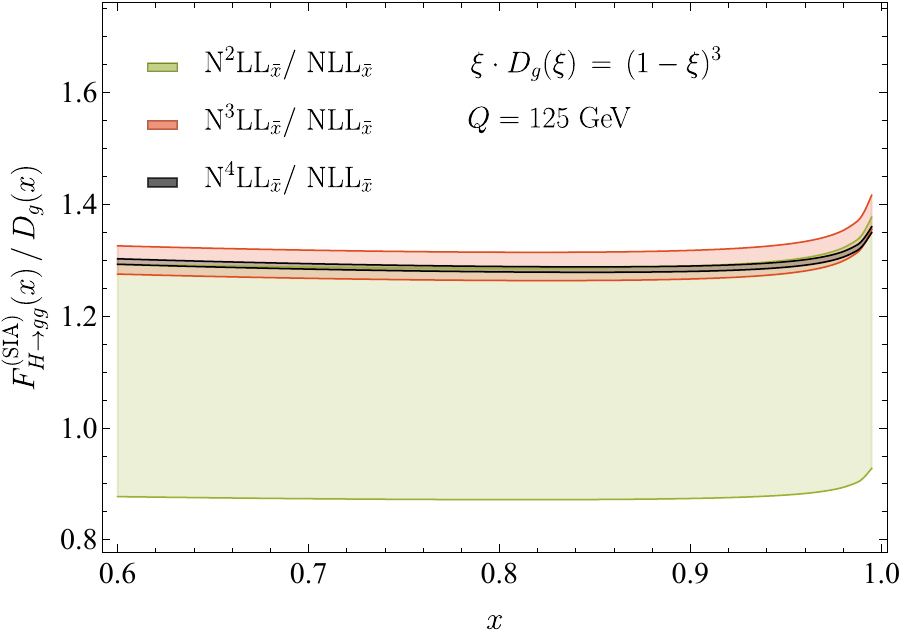}\\
	\caption{The ratio of N$^n$LL result and NLL result for $H \to gg$, with bands representing the variation of the hard scale $\mu_H \in [Q/5,5Q]$ for the numerators.}
	\label{fig:SIAN4LLvsNLLg}
\end{figure}

Figure~\ref{fig:SIAN4LLg} shows the predictions from NLL to N$^4$LL at the center of mass energy of 125 GeV, where good perturbative convergence is observed through to N$^4$LL. 
The hard scale $\mu_H = Q$ is varied around the canonical scale by a factor of 5 as the above case as an estimate of the theory uncertainty. 
The scale uncertainty decreases from NLL to N$^4$LL after the normalization with $K_{H \to gg}$. 
Figure~\ref{fig:SIAN4LLvsNLLg} shows more clearly the comparison of the higher order resummations. 
{For $x=0.9$ at the central scale, the N$^2$LL result modifies NLL result by $21.09\%$, the N$^3$LL result modifies N$^2$LL result by $4.587\%$, and the N$^4$LL result modifies N$^3$LL result by $1.068\%$}.

\subsubsection{$H \to b \bar b $}

The resummation formalism follows the same structure for the above two cases. The new ingredient needed is the $H \to b \bar b $ form factor, which has been calculated to three-loop in \cite{Gehrmann:2014vha}, and to four-loop in \cite{Chakraborty:2022yan}.
The anomalous dimension $\gamma_V$ for $H \to b \bar b $ is the same as the one for $\gamma^* \to q \bar q $. 
The jet function and the cusp anomalous dimension are also the same. 
We normalize the distribution with the $H \to b \bar b $ total cross section (inclusive decay rate) to N$^3$LO \cite{Chetyrkin:1996sr,Baikov:2005rw,Mondini:2019gid}, in order to show perturbative convergence. 
\beeq
K_{H \to b \bar b} &&= 1 +a_s C_F \left(6 L+17\right)
 \\ 
&&
+
a_s^2 \bigg[C_A
   C_F \left(11 L^2+\frac{284 L}{3}-22 \zeta_2-62
   \zeta_3+\frac{893}{4}\right)
\nn \\ 
&&
+
   C_F^2 \left(18 L^2+105 L-36
   \zeta_2-36 \zeta_3+\frac{691}{4}\right)
   \nn \\ 
&&
+C_F n_f \left(-2
   L^2-\frac{44 L}{3}+4 \zeta_2+8
   \zeta_3-\frac{65}{2}\right)\bigg]
\nn \\ 
&&
+ 
a_s^3 \bigg[C_A^2 C_F \bigg(\frac{242 L^3}{9}+\frac{3430
   L^2}{9}+L \left(-\frac{484 \zeta_2}{3}-\frac{1364
   \zeta_3}{3}+\frac{55112}{27}\right)
\nn \\ 
&&
\qquad\qquad
   -\frac{6860 \zeta_2}{9}-\frac{4658
   \zeta_3}{3}+\frac{100 \zeta_5}{3}+\frac{3894493}{972}\bigg)
\nn \\ 
&&
+ 
C_A
   C_F^2 \bigg(66 L^3+766 L^2+L \left(-396 \zeta_2-636
   \zeta_3+\frac{6183}{2}\right)
\nn \\ 
&&
\qquad\qquad
   -1532 \zeta_2-2178 \zeta_3+580
   \zeta_5+\frac{13153}{3}\bigg)
\nn \\ 
&&
+C_A C_F n_f
   \bigg(-\frac{88 L^3}{9}-\frac{1142 L^2}{9}+L \left(\frac{176
   \zeta_2}{3}+\frac{280 \zeta_3}{3}-\frac{16558}{27}\right)
\nn \\ 
&&
\qquad\qquad
   -\frac{48
   \zeta_2^2}{5}+\frac{2284 \zeta_2}{9}+\frac{704 \zeta_3}{3}+\frac{80
   \zeta_5}{3}-\frac{267800}{243}\bigg)
\nn \\ 
&&
+ 
   C_F^3 \bigg(36 L^3+324 L^2+L
   \left(-216 \zeta_2-216 \zeta_3+\frac{2433}{2}\right)
\nn \\ 
&&
\qquad\qquad
   -648 \zeta_2-956
   \zeta_3+360 \zeta_5+\frac{23443}{12}\bigg)
\nn \\ 
&&
+ 
   C_F^2 n_f
   \bigg(-12 L^3-130 L^2+L (72 \zeta_2+144 \zeta_3-562)
\nn \\ 
&&
\qquad\qquad
   +\frac{48
   \zeta_2^2}{5}+260 \zeta_2+520 \zeta_3-160
   \zeta_5-\frac{2816}{3}\bigg)
\nn \\ 
&&
+ 
   C_F n_f^2 \left(\frac{8
   L^3}{9}+\frac{88 L^2}{9}+L \left(-\frac{16 \zeta_2}{3}-\frac{32
   \zeta_3}{3}+\frac{1100}{27}\right)-\frac{176 \zeta_2}{9}-16
   \zeta_3+\frac{15511}{243}\right)\bigg]  \, .    \nn
\eeeq

\begin{figure}[ht]
	\centering
	\setlength{\abovecaptionskip}{-0.0cm}
	\setlength{\belowcaptionskip}{-0.0cm} 
	\includegraphics[width=12.2cm]{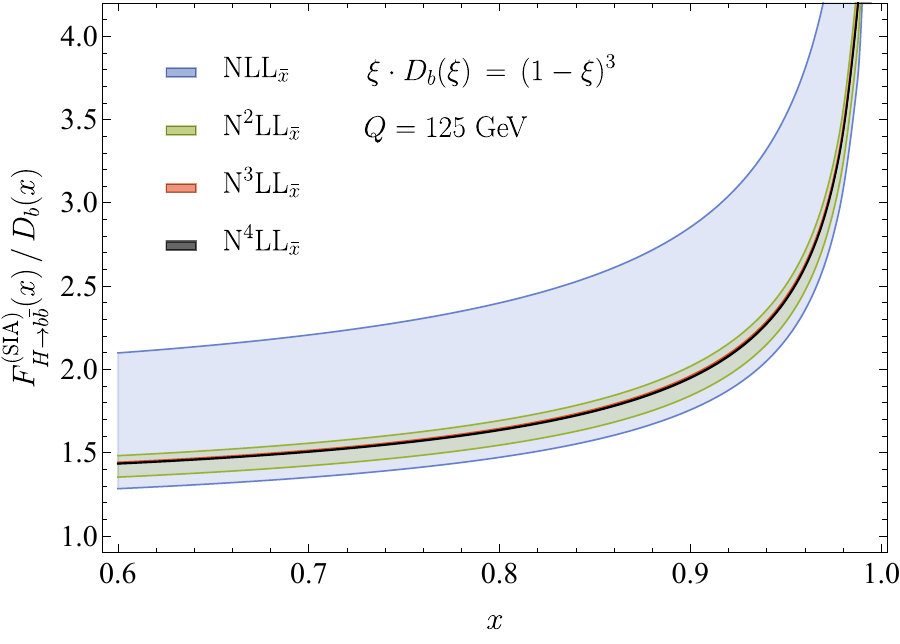}\\
	\caption{SIA cross section for $H \to b \bar b$ to N$^4$LL in the large-$x$ limit, with a typical form for the fragmentation function $D_b^{}(\xi) = (1-\xi)^3/\xi$ near the end-point. Bands represent variation of the hard scale $\mu_H \in [Q/5,5Q]$. We set the number of light flavour quarks $n_f=5$.}
	\label{fig:SIAN4b}
\end{figure}

\begin{figure}[ht]
	\centering
	\setlength{\abovecaptionskip}{-0.0cm}
	\setlength{\belowcaptionskip}{-0.0cm} 
	\includegraphics[width=12.2cm]{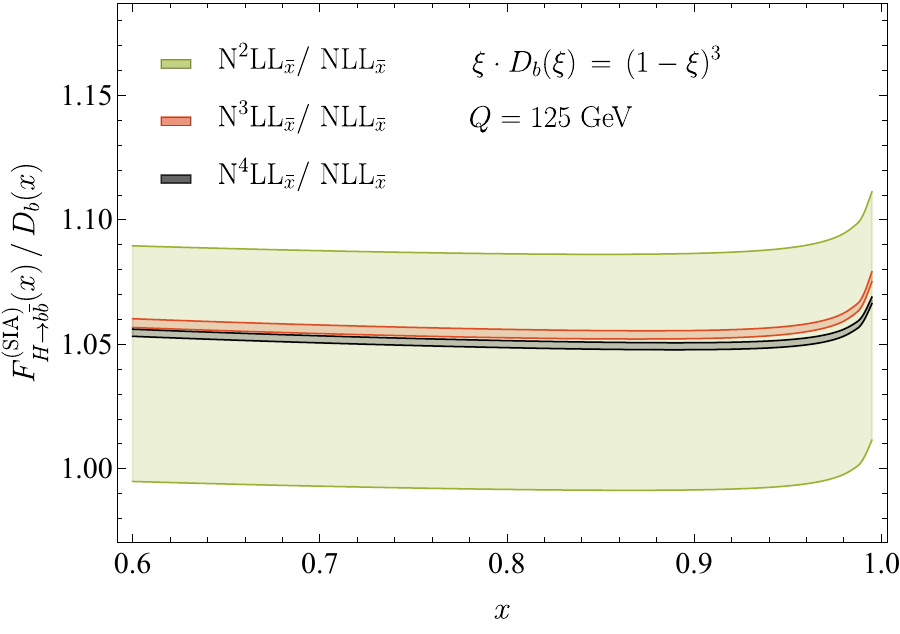}\\
	\caption{The ratio of N$^n$LL result and NLL result for $H \to b \bar b$, with bands representing the variation of the hard scale $\mu_H \in [Q/5,5Q]$ for the numerators.}
	\label{fig:SIAN4LLvsNLLb}
\end{figure}

Figure~\ref{fig:SIAN4b} shows the predictions from NLL to N$^4$LL for $H \to b \bar b$ with the center of mass energy $Q=125$ GeV, where good perturbative convergence is observed through to N$^4$LL, normalized with $K_{H \to b \bar b}$. 
The hard scale $\mu_H = Q$ is varied around the canonical scale by a factor of 5 as the above case, and the scale uncertainty decreases from NLL to N$^4$LL. 
Figure~\ref{fig:SIAN4LLvsNLLb} shows more clearly the comparison of the higher order resummations. 
{For $x=0.9$ at the central scale, the N$^2$LL result modifies NLL result by $4.308\%$, the N$^3$LL result modifies N$^2$LL result by $0.881\%$, and the N$^4$LL result modifies N$^3$LL result by $-0.156\%$}.

\subsection{Fixed-Order Expansion to N$^3$LO}

\subsubsection{$H \to gg$}

The fragmentation coefficient function for $H \to gg$ decay has been calculated to NNLO in \cite{Moch:2007tx,Almasy:2011eq}, with the leading power of large-$x$ limit given by
\beeq
C_{\rm SIA}^{(1)}[g] &=&  C_A \left[4 \mathL_1 -\frac{11}{3} \mathL_0 
+
\bigg( 8 \zeta_{2}+\frac{67}{9} \bigg) \mathL_{-1}
\right]
+n_f \left[\frac{2}{3}  \mathL_0  -\frac{10}{9} \mathL_{-1}  \right]
\, ,   \\
%%%%
C_{\rm SIA}^{(2)}[g] &=&  C_A^2 \bigg[ 8 \mathL_3-\frac{88}{3} \mathL_2 +  \left(8 \zeta_2+\frac{778}{9}\right) \mathL_1+ \bigg( \frac{44 \zeta_2}{3}+32 \zeta_3-\frac{2570}{27} \bigg) \mathL_0  
\nn\\
%%%%
&&
+
\bigg(  \frac{101 \zeta_{2}^2}{5}+\frac{830 \zeta_{2}}{9}-\frac{242 \zeta_3}{3}+\frac{30425}{162} \bigg) \mathL_{-1}
\bigg]     
  \nn\\
%%%%
&+&
C_A n_f
   \left[\frac{16 }{3} \mathL_2 -\frac{56}{3} \mathL_1 + \bigg(-\frac{8 \zeta_2}{3}+\frac{224}{9} \bigg) \mathL_0 
   +
\bigg( -8 \zeta_{2}-\frac{28 \zeta_3}{3}-\frac{4112}{81} \bigg) \mathL_{-1}
 \right]
   \nn\\
%%%%
&+&
C_F n_f \left[   2  \mathL_0   +
\bigg(  24 \zeta_3-\frac{63}{2}\bigg) \mathL_{-1}
\right]
+n_f^2 \left[\frac{8}{9} \mathL_1 -\frac{40}{27} \mathL_0 
 +
\bigg( \frac{100}{81}-\frac{8 \zeta_{2}}{9} \bigg) \mathL_{-1}
\right] . \qquad
\eeeq
All the terms to NNLO can be reproduced from the factorization formula. The three-loop result can be predicted from the resummation formula, 
\beeq
C_{\rm SIA}^{(3)}[g] &=& C_A^3 \bigg[ 8\mathL_5-\frac{550\mathL_4}{9}    +(340-32 \zeta_2) \mathL_3  +   
   \left(\frac{836 \zeta_2}{3}+256 \zeta_3-\frac{9623}{9}\right) \mathL_2
    \nn\\
%%%%
&&
  + \left(-\frac{124 \zeta_2^2}{5}-\frac{3716   \zeta_2}{9}-\frac{2816 \zeta_3}{3}+\frac{192268}{81}\right) \mathL_1
     \nn \\
%%%%
&&
  + \left(-\frac{11 \zeta_2^2}{15}-\frac{128 \zeta_2 \zeta_3}{3}+\frac{50858 \zeta_2}{81}+\frac{40454
   \zeta_3}{27}+80 \zeta_5-\frac{1616486}{729} \right) \mathL_0 
    \nn\\
%%%%
&&
  + \bigg(     
\frac{736 \zeta_2^3}{105}+\frac{721 \zeta_2^2}{27}-132 \zeta_2 \zeta_3+\frac{89950 \zeta_2}{81}+\frac{464 \zeta_3^2}{3}-\frac{90632 \zeta_3}{27}
    \nn\\
%%%%
&& \quad 
+\frac{4048 \zeta_5}{9}+\frac{14160613}{2916}
    \bigg) \mathL_{-1}
   \bigg]
    \nn\\
%%%%
&+&
C_A^2 n_f \bigg[  \frac{100\mathL_4}{9}-\frac{256\mathL_3}{3} + \left(\frac{3106}{9}-\frac{152 \zeta_2}{3}\right) \mathL_2 + \left(\frac{1696 \zeta_2}{9}+\frac{128
   \zeta_3}{3}-\frac{67730}{81}\right) \mathL_1
       \nn\\
%%%%
&&
  + \left(\frac{98 \zeta_2^2}{15}-\frac{22412 \zeta_2}{81}-\frac{1816
   \zeta_3}{9}+\frac{1234307}{1458} \right) \mathL_0 
    \nn\\
%%%%
&&
  + \bigg(     
\frac{11074 \zeta_2^2}{135}-232 \zeta_2 \zeta_3+\frac{2770 \zeta_2}{81}+\frac{6794 \zeta_3}{27}+\frac{1064 \zeta_5}{9}-\frac{1988899}{972}
    \bigg) \mathL_{-1}
   \bigg] 
    \nn\\
%%%%
&+&
C_A C_F n_f \bigg[16\mathL_2 + \left(128 \zeta_3-\frac{598}{3}\right)\mathL_1
 + \left( -\frac{32\zeta_2^2}{5}-8 \zeta_2-\frac{1624 \zeta_3}{9}+\frac{7810}{27} \right)\mathL_0   
    \nn\\
%%%%
&&
  + \bigg(     
\frac{176 \zeta_2^2}{45}+256 \zeta_2 \zeta_3-\frac{1472 \zeta_2}{9}+\frac{6664 \zeta_3}{9}+200 \zeta_5-\frac{411137}{324}
    \bigg) \mathL_{-1}
   \bigg] 
    \nn\\
%%%%
&+&
C_A n_f^2
   \bigg[ \frac{16 \mathL_3}{3}-\frac{292\mathL_2}{9} + \left(\frac{6652}{81}-\frac{176 \zeta_2}{9}\right)\mathL_1+ \left(\frac{344 \zeta_2}{9}-\frac{152
   \zeta_3}{27}-\frac{138493}{1458} \right) \mathL_0 
    \nn\\
%%%%
&&
  + \bigg(     
-\frac{592 \zeta_2^2}{45}-\frac{5732 \zeta_2}{81}+\frac{404 \zeta_3}{9}+\frac{174769}{972}
    \bigg) \mathL_{-1}
   \bigg]
    \nn\\
%%%%
&+&
C_F^2 n_f \left[ -\mathL_0      + \bigg(    \frac{740 \zeta_3}{3}-400 \zeta_5+\frac{751}{9}   \bigg) \mathL_{-1}     \right]
    \nn\\
%%%%
&+&
C_F n_f^2 \bigg[\frac{20\mathL_1}{3}+\left(\frac{80 \zeta_3}{3}-\frac{350}{9} \right)\mathL_0     
  + \bigg(     -\frac{32 \zeta_2^2}{45}-\frac{244 \zeta_2}{9}-\frac{1144 \zeta_3}{9}+\frac{28945}{162}  \bigg) \mathL_{-1}  
  \bigg]
    \nn\\
%%%%
&+&
   n_f^3 \bigg[\frac{8\mathL_2}{9}-\frac{80\mathL_1}{27} + \left(-\frac{16 \zeta_2}{9}+\frac{200}{81} \right)\mathL_0
   + \bigg(  \frac{80 \zeta_2}{27}-\frac{1000}{729}    \bigg) \mathL_{-1}  
   \bigg]  \, ,
\eeeq
%%%%
where the three-loop $\delta(1-x)$ terms include the contributions from the three-loop gluon jet function, including the
scale-independent term from \cite{Banerjee:2018ozf} given in Eq.~(\ref{three-loop-jet-constant-gluon}), and the three-loop $H \to g g $ form factor \cite{Gehrmann:2006wg,Moch:2005tm,Baikov:2009bg,Lee:2010cga,Gehrmann:2010ue} given in Eq.~(\ref{three_loop_form_factor_gluon}).

\subsubsection{$H \to b \bar b $}

The fragmentation coefficient function for $H \to b \bar b $ decay has been calculated to NLO in \cite{Corcella:2004xv}. 
The fixed-order expansion of coefficient function has also been presented in  \cite{Blumlein:2006pj} based on the N$^3$LL accuracy, which predicts the full large-$x$ terms to NNLO and partial results at N$^3$LO.
We present the large-$x$ coefficient function to N$^3$LO from the N$^4$LL prediction of the factorization formula. 
 We find all terms agree and provide the full N$^3$LO result at large-$x$. 
\beeq
C_{\rm SIA}^{(1)}[b] &=& C_F \left[ 4 \mathL_1-3\mathL_0 
+ \left( 8 \zeta_{2}+3  \right) \mathL_{-1}  
\right]   \, ,   \\ 
%%%%
C_{\rm SIA}^{(2)}[b] &=&  C_A C_F \bigg[
   -\frac{22 \mathL_2}{3} 
   +  \left(\frac{367}{9}-8 \zeta_2\right) \mathL_1
   +  \left(\frac{44 \zeta_2}{3}+40\zeta_3-\frac{3155}{54}\right) \mathL_0
   \nn  \\
%%%%
&& \quad
   +  \left( -\frac{49 \zeta_{2}^2}{5}+\frac{95 \zeta_{2}}{3}+\frac{32 \zeta_{3}}{3}+\frac{1691}{24} \right) \mathL_{-1}   
   \bigg]
\nn  \\
%%%%
&+&
   C_F^2 \bigg[
   8 \mathL_3 -18\mathL_2
   + (16 \zeta_2+21) \mathL_1
   +  \left(-8 \zeta_3-\frac{21}{2}\right) \mathL_0
   \nn  \\
%%%%
&& \quad
   +  \left( 30 \zeta_{2}^2+31 \zeta_{2}-78 \zeta_{3}+\frac{109}{8} \right) \mathL_{-1}
   \bigg]
\nn  \\
%%%%
&+&
   C_F n_f \left[
   \frac{4 \mathL_2}{3}-\frac{58\mathL_1}{9}
   +  \left(-\frac{8 \zeta_2}{3}+\frac{247}{27}\right) \mathL_0
   +  \left( -\frac{14 \zeta_{2}}{3}+\frac{4 \zeta_{3}}{3}-\frac{91}{12} \right) \mathL_{-1}
   \right]    \, , \qquad   \\
%%%%
C_{\rm SIA}^{(3)}[b] &=&   
   C_A^2 C_F \bigg[ \frac{484 \mathL_3}{27}
   + \left(\frac{88\zeta_2}{3}-\frac{4649}{27}\right) \mathL_2
   +  \left(\frac{176 \zeta_2^2}{5}-\frac{680\zeta_2}{3}-264 \zeta_3+\frac{50689}{81}\right) \mathL_1
\nn  \\
%%%%
&& \quad
   + \left(-\frac{652\zeta_2^2}{15}-\frac{176 \zeta_2 \zeta_3}{3}+\frac{32126\zeta_2}{81}
   +\frac{21032 \zeta_3}{27}-232 \zeta_5-\frac{599375}{729}
   \right) \mathL_0
\nn  \\
%%%%
&& \quad
   + \bigg(
\frac{15704 \zeta_2^3}{315}-\frac{3247 \zeta_2^2}{135}-576 \zeta_2 \zeta_3-\frac{9386 \zeta_2}{81}-\frac{248 \zeta_3^2}{3}+\frac{9115 \zeta_3}{81}
\nn  \\
&& \qquad
-\frac{56 \zeta_5}{3}+\frac{2757769}{1944}
   \bigg) \mathL_{-1}
   \bigg]
\nn  \\
%%%%
&+&
   C_A C_F^2 \bigg[
   -\frac{220 \mathL_4}{9} +
    \left(\frac{1732}{9}-32\zeta_2\right) \mathL_3 
    + 
    \left(\frac{460 \zeta_2}{3}+240\zeta_3-\frac{10009}{18}\right) \mathL_2
\nn  \\
%%%%
&& \quad
   +  \left(-\frac{196 \zeta_2^2}{5}-\frac{352 \zeta_2}{3}-\frac{592\zeta_3}{3}+\frac{13783}{18}\right) \mathL_1
\nn  \\
%%%%
&& \quad
   +  \left( 39 \zeta_2^2+80 \zeta_2\zeta_3+\frac{3365 \zeta_2}{27}+\frac{1988 \zeta_3}{9}-120 \zeta_5-\frac{31151}{72} \right) \mathL_0
\nn  \\
%%%%
&& \quad
   + \bigg(
-\frac{8072 \zeta_2^3}{63}+\frac{988 \zeta_2^2}{27}+\frac{7036 \zeta_2 \zeta_3}{9}+\frac{21142 \zeta_2}{27}+\frac{1016 \zeta_3^2}{3}
\nn  \\
%%%%
&& \quad
-\frac{3431 \zeta_3}{3}-\frac{2792 \zeta_5}{9}+\frac{731}{36}
   \bigg) \mathL_{-1}
   \bigg]
\nn  \\
%%%%
&+&
   C_A C_F n_f \bigg[ 
   -\frac{176 \mathL_3}{27}
   + \left(\frac{1552}{27}-\frac{16 \zeta_2}{3}\right) \mathL_2
   +  \left(\frac{512 \zeta_2}{9}+16 \zeta_3-\frac{15062}{81}\right) \mathL_1
\nn  \\
%%%%
&& \quad
   +  \left( \frac{208 \zeta_2^2}{15}-\frac{9920 \zeta_2}{81}-\frac{776 \zeta_3}{9}+\frac{160906}{729}\right) \mathL_0
\nn  \\
%%%%
&& \quad
   + \bigg(
-\frac{836 \zeta_2^2}{27}+\frac{232 \zeta_2 \zeta_3}{3}+\frac{1850 \zeta_2}{81}-\frac{9746 \zeta_3}{81}+\frac{8 \zeta_5}{3}-\frac{147659}{486}
   \bigg) \mathL_{-1}
   \bigg]
\nn  \\
%%%%
&+&
   C_F^3 \bigg[
   8 \mathL_5
   -30 \mathL_4 
   +60 \mathL_3
   + \left(72 \zeta_2+16\zeta_3-\frac{153}{2}\right) \mathL_2
\nn  \\
%%%%
&& \quad
   + \left(-\frac{104 \zeta_2^2}{5}+4 \zeta_2-360 \zeta_3+\frac{181}{2}\right) \mathL_1
\nn  \\
%%%%
&& \quad
   +  \left( 6 \zeta_2^2-64 \zeta_2\zeta_3-45 \zeta_2+178 \zeta_3+432 \zeta_5-\frac{479}{8} \right) \mathL_0
\nn  \\
%%%%
&& \quad
   + \bigg(
\frac{26864 \zeta_2^3}{315}+\frac{196 \zeta_2^2}{5}-272 \zeta_2 \zeta_3-\frac{50 \zeta_2}{3}-\frac{304 \zeta_3^2}{3}
\nn  \\
%%%%
&& \quad
-1262 \zeta_3+760 \zeta_5+\frac{12371}{24}
   \bigg) \mathL_{-1}
   \bigg]
\nn  \\
%%%%
&+&
   C_F^2 n_f \bigg[
   \frac{40\mathL_4}{9}
   -\frac{280 \mathL_3}{9}
   +  \left(\frac{827}{9}-\frac{64 \zeta_2}{3}\right) \mathL_2
   +  \left(\frac{64 \zeta_2}{3}+\frac{112 \zeta_3}{3}-\frac{1343}{9}\right) \mathL_1
\nn  \\
%%%%
&& \quad
   +  \left(  -16 \zeta_2^2-\frac{722 \zeta_2}{27}-60\zeta_3+\frac{16423}{108} \right) \mathL_0
   \nn  \\
%%%%
&& \quad
   + \bigg(
\frac{1528 \zeta_2^2}{135}-\frac{1072 \zeta_2 \zeta_3}{9}-\frac{3280 \zeta_2}{27}+\frac{1300 \zeta_3}{3}-\frac{592 \zeta_5}{9}-\frac{4339}{36}
   \bigg) \mathL_{-1}
   \bigg]
\nn  \\
%%%%
&+&
   C_F n_f^2 \bigg[
   \frac{16 \mathL_3}{27}
   -\frac{116 \mathL_2}{27}
   +    \left(\frac{940}{81}-\frac{32 \zeta_2}{9}\right) \mathL_1
   +  \left(   \frac{232 \zeta_2}{27}-\frac{32 \zeta_3}{27}-\frac{8714}{729} \right) \mathL_0
   \nn  \\
%%%%
&& \quad
   + \bigg(
\frac{112 \zeta_2^2}{27}-\frac{52 \zeta_2}{27}-\frac{8 \zeta_3}{81}+\frac{4933}{486}
   \bigg) \mathL_{-1}
   \bigg] 
   \, ,
\eeeq
where the three-loop result includes contributions from the same quark jet function as in the $\gamma^* \to q \bar q$ case as well as the three-loop hard function from the $H \to b \bar b $ form factor \cite{Gehrmann:2006wg,Moch:2005tm,Baikov:2009bg,Lee:2010cga,Gehrmann:2010ue} given in Eq.~(\ref{three_loop_form_factor_bottom}).

\section{Summary}
\label{sec:summary}

Semi-inclusive $e^+e^-$ annihilation (SIA) plays a crucial role in extracting fragmentation functions, which are essential for understanding parton-to-hadron transitions in collider events. Enhancing the theoretical precision of SIA cross-section calculations is vital for advancing our comprehension of both perturbative and non-perturbative QCD aspects in final-state hadron measurements.

In this work, we have addressed the threshold corrections from soft gluons to the semi-inclusive distributions for $\gamma^* \to $ \emph{hadrons} and $H \to $ \emph{hadrons}. We employed a momentum-space resummation technique based on effective field theory, achieving next-to-next-to-next-to-next-to-leading logarithmic (N$^4$LL) accuracy for $\gamma^* \to q \bar q$, $H \to gg$, and $H \to b\bar b$ processes. This represents a significant advancement over previous studies, extending the resummation accuracy by one logarithmic order.

Our results demonstrate improved perturbative convergence and reduced scale uncertainty from NLL to N$^4$LL across all examined cases. This enhanced precision is particularly valuable for extracting fragmentation functions from experimental data.

Historically, SIA measurements for $e^+e^- \to \gamma^*/Z \to q \bar q$ have primarily constrained quark fragmentation functions. However, the extraction of heavy-quark and gluon fragmentation functions from this process has been more challenging. We highlight the potential of future $e^+e^-$ Higgs factories to significantly improve measurements of these quantities through the $H \to b\bar b$ and $H \to gg$ processes.

The N$^4$LL resummations presented in this paper, along with their fixed-order expansions, are expected to  benefit the extraction of light-quark and gluon fragmentation functions. When combined with complementary measurements from semi-inclusive deep inelastic scattering and jet fragmentation at the LHC, these results will contribute to a more comprehensive understanding of QCD dynamics in collider physics.

Looking ahead, several promising avenues for future research emerge:
\begin{enumerate}
\item  Phenomenological studies incorporating our N$^4$LL results into global fits of fragmentation functions, quantifying the impact on their precision and uncertainties.

\item  Extension of the resummation techniques developed here to related processes, such as single-inclusive hadron production in proton-proton collisions.

\item  Application of these high-precision calculations to improve the extraction of the strong coupling constant from energy correlators and other jet substructure observables.

\item  Exploration of subleading power corrections, which may become increasingly relevant as the accuracy of the leading-power resummation improves.
\end{enumerate}

In conclusion, this work represents a step forward in the theoretical understanding of semi-inclusive $e^+e^-$ annihilation and lays the groundwork for more precise determinations of fragmentation functions. These advances will play a crucial role in enhancing our ability to make accurate predictions for a wide range of hadronic observables in current and future collider experiments.

\section*{Acknowledgements}
We thank Tong-Zhi Yang for useful discussions. This work is supported by the National Science Foundation of China under contract No.~11975200 and No.~12425505. H.X.Z. is also supported by the Asian Young Scientist fellowship.

\appendix

\section{Perturbative Ingredients}

\label{sec:ingredients}

We present in this section the perturbative ingredients needed for the threshold resummation for semi-inclusive distributions in $\gamma^* \to q \bar q $,  $H \to gg $ and $H \to b \bar b $ to N$^4$LL.

\subsection{Hard Function and Anomalous Dimension}

The hard Wilson coefficient $C_V$ can be extracted from the time-like on-shell form factor. It has been calculated to the four-loop order in 
\cite{Kramer:1986sg,Matsuura:1987wt,Matsuura:1988sm,Harlander:2000mg,Ravindran:2004mb,Gehrmann:2005pd,Gehrmann:2006wg,Heinrich:2007at,
Moch:2005id,Moch:2005tm,Baikov:2009bg,Lee:2010cga,Gehrmann:2010ue,vonManteuffel:2015gxa,Lee:2021uqq,Lee:2022nhh}. 
For N$^4$LL resummation, we need the fixed-order hard function $C_V$ to three-loop and its anomalous dimension $\gamma_V$ to four-loop. 
The perturbative expansions are denoted as
\begin{eqnarray}
C_V &= & \sum_{n=0} a_s^n  C_V^{(n)} \qquad \text{and} \qquad \gamma_V = \sum_{n=0}  a_s^{n+1} \,\gamma_V^{(n)} \, . 
\end{eqnarray}
The expansion coefficients are classified with their color structure. For ingredients required for N$^4$LL resummation, we have for $SU(N_c)$
\beeq
\label{eq:casimirs}
C_F &=& (N_c^2-1)/(2 N_c)
\, ,\nonumber\\
C_A &=& N_c
\, ,\nonumber\\
N_F &=& N_c
\, ,\nonumber\\
N_A &=& N_c^2-1
\, ,\nonumber\\
d_F^{abc}d_F^{abc} / N_F &=&  \frac{(N_c^2-4)(N_c^2-1)}{16 N_c^2}
\, ,\nonumber\\
d_F^{abcd}d_F^{abcd} / N_F &=& \frac{(N_c^2-1)(N_c^4-6N_c^2+18)}{96 N_c^3}
\, ,\nonumber\\
d_F^{abcd}d_A^{abcd} / N_F &=& \frac{(N_c^2-1)(N_c^2+6)}{48}
\, ,\nonumber\\
d_A^{abcd}d_A^{abcd} / N_F &=& \frac{N_c(N_c^2-1)(N_c^2+36)}{24}
\, ,
\eeeq
where $N_c$ is the number of color, and the subscripts $A$ and $F$ represent the adjoint (gluon) or fundamental (quark) representation, respectively.

The required coefficients for the quark form factor for $e^+e^- \to \gamma^* \to q \bar q $ are, 
\begin{eqnarray}
\label{hard_fn}
C_V^{(0)}[q]&=&  1 \, , \\
C_V^{(1)}[q]&=& C_F \left[-L^2+3 L+\zeta_2-8\right] \, ,  \\
%%%%
C_V^{(2)}[q]&=& C_A C_F \bigg [
\frac{11 L^3}{9}+ \left(2 \zeta_2-\frac{233}{18}\right)L^2 +
   \left(\frac{22 \zeta_2}{3}-26 \zeta_3+\frac{2545}{54}\right)L
   \nonumber\\
 && \hspace{+6.5cm}  +\frac{44 \zeta_2^2}{5}-\frac{337
   \zeta_2}{18}+\frac{313 \zeta_3}{9}-\frac{51157}{648} \bigg]
 \nonumber\\
 &+& 
   C_F n_f \bigg[
   -\frac{2 L^3}{9}+\frac{19 L^2}{9}+ \left(-\frac{4
   \zeta_2}{3}-\frac{209}{27}\right)L +\frac{23 \zeta_2}{9}+\frac{2
   \zeta_3}{9}+\frac{4085}{324}
   \bigg] \nonumber\\ 
   &+& C_F^2  \bigg[
   \frac{L^4}{2}-3 L^3+ \left(\frac{25}{2}-\zeta_2\right) L^2 + \left(-9 \zeta_2+24
   \zeta_3-\frac{45}{2}\right) L
   \nonumber\\
   &&	 \hspace{+6.5cm} 	-\frac{83 \zeta_2^2}{10}+21 \zeta_2-30\zeta_3+\frac{255}{8} \bigg] \, , 
  \\
 \nonumber \\
C_V^{(3)}[q]&=&
C_F^3 \bigg[
-\frac{1}{6}L^6+\frac{3}{2}L^5+\bigg(-\frac{17}{2}+\frac{1}{2}\zeta_2\bigg)L^4+\bigg(9\zeta_2+27-24\zeta_3\bigg)L^3\nonumber \\ 
%%%%%%
&&\hspace{1cm}
+\bigg(102\zeta_3-\frac{507}{8}-\frac{105}{2}\zeta_2+\frac{83}{10}\zeta_2^2\bigg)L^2\nonumber \\ 
%%%%%%
&&\hspace{1cm}+\bigg(-214\zeta_3-240\zeta_5-8\zeta_2\zeta_3+\frac{357}{2}\zeta_2+\frac{207}{10}\zeta_2^2+\frac{785}{8}\bigg)L\nonumber \\
%%%%%%
&&\hspace{1cm}-\frac{413}{5}\zeta_2^2+664\zeta_5-\frac{6451}{24}\zeta_2+\frac{37729}{630}\zeta_2^3-470\zeta_3+250\zeta_2\zeta_3-\frac{2539}{12}+16\zeta_3^2 \bigg]
\nonumber \\ 
%%%%%%
&+& 
C_F^2C_A \bigg[
-\frac{11}{9}L^5+\bigg(\frac{299}{18}-2\zeta_2\bigg)L^4+\bigg(-\frac{2585}{27}+26\zeta_3-\frac{1}{9}\zeta_2\bigg)L^3\nonumber \\ 
%%%%%%
&&\hspace{1cm}+\bigg(\frac{206317}{648}-\frac{1807}{9}\zeta_3+\frac{502}{9}\zeta_2-\frac{34}{5}\zeta_2^2\bigg)L^2\nonumber \\ 
%%%%%%
&&\hspace{1cm}+\bigg(-\frac{13805}{24}+120\zeta_5+\frac{2441}{3}\zeta_3-\frac{11260}{27}\zeta_2-10\zeta_2\zeta_3+\frac{162}{5}\zeta_2^2\bigg)L\nonumber \\ 
&&\hspace{1cm}+\frac{415025}{648}-\frac{2756}{9}\zeta_5-\frac{18770}{27}\zeta_3+\frac{296}{3}\zeta_3^2+\frac{538835}{648}\zeta_2-\frac{3751}{9}\zeta_2\zeta_3\nonumber \\ 
%%%%%%
&&\hspace{1cm}-\frac{4943}{270}\zeta_2^2-\frac{12676}{315}\zeta_2^3 \bigg]\nonumber \\ 
%%%%%%
&+&
C_F^2 n_f \bigg[
\frac{2}{9}L^5-\frac{25}{9}L^4+\bigg(\frac{410}{27}+\frac{10}{9}\zeta_2\bigg)L^3+\bigg(-\frac{12815}{324}+\frac{70}{9}\zeta_3-\frac{112}{9}\zeta_2\bigg)L^2\nonumber \\ 
%%%%%%
&&\hspace{1cm}+\bigg(\frac{3121}{108}-\frac{610}{9}\zeta_3+\frac{1618}{27}\zeta_2+\frac{28}{5}\zeta_2^2\bigg)L\nonumber \\ 
%%%%%%
&&\hspace{1cm}+\frac{41077}{972}-\frac{416}{9}\zeta_5+\frac{13184}{81}\zeta_3-\frac{31729}{324}\zeta_2-\frac{38}{9}\zeta_2\zeta_3-\frac{331}{27}\zeta_2^2 \bigg]\nonumber \\ 
%%%%%%
&+&
C_FC_A^2 \bigg[
-\frac{121}{54}L^4+\bigg(\frac{2869}{81}-\frac{44}{9}\zeta_2\bigg)L^3+\bigg(-\frac{18682}{81}+88\zeta_3+\frac{26}{9}\zeta_2-\frac{44}{5}\zeta_2^2\bigg)L^2\nonumber \\ 
%%%%%%
&&\hspace{1cm}+\bigg(\frac{1045955}{1458}+136\zeta_5-\frac{17464}{27}\zeta_3+\frac{17366}{81}\zeta_2+\frac{88}{3}\zeta_2\zeta_3-\frac{94}{3}\zeta_2^2\bigg)L\nonumber \\ 
%%%%%%
&&\hspace{1cm}-\frac{51082685}{52488}-\frac{434}{9}\zeta_5+\frac{505087}{486}\zeta_3-\frac{1136}{9}\zeta_3^2-\frac{412315}{729}\zeta_2+\frac{416}{3}\zeta_2\zeta_3\nonumber \\ 
%%%%%%
&&\hspace{1cm}+\frac{22157}{270}\zeta_2^2-\frac{6152}{189}\zeta_2^3 \bigg] \nonumber \\ 
%%%%%%
&+&
C_FC_A n_f \bigg[
\frac{22}{27}L^4+\bigg(-\frac{974}{81}+\frac{8}{9}\zeta_2\bigg)L^3+\bigg(\frac{5876}{81}-8\zeta_3+\frac{16}{3}\zeta_2\bigg)L^2\nonumber \\ 
%%%%%%
&&\hspace{1cm}+\bigg(-\frac{154919}{729}+\frac{724}{9}\zeta_3-\frac{5864}{81}\zeta_2+\frac{44}{15}\zeta_2^2\bigg)L\nonumber \\ 
%%%%%%
&&\hspace{1cm}+\frac{1700171}{6561}-\frac{4}{3}\zeta_5-\frac{4288}{27}\zeta_3+\frac{115555}{729}\zeta_2+\frac{4}{3}\zeta_2\zeta_3+\frac{2}{27}\zeta_2^2 \bigg]\nonumber \\ 
%%%%%%
&+&
C_F n_f^2 \bigg[
-\frac{2}{27}L^4+\frac{76}{81}L^3+\bigg(-\frac{406}{81}-\frac{8}{9}\zeta_2\bigg)L^2+\bigg(\frac{9838}{729}+\frac{16}{27}\zeta_3+\frac{152}{27}\zeta_2\bigg)L\nonumber \\ 
%%%%%%
&&\hspace{1cm}-\frac{190931}{13122}-\frac{416}{243}\zeta_3-\frac{824}{81}\zeta_2-\frac{188}{135}\zeta_2^2 \bigg]
%%%%%%
\nonumber \\ 
%%%%%%
&+&
{N_{F,V}} \frac{d_F^{abc}d_F^{abc} }{N_F} \bigg[ -\frac{16 \zeta_2^2}{5}+80 \zeta_2+\frac{112 \zeta_3}{3}-\frac{640 \zeta_5}{3}+32 \bigg] \, .
\label{hard_fn_three_loop}
\end{eqnarray}
%%%
%%%
And the anomalous dimension $\gamma_V$ for the quark form factor, 
\begin{align}
\label{gamma_hard_quark}
\gamma_V^{(0)}[q]&=  -6 C_F \, , \\
\gamma_V^{(1)}[q]&= C_A C_F \left(-22 \zeta_2+52 \zeta_3-\frac{961}{27}\right)
+ C_F n_f \left(4 \zeta_2+\frac{130}{27}\right) + C_F^2 \big(24   \zeta_2-48 \zeta_3-3\big)  	 \, , \\
%%%
%%%
\gamma_V^{(2)}[q]&= 
C_A C_F n_f \left(\frac{88 \zeta_2^2}{5}+\frac{5188 \zeta_2}{81}-\frac{1928
   \zeta_3}{27}-\frac{17318}{729}\right)
 \nonumber\\
 &
 + C_A C_F^2 \left(\frac{1976 \zeta_2^2}{15}-32 \zeta_2 \zeta_3+\frac{820
   \zeta_2}{3}-\frac{1688 \zeta_3}{3}-240 \zeta_5-\frac{151}{2}\right)
 \nonumber\\
 &
 +C_A^2 C_F \left(-\frac{332 \zeta_2^2}{5}-\frac{176 \zeta_2 \zeta_3}{3}-\frac{14326
   \zeta_2}{81}+\frac{7052 \zeta_3}{9}-272 \zeta_5-\frac{139345}{1458} \right)
 \nonumber\\
 &
+C_F^2 n_f \left(-\frac{112 \zeta_2^2}{3}-\frac{52 \zeta_2}{3}+\frac{512
   \zeta_3}{9}+\frac{2953}{27} \right)
 +C_F n_f^2 \left(-\frac{40 \zeta_2}{9}-\frac{16 \zeta_3}{27}+\frac{4834}{729}\right)
 \nonumber\\
 &
 + C_F^3 \left( -\frac{576 \zeta_2^2}{5}+64 \zeta_2 \zeta_3-36 \zeta_2-136 \zeta_3+480
   \zeta_5 -29 \right)  \, , 
 \\
\nonumber \\
 %%%%%%
\gamma_V^{(3)} [q]&= C_F n_f^3 \left(-\frac{128 \zeta_2^2}{135}-\frac{16 \zeta_2}{27}-\frac{1424 \zeta_3}{243}+\frac{37382}{6561}\right) 
 \label{gamma_hard_quark_3_loop}
\nonumber\\ 
%%%
&+C_F^3 n_f \left(\frac{117344 \zeta
_2^3}{315}+\frac{668 \zeta_2^2}{5}+\frac{512 \zeta_3 \zeta_2}{3}-322 \zeta_2-368 \zeta_3^2+\frac{1120 \zeta_3}{9}-\frac{3872 
\zeta_5}{3}-\frac{27949}{108}\right) \nonumber\\ 
%%%
&+C_A^2 C_F n_f \bigg(-\frac{24184 \zeta_2^3}{315}+\frac{17164 
\zeta_2^2}{45} +\frac{3584 \zeta_3 \zeta_2}{9}
\nonumber\\ 
& \qquad\qquad +\frac{445117 \zeta_2}{729}-\frac{6916 \zeta_3^2}{9}-\frac{140632 \zeta_3}{243}-\frac{6088 \zeta_5}{27}-\frac{326863}{1944}\bigg) \nonumber\\
%%%
&+ C_F^2 n_f^2 \bigg(\frac{8032 \zeta
_2^2}{135}+\frac{224 \zeta_3 \zeta_2}{9}-\frac{1972 \zeta
_2}{27} +\frac{4232 \zeta_3}{81}-\frac{1040 \zeta
_5}{9}-\frac{9965}{486}\bigg) \nonumber\\ 
%%%
&+C_A C_F n_f^2 \left(-\frac{152 \zeta_2^2}{15}-\frac{256 \zeta_3 \zeta
_2}{9}-\frac{41579 \zeta_2}{729}-\frac{14872 \zeta
_3}{243}+\frac{1184 \zeta_5}{9}+\frac{97189}{17496}\right) \nonumber\\
%%%
&+
C_A C_F^2 n_f \bigg(-\frac{5744 \zeta_2^3}{35}-\frac{105488 \zeta_2^2}{135}-\frac{3904 \zeta_3 \zeta_2}{9}-\frac{673 \zeta_2}{27}+\frac{3400 \zeta_3^2}{3}
\nonumber\\
&\qquad\qquad 
+\frac{23518 \zeta_3}{81}+\frac{4472 \zeta_5}{3}+\frac{1092511}{972}\bigg) \nonumber\\ 
%%%
&+\frac{ d_F^{abcd}d_F^{abcd}  n_f }{N_F} \bigg(-\frac{9472 \zeta_2^3}{315}+\frac{320 \zeta_2^2}{3}-128 \zeta_3 \zeta_2-\frac{4544 \zeta_2}{3}-\frac{1216 \zeta_3^2}{3}
\nonumber\\
&\qquad\qquad
+\frac{5312 \zeta_3}{9}+\frac{21760 \zeta_5}{9}+384\bigg)\nonumber\\ 
%%%
&+C_A^2 C_F^2 \bigg(\frac{43976 \zeta_2^3}{35}+\frac{128}{5} 
\zeta_3 \zeta_2^2+\frac{48680 \zeta_2^2}{27}-\frac{4192 \zeta_3 
\zeta_2}{9}+4208 \zeta_5 \zeta_2
\nonumber\\ 
&\qquad\qquad+\frac{93542 \zeta
_2}{27}+\frac{14204 \zeta_3^2}{3}-\frac{259324 \zeta
_3}{27}-\frac{10708 \zeta_5}{9}-17220 \zeta_7-\frac{29639}{18}\bigg) \nonumber\\ 
%%%
&+C_F^4 \bigg(\frac{33776 \zeta
_2^3}{35}-\frac{256}{5} \zeta_3 \zeta_2^2+\frac{1368 \zeta
_2^2}{5}+240 \zeta_3 \zeta_2+768 \zeta_5 \zeta_2+900 \zeta
_2 
\nonumber\\ 
&\qquad\qquad +2304 \zeta_3^2-4008 \zeta_3+5040 \zeta_5-11760 \zeta_7-\frac{4873}{12}\bigg) \nonumber\\ 
%%%
&+\frac{d_F^{abcd}d_A^{abcd}  }{N_F}
\bigg(-\frac{27808 \zeta_2^3}{315}+\frac{736}{5} \zeta_3 \zeta
_2^2-\frac{224 \zeta_2^2}{15}+1792 \zeta_3 \zeta_2-1024 \zeta_5 
\zeta_2+\frac{2176 \zeta_2}{3}
\nonumber\\ 
&\qquad\qquad+\frac{3344 \zeta_3^2}{3}+\frac{7808 
\zeta_3}{9}+\frac{1840 \zeta_5}{9}-3484 \zeta_7-192\bigg) \nonumber\\ 
%%%
&+C_A^3 C_F \bigg(\frac{77152 \zeta
_2^3}{315}+\frac{4132}{15} \zeta_3 \zeta_2^2-\frac{186742 \zeta
_2^2}{135}-\frac{15400 \zeta_3 \zeta_2}{9}-\frac{1648 \zeta_5 
\zeta_2}{3}
\nonumber\\ 
&\qquad\qquad-\frac{1062149 \zeta_2}{729}-\frac{5126 \zeta_3^2}{9}+\frac{1751224 \zeta_3}{243}
-\frac{175166 \zeta_5}{27}+\frac{45511 \zeta_7}{6}-\frac{7179083}{26244}\bigg) \nonumber\\ 
%%%
&+C_A C_F^3 \bigg(-\frac{634376 \zeta
_2^3}{315}-\frac{512}{5} \zeta_3 \zeta_2^2-\frac{8668 \zeta
_2^2}{5}+\frac{3976 \zeta_3 \zeta_2}{3}-4128 \zeta_5 \zeta_2-2334 
\zeta_2
\nonumber\\ 
&\qquad\qquad -6440 \zeta_3^2+6520 \zeta_3+1952 \zeta_5+21840 \zeta
_7+\frac{2085}{2}\bigg) \, . 
\end{align}
%%%%
The coefficients for the gluon form factor needed for the $e^+e^- \to H \to gg $ process are
%%%%
\begin{eqnarray}
C_V^{(0)}[g] &=&  1 \, ,  \\
%%%%
C_V^{(1)}[g] &=& C_A \bigg[ -L^2 + \zeta_2  \bigg],  \\
%%%%
C_V^{(2)}[g] &=&
C_A^2 \bigg[
\frac{1}{2}L^4+\frac{11}{9}L^3+\bigg(-\frac{67}{9}+\zeta_2\bigg)L^2+\bigg(\frac{80}{27}-2\zeta_3-\frac{22}{3}\zeta_2\bigg)L\nonumber \\ &&\hspace{2.5cm}+\frac{5105}{162}-\frac{143}{9}\zeta_3+\frac{67}{6}\zeta_2+\frac{1}{2}\zeta_2^2 \bigg]
\nonumber \\ 
&+&
C_A n_f \bigg[
-\frac{2}{9}L^3+\frac{10}{9}L^2+\bigg(\frac{52}{27}+\frac{4}{3}\zeta_2\bigg)L-\frac{916}{81}-\frac{46}{9}\zeta_3-\frac{5}{3}\zeta_2 \bigg]
\nonumber \\ 
&+&
C_F n_f \bigg[
2L -\frac{67}{6}+8\zeta_3 \bigg] \, ,   \\ 
%%%%
C_V^{(3)}[g] &=& 
C_A^3 \bigg[
-\frac{1}{6}L^6-\frac{11}{9}L^5+\bigg(\frac{281}{54}-\frac{3}{2}\zeta_2\bigg)L^4+\bigg(\frac{11}{3}\zeta_2+\frac{1540}{81}+2\zeta_3\bigg)L^3\nonumber \\ &&\hspace{2cm}+\bigg(\frac{143}{9}\zeta_3-\frac{6740}{81}+\frac{685}{18}\zeta_2-\frac{73}{10}\zeta_2^2\bigg)L^2\nonumber \\ &&\hspace{2cm}+\bigg(\frac{2048}{27}\zeta_3+16\zeta_5+\frac{34}{3}\zeta_2\zeta_3-\frac{13420}{81}\zeta_2+\frac{176}{5}\zeta_2^2-\frac{373975}{1458}\bigg)L\nonumber \\ &&\hspace{2cm}-\frac{1939}{270}\zeta_2^2+\frac{2222}{9}\zeta_5+\frac{105617}{729}\zeta_2-\frac{24389}{1890}\zeta_2^3-\frac{152716}{243}\zeta_3-\frac{605}{9}\zeta_2\zeta_3\nonumber \\ &&\hspace{2cm}+\frac{29639273}{26244}-\frac{104}{9}\zeta_3^2 \bigg]
\nonumber \\ 
&+&
C_A^2 n_f \bigg[
\frac{2}{9}L^5-\frac{8}{27}L^4+\bigg(-\frac{734}{81}-\frac{2}{3}\zeta_2\bigg)L^3+\bigg(\frac{377}{27}+\frac{118}{9}\zeta_3-\frac{103}{9}\zeta_2\bigg)L^2\nonumber \\ &&\hspace{2cm}+\bigg(\frac{133036}{729}+\frac{28}{9}\zeta_3+\frac{3820}{81}\zeta_2-\frac{48}{5}\zeta_2^2\bigg)L\nonumber \\ &&\hspace{2cm}-\frac{3765007}{6561}+\frac{428}{9}\zeta_5-\frac{460}{81}\zeta_3-\frac{14189}{729}\zeta_2-\frac{82}{9}\zeta_2\zeta_3+\frac{73}{45}\zeta_2^2 \bigg]
\nonumber \\ 
&+&
C_A n_f^2 \bigg[
-\frac{2}{27}L^4+\frac{40}{81}L^3+\bigg(\frac{116}{81}+\frac{8}{9}\zeta_2\bigg)L^2+\bigg(-\frac{14057}{729}-\frac{128}{27}\zeta_3-\frac{80}{27}\zeta_2\bigg)L\nonumber \\ &&\hspace{2cm}+\frac{611401}{13122}+\frac{4576}{243}\zeta_3+\frac{4}{9}\zeta_2+\frac{4}{27}\zeta_2^2 \bigg]
\nonumber \\ 
&+&
C_F n_f^2 \bigg[
\frac{4}{3}L^2+\bigg(-\frac{52}{3}+\frac{32}{3}\zeta_3\bigg)L+\frac{4481}{81}-\frac{112}{3}\zeta_3-\frac{20}{9}\zeta_2-\frac{16}{45}\zeta_2^2 \bigg]\nonumber \\ 
&+&
C_F C_A n_f \bigg[
-\frac{8}{3}L^3+\bigg(13-16\zeta_3\bigg)L^2+\bigg(\frac{3833}{54}-\frac{376}{9}\zeta_3+6\zeta_2+\frac{16}{5}\zeta_2^2\bigg)L\nonumber \\ 
&&\hspace{2cm}-\frac{341219}{972}+\frac{608}{9}\zeta_5+\frac{14564}{81}\zeta_3-\frac{68}{9}\zeta_2+\frac{64}{3}\zeta_2\zeta_3-\frac{64}{45}\zeta_2^2 \bigg]
\nonumber \\ 
&+&
C_F^2 n_f \bigg[
-2L + \frac{304}{9}-160\zeta_5+\frac{296}{3}\zeta_3 \bigg] \, .
\label{three_loop_form_factor_gluon}
\end{eqnarray}
%%%%
The coefficients of the anomalous dimension $\gamma_V$ for the gluonic form factor are
\begin{eqnarray}
\label{gamma_hard_gluon}
\gamma_V^{(0)} [g]&=&  -\frac{22 C_A}{3} + \frac{4 n_f}{3} \,, \\
\gamma_V^{(1)} [g]&=&  4 C_F n_f + C_A n_f \left(\frac{256}{27}-\frac{4 \zeta_2}{3}\right)   + C_A^2 \left(\frac{22 \zeta_2}{3}+4 \zeta_3-\frac{1384}{27}\right)  \,, \\
%%%%
\gamma_V^{(2)} [g]&=&  -2 C_F^2 n_f -\frac{22}{9} C_F n_f^2 + C_A C_F n_f \left(-\frac{32 \zeta_ 2^2}{5}-4 \zeta_ 2-\frac{304 \zeta_
3}{9}+\frac{2434}{27}\right)  \nonumber\\ 
&+&
C_A n_f^2 \left(\frac{40 \zeta_ 2}{27}-\frac{112 \zeta_3}{27}-\frac{269}{729}\right)  +
C_A^2 n_f \left(\frac{328 \zeta_ 2^2}{15}-\frac{2396 \zeta_ 2}{81}+\frac{712
\zeta_ 3}{27}+\frac{30715}{729}\right)  \nonumber\\
&+& 
C_A^3 \left(-\frac{1276 \zeta_ 2^2}{15}-\frac{80 \zeta_ 3 \zeta_2}{3}+\frac{12218 \zeta_ 2}{81}+\frac{244 \zeta_ 3}{3}-32 \zeta_5-\frac{194372}{729}\right) \,, \\
%%%%
\nonumber\\
%%%%
\gamma_V^{(3)} [g]&=& 	
-46 C_F^3 n_f -\frac{308 C_F n_f^3}{243}
+	\frac{d_F^{abcd}d_F^{abcd}  }{N_A}	n_f^2 
\left(\frac{1408}{9}-\frac{1024 \zeta_3}{3}\right) \nonumber\\ 
&+& 
C_A n_f^3 \left(-\frac{256 \zeta
_2^2}{135}+\frac{16 \zeta_2}{81}+\frac{400 \zeta
_3}{243}+\frac{15890}{6561}\right) +
C_F^2 n_f^2 \left(\frac{352 \zeta_3}{9}-\frac{676}{27}\right) \nonumber\\ 
&+& 
C_A^3 n_f \bigg(-\frac{148976 \zeta_2^3}{945}+
\frac{69502 \zeta_2^2}{135}-148 \zeta_3 \zeta_2-\frac{155273 \zeta
_2}{729}
 \nonumber\\
  && \hspace{+5cm} +\frac{596 \zeta_3^2}{9} + 
\frac{260822 \zeta
_3}{243}-\frac{16066 \zeta_5}{27}+\frac{421325}{1944}\bigg) \nonumber\\ 
&+& 
C_A C_F n_f^2 \left(-\frac{128 \zeta_2^2}{45}-\frac{32 \zeta_3 \zeta_2}{3}+\frac{172 \zeta_2}{9}+\frac{1688 \zeta_3}{81}-\frac{304 \zeta
_5}{9}-\frac{1199}{18}\right) \nonumber\\
&+& 
 C_A^2 C_F n_f \bigg(-\frac{5632 \zeta_2^3}{315}+\frac{1196 \zeta_2^2}{45}+176 \zeta_3 \zeta_2-\frac{3023 \zeta_2}{9}
  \nonumber\\
  && \hspace{+6cm} -152 \zeta_3^2
 -\frac{29606 \zeta_3}{81}-\frac{8 \zeta
_5}{9}+\frac{903983}{972}\bigg) 
\nonumber\\ 
&+& 
C_A^2 n_f^2 \left(-\frac{3128 \zeta_2^2}{135}+32 \zeta_3 \zeta_2+\frac{13483 \zeta_2}{729}-\frac{37354 \zeta_3}{243}+\frac{1024 \zeta_5}{9}-\frac{611939}{17496}\right) 
\nonumber\\ 
&+& 
C_A C_F^2 n_f \left(\frac{320 \zeta_2^3}{7}-\frac{148 \zeta_2^2}{5}+2 \zeta_2+80 \zeta_3^2-\frac{1592 \zeta_3}{3}+\frac{1600 \zeta
_5}{3}-\frac{685}{12}\right) 
\nonumber\\ 
&+& \frac{d_F^{abcd}d_A^{abcd}  }{N_A} n_f \bigg(\frac{14464 \zeta_2^3}{315}-\frac{2464 \zeta_2^2}{15}-1216 \zeta_3 \zeta_2+64 \zeta_2
 \nonumber\\
  && \hspace{+5.5cm} -\frac{1216 \zeta_3^2}{3}
  -\frac{2560 \zeta_3}{9}+\frac{30880 \zeta_5}{9}-\frac{448}{9}\bigg)
 \nonumber\\ 
 &+& \frac{d_A^{abcd}d_A^{abcd}  }{N_A} \bigg(-\frac{39776 \zeta_2^3}{315}+\frac{736}{5} \zeta_3 \zeta_2^2+\frac{1808 \zeta_2^2}{15}+2336 \zeta_3 \zeta_2-1024 \zeta_5 \zeta_2 \nonumber\\
  && \hspace{+2.5cm} 
-64 \zeta_2  + \frac{3344 \zeta_3^2}{3}+\frac{12512 \zeta_3}{9}-\frac{2720 \zeta_5}{9}-3484 \zeta_7-\frac{128}{9}\bigg)
\nonumber\\ 
&+& 
C_A^4 \bigg(\frac{674696 \zeta_2^3}{945}+\frac{2212}{15} \zeta_3 \zeta_2^2-\frac{249448 \zeta_2^2}{135}-\frac{1588 \zeta_3 \zeta_2}{3}+\frac{896 \zeta_5 \zeta_2}{3} \nonumber\\
&& \hspace{+0.5cm} + 
\frac{1051411 \zeta_2}{729}+\frac{286 \zeta_3^2}{9}-\frac{36380 \zeta_3}{243}-\frac{19232 \zeta_5}{27}+\frac{2671 \zeta
_7}{6}-\frac{10672040}{6561}\bigg) \, .
\end{eqnarray} \\
%%%%%%
The $H \to b \bar b $ form factor to three-loop, 
\beeq
C_V^{(0)}[b] &=& 1  \, , \\
%%%%%%
C_V^{(1)}[b]  &=& C_F \left[-L^2+\zeta_2-2\right] \, , 
\\
%%%%%%
C_V^{(2)}[b]  &=&C_A C_F \bigg[\frac{11 L^3}{9}+L^2 \left(2 \zeta_2-\frac{67}{9}\right)+L \left(\frac{22
   \zeta_2}{3}-26 \zeta_3+\frac{242}{27}\right)
\nn\\
%%%%%%
&&\qquad
   +\frac{44 \zeta_2^2}{5}-\frac{103
   \zeta_2}{18}+\frac{151 \zeta_3}{9}-\frac{467}{81}\bigg]
\nn\\
%%%%%%
&+&
   C_F^2 \bigg[\frac{L^4}{2}+L^2
   (2-\zeta_2)+L (24 \zeta_3-12 \zeta_2)-\frac{83 \zeta_2^2}{10}+14 \zeta_2-30   \zeta_3+6\bigg]
   \nn\\
%%%%%%
&+&
   C_F n_f \bigg[-\frac{2 L^3}{9}+\frac{10 L^2}{9}+L \left(-\frac{4
   \zeta_2}{3}-\frac{56}{27}\right)+\frac{5 \zeta_2}{9}+\frac{2 \zeta_3}{9}+\frac{200}{81}\bigg]
    \, , \\
%%%%%%
C_V^{(3)}[b]  &=& C_A^2 C_F \bigg[-\frac{121 L^4}{54}+L^3 \left(\frac{1780}{81}-\frac{44
   \zeta_2}{9}\right)+L^2 \left(-\frac{44 \zeta_2^2}{5}+\frac{26 \zeta_2}{9}+88
   \zeta_3-\frac{11939}{162}\right)
   \nn\\
%%%%%%
&&
   +L \left(-\frac{94 \zeta_2^2}{3}+\frac{88 \zeta_2
   \zeta_3}{3}+\frac{9644 \zeta_2}{81}-\frac{13900 \zeta_3}{27}+136 \zeta_5+\frac{10289}{1458}\right)
   %%%%%%
   -\frac{6152 \zeta_2^3}{189}+\frac{10093 \zeta_2^2}{135}
   \nn\\
%%%%%%
&&
   +\frac{326 \zeta_2 \zeta_3}{3}-\frac{264515 \zeta_2}{1458}-\frac{1136 \zeta_3^2}{9}+\frac{107648
   \zeta_3}{243}+\frac{106 \zeta_5}{9}+\frac{5964431}{26244}\bigg]
      \nn\\
%%%%%%
&+&
  C_A C_F^2
   \bigg[-\frac{11 L^5}{9}+L^4 \left(\frac{67}{9}-2 \zeta_2\right)+L^3 \left(-\frac{55
   \zeta_2}{9}+26 \zeta_3-\frac{308}{27}\right)
      \nn\\
%%%%%%
&&
+L^2 \left(-\frac{34 \zeta_2^2}{5}+\frac{689
   \zeta_2}{18}-\frac{943 \zeta_3}{9}+\frac{1673}{81}\right)
      \nn\\
%%%%%%
&&
   +L \left(6 \zeta_2^2-10 \zeta_2
   \zeta_3-\frac{7012 \zeta_2}{27}+\frac{1660 \zeta_3}{3}+120 \zeta_5+\frac{614}{27}\right)
      \nn\\
%%%%%%
&&
-\frac{12676 \zeta_2^3}{315}-\frac{893 \zeta_2^2}{270}-\frac{3049 \zeta_2 \zeta_3}{9}+\frac{31819 \zeta_2}{81}+\frac{296 \zeta_3^2}{3}-\frac{4820
   \zeta_3}{27}-\frac{1676 \zeta_5}{9}-\frac{9335}{81}\bigg]
      \nn\\
%%%%%%
&+&
C_A C_F n_f
   \bigg[\frac{22 L^4}{27}+L^3 \left(\frac{8 \zeta_2}{9}-\frac{578}{81}\right)+L^2 \left(\frac{16
   \zeta_2}{3}-8 \zeta_3+\frac{1727}{81}\right)
      \nn\\
%%%%%%
&&
+L \left(\frac{44 \zeta_2^2}{15}-\frac{3272
   \zeta_2}{81}+\frac{724 \zeta_3}{9}-\frac{7499}{729}\right)
      \nn\\
%%%%%%
&&
   -\frac{476 \zeta_2^2}{135}+\frac{4
   \zeta_2 \zeta_3}{3}+\frac{33259 \zeta_2}{729}-\frac{2860 \zeta_3}{27}-\frac{4 \zeta_5}{3}-\frac{521975}{13122}\bigg]
      \nn\\
%%%%%%
&+&
C_F^3 \bigg[-\frac{L^6}{6}+L^4
   \left(\frac{\zeta_2}{2}-1\right)+L^3 (12 \zeta_2-24 \zeta_3)+L^2 \left(\frac{83
   \zeta_2^2}{10}-14 \zeta_2+30 \zeta_3-6\right)
      \nn\\
%%%%%%
&&
+L \left(\frac{228 \zeta_2^2}{5}-8 \zeta_2
   \zeta_3+42 \zeta_2+20 \zeta_3-240 \zeta_5-50\right)
      \nn\\
%%%%%%
&&
+\frac{37729 \zeta_2^3}{630}-77
   \zeta_2^2+178 \zeta_2 \zeta_3-\frac{353 \zeta_2}{3}+16 \zeta_3^2-654 \zeta_3+424 \zeta_5+\frac{575}{3}\bigg]
      \nn\\
%%%%%%
&+&
C_F^2 n_f \bigg[\frac{2 L^5}{9}-\frac{10 L^4}{9}+L^3
   \left(\frac{10 \zeta_2}{9}+\frac{50}{27}\right)+L^2 \left(-\frac{67 \zeta_2}{9}+\frac{70
   \zeta_3}{9}+\frac{725}{162}\right)
      \nn\\
%%%%%%
&&
+L \left(\frac{28 \zeta_2^2}{5}+\frac{880
   \zeta_2}{27}-\frac{832 \zeta_3}{9}-\frac{1415}{54}\right)
      \nn\\
%%%%%%
&&
   -\frac{61 \zeta_2^2}{27}-\frac{38
   \zeta_2 \zeta_3}{9}-\frac{6131 \zeta_2}{162}+\frac{11996 \zeta_3}{81}-\frac{416 \zeta_5}{9}+\frac{35875}{972}\bigg]
      \nn\\
%%%%%%
&+&
C_F n_f^2 \bigg[-\frac{2 L^4}{27}+\frac{40 L^3}{81}+L^2
   \left(-\frac{8 \zeta_2}{9}-\frac{100}{81}\right)+L \left(\frac{80 \zeta_2}{27}+\frac{16
   \zeta_3}{27}+\frac{928}{729}\right)
      \nn\\
%%%%%%
&&
-\frac{188 \zeta_2^2}{135}-\frac{212 \zeta_2}{81}-\frac{200
   \zeta_3}{243}+\frac{2072}{6561}\bigg] \,. 
\label{three_loop_form_factor_bottom}
\eeeq

The anomalous dimension $\gamma_V$ for $H \to b \bar b $ is the same as the one for $\gamma^* \to q \bar q $, given in Eqs.(\ref{gamma_hard_quark})-(\ref{gamma_hard_quark_3_loop}).

\subsection{Jet Function and Anomalous Dimension}

The jet function $J_{\rm SIA}$ is the same in many collider processes with final state jets, in particular it is the same as in the threshold factorization for the DIS structure function. The jet functions for both gluons and quarks have been calculated to three-loop in \cite{Bauer:2003pi,Bosch:2004th,Becher:2006qw,Becher:2009th,Becher:2010pd,Bruser:2018rad,Banerjee:2018ozf}, which are nicely summarized in the appendix in \cite{Ju:2023dfa}.  And we extract the four-loop anomalous dimension $\gamma_J$ from the RG consistency relation. 
We denote the perturbative coefficients as
%%%%
\begin{align}
\widetilde J &=  \sum_{n=0} a_s^n  \widetilde J^{(n)} \qquad \text{and} \qquad \gamma_J = \sum_{n=0}  a_s^{n+1} \,\gamma_J^{(n)} \, .
\end{align}
%%%
The perturbative coefficients for the Laplace-space quark jet function are \cite{Becher:2006qw,Bruser:2018rad}
\begin{align}
\label{quark_jet_fn}
\widetilde J^{(0)}[q]&=  1 \,, \\
%%%%
\widetilde J^{(1)}[q]&=  C_F \left[2 L^2-3 L- 4\zeta_2 +7\right] \,,	\\
%%%%
\widetilde J^{(2)}[q]&=  C_A C_F \bigg[-\frac{22 L^3}{9}+L^2 \left(\frac{367}{18}-4 \zeta_2\right)+L
   \left(\frac{22 \zeta_2}{3}+40 \zeta_3-\frac{3155}{54}\right) 
   \nn \\
& \qquad\qquad
   -\frac{37 \zeta_2^2}{5}-\frac{155
   \zeta_2}{6}-18 \zeta_3+\frac{53129}{648} \bigg]
 \nonumber\\
 &
+ C_F n_f \bigg[\frac{4 L^3}{9}-\frac{29 L^2}{9}+L \left(\frac{247}{27}-\frac{4
   \zeta_2}{3}\right)+\frac{13 \zeta_2}{3}-\frac{4057}{324} \bigg]
 \nonumber\\
 &
+C_F^2 \bigg[2 L^4-6 L^3+L^2 \left(\frac{37}{2}-8 \zeta_2\right)+L \left(24 \zeta_2-24
   \zeta_3-\frac{45}{2}\right)
   \nn \\
& \qquad\qquad
   +\frac{122 \zeta_2^2}{5}-\frac{97 \zeta_2}{2}-6
   \zeta_3+\frac{205}{8}\bigg]  \,,
\\
%%%%
\widetilde J^{(3)}[q]&=C_F^3 \bigg[\frac{4 L^6}{3}-6 L^5+L^4 (23-8 \zeta_2)+L^3 \bigg(48 \zeta_2-48 \zeta_3-\frac{99}{2}\bigg)  
\nonumber\\
 &
 +L^2 \bigg(\frac{244
   \zeta_2^2}{5}-151 \zeta_2+60 \zeta_3+\frac{349}{4}\bigg) 
    \nonumber\\
 &
 +L \bigg(-\frac{894 \zeta_2^2}{5}+128 \zeta_2 \zeta_3+\frac{435
   \zeta_2}{2}-218 \zeta_3+240 \zeta_5-\frac{815}{8}\bigg)\bigg]
   %%%%
       \nonumber\\
 + & \,
 C_A C_F^2 \bigg[-\frac{44 L^5}{9}+L^4 \bigg(\frac{433}{9}-8 \zeta_2\bigg)+L^3
   \bigg(\frac{328 \zeta_2}{9}+80 \zeta_3-\frac{10537}{54}\bigg)
   %%%%
       \nonumber\\
 &
   +L^2 \bigg(\frac{6
   \zeta_2^2}{5}-\frac{2045 \zeta_2}{9}-68 \zeta_3+\frac{157943}{324}\bigg)
      %%%%
       \nonumber\\
 &
 +L
   \bigg(-\frac{923 \zeta_2^2}{15}-176 \zeta_2 \zeta_3+\frac{35075
   \zeta_2}{54}+\frac{290 \zeta_3}{3}-120 \zeta_5-\frac{151405}{216}\bigg)\bigg]
      %%%%
       \nonumber\\
 + & \,
   C_F^2 n_f \bigg[\frac{8 L^5}{9}-\frac{70 L^4}{9}+L^3 \bigg(\frac{875}{27}-\frac{40
   \zeta_2}{9}\bigg)+L^2 \bigg(\frac{302 \zeta_2}{9}-\frac{15775}{162}\bigg)
      %%%%
       \nonumber\\
 &
   +L
   \bigg(\frac{16 \zeta_2^2}{3}-\frac{2833 \zeta_2}{27}+\frac{32
   \zeta_3}{9}+\frac{7325}{36}\bigg)\bigg]
         %%%%
       \nonumber\\
 + & \,
   C_A^2 C_F \bigg[\frac{121 L^4}{27}+L^3 \bigg(\frac{88
   \zeta_2}{9}-\frac{4649}{81}\bigg)+L^2 \bigg(\frac{88 \zeta_2^2}{5}-\frac{778
   \zeta_2}{9}-132 \zeta_3+\frac{50689}{162}\bigg)
      %%%%
       \nonumber\\
 &
   +L \bigg(-\frac{212
   \zeta_2^2}{15}-\frac{176 \zeta_2 \zeta_3}{3}+\frac{18179 \zeta_2}{81}+\frac{6688
   \zeta_3}{9}-232 \zeta_5-\frac{599375}{729}\bigg)\bigg]
         %%%%
       \nonumber\\
 + & \,
   C_A C_F n_f \bigg[-\frac{44 L^4}{27}+L^3 \bigg(\frac{1552}{81}-\frac{16
   \zeta_2}{9}\bigg)+L^2 \bigg(\frac{56 \zeta_2}{3}+8 \zeta_3-\frac{7531}{81}\bigg)
      %%%%
       \nonumber\\
 &
 +L
   \bigg(\frac{128 \zeta_2^2}{15}-\frac{5264 \zeta_2}{81}-\frac{1976
   \zeta_3}{27}+\frac{160906}{729}\bigg)\bigg]
         %%%%
       \nonumber\\
+ & \,
   C_F n_f^2 \bigg[\frac{4 L^4}{27}-\frac{116 L^3}{81}+L^2
   \bigg(\frac{470}{81}-\frac{8 \zeta_2}{9}\bigg)+L \bigg(\frac{116 \zeta_2}{27}-\frac{64
   \zeta_3}{27}-\frac{8714}{729}\bigg)\bigg]
         %%%%
       \nonumber\\
 + & \,
   \widetilde J^{(3)}[q]\big|_{L=0} \,, 
\end{align}
%%%%
where the scale-independent three-loop constant term is obtained by the Laplace transform of the three-loop jet function calculated in \cite{Bruser:2018rad}
%%%%
\beeq
\label{three-loop-jet-constant-quark}
\widetilde J^{(3)}[q]\big|_{L=0} &=&
C_F n_f^2 \bigg(\frac{28 \zeta_2^2}{45}-\frac{154 \zeta_2}{27}+\frac{1072 \zeta_3}{243}+\frac{124903}{13122}\bigg)                                                                                                                                                                                                                                                                                                       +
 C_A^2 C_F \bigg(\frac{25472 \zeta_2^3}{945}-\frac{3002 \zeta_2^2}{45}
  \nn \\
&& 
+\frac{1000 \zeta_2 \zeta_3}{9}-\frac{473129 \zeta_2}{1458}+\frac{1528 \zeta_3^2}{9}-\frac{160057 \zeta_3}{243}
 -\frac{380 \zeta_5}{9}+\frac{50602039}{52488}\bigg)                                                                                                                                                                                                                                                                                                       \nn \\   &+& C_A C_F n_f \bigg(\frac{16 \zeta_2^2}{15}-\frac{88 \zeta_2 \zeta_3}{9}+\frac{68869 \zeta_2}{729}+\frac{4310 \zeta_3}{81}+\frac{16 \zeta_5}{3}-\frac{2942843}{13122}\bigg)                                                                                                                                                                                                                                                                                                       \nn \\   &+& C_F^2 n_f \bigg(-\frac{6592 \zeta_2^2}{135}-\frac{32 \zeta_2 \zeta_3}{9}+\frac{13951 \zeta_2}{81}+\frac{5966 \zeta_3}{81}+\frac{16 \zeta_5}{3}-\frac{261587}{972}\bigg)                                                                                                                                                                                                                                                                                                       \nn \\   &+& C_F^3 \bigg(-\frac{59384 \zeta_2^3}{315}+\frac{1664 \zeta_2^2}{5}-112 \zeta_2 \zeta_3-\frac{1229 \zeta_2}{6}-\frac{272 \zeta_3^2}{3}+373 \zeta_3
  \nn \\
&& 
+56 \zeta_5+\frac{1173}{8}\bigg)                                                                                                                                                                                                                                                                                                       +
 C_A C_F^2 \bigg(\frac{3364 \zeta_2^3}{45}+\frac{8048 \zeta_2^2}{27}+\frac{2252 \zeta_2 \zeta_3}{9}-\frac{79213 \zeta_2}{81}
  \nn \\
&& 
-\frac{56 \zeta_3^2}{3}-\frac{17704 \zeta_3}{27}
+\frac{1616 \zeta_5}{9}+\frac{206197}{324}\bigg)     \, .                                                                                                                                                                                                                                                                                                   
\eeeq
Anomalous dimensions for the quark jet are \cite{Becher:2006qw,Das:2019btv,Duhr:2022cob}
\begin{align}
\label{gamma_quark_jet}
\gamma_J^{(0)}[q]&=  -3 C_F \, , \\
\gamma_J^{(1)}[q]&=  C_A C_F \left(-\frac{22 \zeta_2}{3}+40 \zeta_3-\frac{1769}{54}\right) 
+ C_F n_f \left(\frac{4 \zeta_2}{3}+\frac{121}{27}\right)+C_F^2 \left(12 \zeta_2-24 \zeta_3-\frac{3}{2}\right) \, , \\
%%%%
\gamma_J^{(2)}[q]&= C_A C_F n_f \left( \frac{92 \zeta_2^2}{5}+\frac{1180 \zeta_2}{81}-\frac{1328
   \zeta_3}{27}-\frac{2738}{729} \right) 
 \nonumber\\
 &
+ C_A C_F^2 \left(\frac{988 \zeta_2^2}{15}-16 \zeta_2 \zeta_3+\frac{410
   \zeta_2}{3}-\frac{844 \zeta_3}{3}-120 \zeta_5-\frac{151}{4}\right) 
 \nonumber\\
 &
+ C_A^2 C_F \left( -\frac{342 \zeta_2^2}{5}-\frac{176 \zeta_2 \zeta_3}{3}-\frac{838
   \zeta_2}{81}+\frac{5500 \zeta_3}{9}-232 \zeta_5-\frac{412907}{2916} \right) 
 \nonumber\\
 &
 + C_F^2 n_f \left( -\frac{328 \zeta_2^2}{15}-\frac{32 \zeta_2}{3}+\frac{104
   \zeta_3}{9}+\frac{2332}{27} \right) 
%%%%%
 + C_F n_f^2  \left( -\frac{40 \zeta_2}{27}-\frac{64 \zeta_3}{27}+\frac{3457}{729} \right) 
 \nonumber\\
 &
+ C_F^3 \left(-\frac{288 \zeta_2^2}{5}+32 \zeta_2 \zeta_3-18 \zeta_2-68 \zeta_3+240
   \zeta_5-\frac{29}{2}\right)  \, ,
\nonumber\\
\gamma_J^{(3)}[q]&= 4483.56 \pm 0.09 \quad ({\rm for }\; n_f=5) \,,
\label{gamma_quark_jet_4_loop}
\\
\gamma_J^{(3)}[q]&= 4371.68 \pm 0.11 \quad ({\rm for }\; n_f=4) \,,
\nonumber\\
\gamma_J^{(3)}[q]&= 3735.40 \pm 0.13 \quad ({\rm for }\; n_f=3) \,.
\nonumber
\end{align}
The four-loop $\gamma_J^{(3)}[q]$ has been extracted from the large-$x$ end-point limit of the splitting function and determined numerically \cite{Das:2019btv,Duhr:2022cob}, based on
\begin{equation}\label{Pqq}
   P_{q\leftarrow q}^{({\rm endpt})}(x)
   = \frac{2\Gamma_{\rm cusp}(\alpha_s)}{(1-x)_+}
   + 2\gamma^\phi(\alpha_s)\,\delta(1-x) \, ,
\end{equation}
where $\gamma^\phi = \gamma_J - \gamma_V$ as defined above.

The perturbative coefficients for the Laplace-space gluon jet function are given by \cite{Becher:2009th,Becher:2010pd,Banerjee:2018ozf}
\begin{align}
\label{gluon_jet_fn}
\widetilde J^{(0)}[g]&=  1 \,, \\
%%%%
\widetilde J^{(1)}[g]&=  C_A \left[2 L^2-\frac{11 L}{3}-4 \zeta_2+\frac{67}{9}\right] + n_f \left[\frac{2 L}{3}-\frac{10}{9}\right]   \, ,  \\
%%%%
\widetilde J^{(2)}[g]&=  
C_A^2 \bigg[2 L^4-\frac{88 L^3}{9}+L^2 \left(\frac{389}{9}-12 \zeta_2\right)+L \left(\frac{110 \zeta_2}{3}+16 \zeta_3-\frac{2570}{27}\right)
   \nn\\
   %%%% 
   &
 \qquad\qquad +17 \zeta_2^2-\frac{724 \zeta_2}{9}-\frac{88 \zeta_3}{3}+\frac{20215}{162}\bigg]
   \nn\\
   %%%% 
   &
   +C_A n_f \left(\frac{16 L^3}{9}-\frac{28 L^2}{3}+L
   \left(\frac{224}{9}-\frac{20 \zeta_2}{3}\right)+\frac{134 \zeta_2}{9}-\frac{8 \zeta_3}{3}-\frac{760}{27}\right)
   \nn\\
   %%%% 
   &
   +C_F n_f \left(2 L+8
   \zeta_3-\frac{55}{6}\right)+n_f^2 \left(\frac{4 L^2}{9}-\frac{40 L}{27}-\frac{4 \zeta_2}{9}+\frac{100}{81}\right)
     \, , 
     \\
%%%%
\widetilde J^{(3)}[g]&= 
C_A^3 \bigg[\frac{4 L^6}{3}-\frac{110 L^5}{9}+L^4 (85-16 \zeta_2)+L^3 \left(\frac{968
   \zeta_2}{9}+32 \zeta_3-\frac{9623}{27}\right)
      \nn\\
   %%%% 
   &
   +L^2 \left(\frac{338 \zeta_2^2}{5}-\frac{4724
   \zeta_2}{9}-\frac{484 \zeta_3}{3}+\frac{85924}{81}\right)
      \nn\\
   %%%% 
   &
   +L \left(-\frac{4411
   \zeta_2^2}{15}-\frac{320 \zeta_2 \zeta_3}{3}+\frac{105356 \zeta_2}{81}+\frac{6316
   \zeta_3}{9}-112 \zeta_5-\frac{1448021}{729}\right)\bigg]
      \nn\\
   %%%% 
   &
   +C_A^2 n_f \bigg[\frac{20
   L^5}{9}-\frac{64 L^4}{3}-L^3 \left(\frac{176 \zeta_2}{9}-\frac{3106}{27}\right)+L^2 \left(\frac{1172
   \zeta_2}{9}-\frac{8 \zeta_3}{3}-\frac{10067}{27}\right)
      \nn\\
   %%%% 
   &	
   +L \left(\frac{898 \zeta_2^2}{15}
   -{\frac{33662\zeta_2}{81}}
   -\frac{1280 \zeta_3}{27}
   + {\frac{1052135}{1458}}
   \right)\bigg]
      \nn\\
   %%%% 
   &
   +C_A C_F n_f
   \bigg[\frac{16 L^3}{3}+L^2 (32 \zeta_3-55)+L \left(-\frac{32 \zeta_2^2}{5}-20 \zeta_2-\frac{1096
   \zeta_3}{9}+\frac{5599}{27}\right)\bigg]
      \nn\\
   %%%% 
   &
   +C_A n_f^2 \left[\frac{4 L^4}{3}-\frac{292
   L^3}{27}+L^2 \left(\frac{3326}{81}-8 \zeta_2\right)+L \left(\frac{1016 \zeta_2}{27}-\frac{256
   \zeta_3}{27}-\frac{116509}{1458}\right)\right]
      \nn\\
   %%%% 
   &
   -C_F^2 n_f L +C_F n_f^2
   \left[\frac{10 L^2}{3}+L (16 \zeta_3-24)\right]+n_f^3 \left[\frac{8 L^3}{27}-\frac{40 L^2}{27}+L
   \left(\frac{200}{81}-\frac{8 \zeta_2}{9}\right)\right]
         \nn\\
   %%%% 
   &
    + \widetilde J^{(3)}[g]\big|_{L=0} \,, 
\end{align}
where the scale-independent three-loop constant term $ \widetilde J^{(3)}[g]\big|_{L=0} $ is obtained by the Laplace transform of the three-loop gluon jet function in \cite{Banerjee:2018ozf} 
%%%
\beeq
\label{three-loop-jet-constant-gluon}
\widetilde J^{(3)}[g]\big|_{L=0} &=&
C_F n_f^2 \bigg(-\frac{10 \zeta_2}{3}-\frac{104 \zeta_3}{3}+\frac{7001}{162}\bigg)                                                                                                                                                                                                                                                                                                       + n_f^3 \bigg(\frac{40 \zeta_2}{27}-\frac{16 \zeta_3}{27}-\frac{1000}{729}\bigg)                                                                                                                                                                                                                                                                                                       \nn \\   &+& C_A n_f^2 \bigg(\frac{836 \zeta_2^2}{135}-\frac{3986 \zeta_2}{81}+\frac{4252 \zeta_3}{243}+\frac{1613639}{26244}\bigg)                                                                                                                                                                                                                                                                                                       
%%%%
+ C_F^2 n_f \bigg(\frac{148 \zeta_3}{3}-80 \zeta_5+\frac{143}{9}\bigg)                                                                                                                                                                                                                                                                                                       \nn \\   &+& C_A^2 n_f \bigg(-\frac{18146 \zeta_2^2}{135}+\frac{200 \zeta_2 \zeta_3}{9}+\frac{388315 \zeta_2}{729}+\frac{1990 \zeta_3}{27}-\frac{272 \zeta_5}{9}-\frac{17323633}{26244}\bigg)                                                                                                                                                                                                                                                                                                       \nn \\   &+& C_A C_F n_f \bigg(\frac{304 \zeta_2^2}{45}-\frac{224 \zeta_2 \zeta_3}{3}+\frac{929 \zeta_2}{9}+\frac{20336 \zeta_3}{81}+\frac{584 \zeta_5}{9}-\frac{389369}{972}\bigg)                                                                                                                                                                                                                                                                                                       \nn \\   &+& C_A^3 \bigg(-\frac{82036 \zeta_2^3}{945}+\frac{27718 \zeta_2^2}{45}+\frac{2596 \zeta_2 \zeta_3}{9}-\frac{1281793 \zeta_2}{729}+\frac{544 \zeta_3^2}{9}-\frac{279556 \zeta_3}{243}
 \nn \\
&& 
+\frac{748 \zeta_5}{3}+\frac{55853711}{26244}\bigg)    \, .                                                                                                                                                                                                                                                                                                                                                                                                                                                                                                                                                                                                
\eeeq

Anomalous dimensions for the gluon jet are given by \cite{Becher:2009th,Das:2020adl,Duhr:2022cob}
%%%%
\begin{align}
\label{gamma_gluon_jet}
\gamma_J^{(0)}[g]&= \frac{2 }{3} n_f -\frac{11}{3} C_A \,,\\
%%%
\gamma_J^{(1)}[g]&= 
C_A^2 \left(\frac{22 \zeta_2}{3}+16 \zeta_3-\frac{1096}{27}\right)+C_A n_f
   \left(\frac{184}{27}-\frac{4 \zeta_2}{3}\right)+2 C_F n_f
    \,,\\
%%%%
\gamma_J^{(2)}[g]&= 
C_A^3 \left(-\frac{1166 \zeta_2^2}{15}-\frac{128 \zeta_2 \zeta_3}{3}+\frac{12434
   \zeta_2}{81}+260 \zeta_3-112 \zeta_5-\frac{331153}{1458}\right)
%%%%
    \nn \\
   & 
   +C_A^2 n_f
   \left(\frac{308 \zeta_2^2}{15}-\frac{2612 \zeta_2}{81}-\frac{8
   \zeta_3}{27}+\frac{42557}{1458}\right)
%%%%
    \nn \\
    &
   +C_A C_F n_f \left(-\frac{32
   \zeta_2^2}{5}-4 \zeta_2-\frac{304 \zeta_3}{9}+\frac{4145}{54}\right)
%%%%
    \nn \\
    &
   +C_A n_f^2
   \left(\frac{40 \zeta_2}{27}-\frac{112 \zeta_3}{27}+\frac{1811}{1458}\right)
   -C_F^2
   n_f-\frac{11 C_F n_f^2}{9} \,,\nn\\
%%%
\gamma_J^{(3)}[g]&= 26138.7 \pm 3.1  \quad ({\rm for }\; n_f=5) \,,
\\
\gamma_J^{(3)}[g]&= 19674.7 \pm 3.0 \quad ({\rm for }\; n_f=4) \,,
\nonumber\\
\gamma_J^{(3)}[g]&= 12196.6 \pm 2.9 \quad ({\rm for }\; n_f=3) \,.
\nonumber
\end{align}
We remind kindly the readers that a factor of $\pi^2$ is missing in the second term in Eq.(132) of \cite{Becher:2009th}, where the three-loop gluon-jet anomalous dimension $\gamma_J^{(2)}[g]$ has first been presented. We also note that though the N$^4$LL requires the jet function constant term only to the three-loop order, the scale-dependent terms described by its anomalous dimension are required to the four-loop order. 

\subsection{Cusp Anomalous Dimension}

The exact three-loop cusp anomalous dimensions have been known for a long time \cite{Korchemsky:1987wg,Korchemsky:1988si,Moch:2004pa}. Much efforts have been paid for obtaining the four-loop results, including the partial results in  \cite{Beneke:1995pq,Lee:2016ixa,Henn:2016men,vonManteuffel:2019wbj,Lee:2019zop,Henn:2019rmi,Boels:2017skl,Moch:2017uml,Moch:2018wjh,Davies:2016jie,Bruser:2019auj,Gracey:1994nn,Grozin:2018vdn}, and finally the full results in \cite{Henn:2019swt}. 
We use also the estimated 5-loop cusp \cite{Herzog:2018kwj} in the N$^4$LL resummation. 
In the resummation results, the uncertainty at N$^4$LL from the estimation of the 5-loop cusp is found to be numerically small and at the sub-milli level. 
We denote the expansion coefficient for the cusp anomalous dimension as 
\begin{align}
 \Gamma_{\rm cusp} = \sum_{n=0}  a_s^{n+1} \,\Gamma_{n} \, .
\end{align}
The analytic results through to the four-loop order are given by
\beeq
   \Gamma_0 &=& 4 C_R \,, \\
   \Gamma_1 &=& 4 C_R \bigg[C_A \left(\frac{67}{9}-2 \zeta_2\right)-\frac{10n_f}{9} \bigg] \,, \\
   \Gamma_2 &=& 4 C_R \bigg[C_A^2 \left(\frac{44 \zeta_2^2}{5}-\frac{268 \zeta_2}{9}+\frac{22 \zeta_3}{3}+\frac{245}{6}\right)+C_An_f \left(\frac{40 \zeta_2}{9}-\frac{28 \zeta_3}{3}-\frac{209}{27}\right)
    \nonumber\\
    %%%%
    && \hspace{+0.65cm}
    +C_Fn_f \left(8 \zeta_3-\frac{55}{6}\right)-\frac{4}{27}n_f^2 \bigg] \,,  \\
    %%%%
  \Gamma_3 &=& 4 C_R \bigg[ C_A^3 \left(-\frac{5008}{105} \zeta_2^3+\frac{902 \zeta_2^2}{5}-\frac{88}{3} \zeta_3 \zeta_2-\frac{22100 \zeta_2}{81}-4 \zeta_3^2+\frac{5236 \zeta_3}{27}-\frac{902 \zeta_5}{9}+\frac{42139}{162}\right)
      \nonumber\\
    %%%%
    && \hspace{+0.65cm}
    +C_A^2n_f \left(-\frac{88}{15} \zeta_2^2+\frac{112}{3} \zeta_3 \zeta_2+\frac{5080 \zeta_2}{81}-\frac{5776 \zeta_3}{27}+\frac{524 \zeta_5}{9}-\frac{24137}{324}\right)
        \nonumber\\
    %%%%
    && \hspace{+0.65cm}
    +C_AC_Fn_f
   \left(-\frac{88}{5} \zeta_2^2-32 \zeta_3 \zeta_2+\frac{110 \zeta_2}{3}+\frac{928 \zeta_3}{9}+40 \zeta_5-\frac{17033}{162}\right)
       \nonumber\\
    %%%%
    && \hspace{+0.65cm}
    +C_An_f^2 \left(-\frac{56}{15} \zeta_2^2-\frac{152 \zeta_2}{81}+\frac{560 \zeta_3}{27}+\frac{923}{324}\right)
    +C_F^2 n_f \left(\frac{148 \zeta_3}{3}-80 \zeta_5+\frac{143}{9}\right)
        \nonumber\\
    %%%%
    && \hspace{+0.65cm}
    +C_Fn_f^2
   \left(\frac{16 \zeta_2^2}{5}-\frac{160 \zeta_3}{9}+\frac{598}{81}\right)+n_f^3 \left(\frac{16 \zeta_3}{27}-\frac{8}{81}\right)   \bigg] 
 \nonumber\\
    %%%%
    &&
   +\frac{ d_R^{abcd}d_A^{abcd} }{N_R} \left[-\frac{7936}{35} \zeta_2^3-128 \zeta_2-384 \zeta_3^2+\frac{128 \zeta_3}{3}+\frac{3520 \zeta_5}{3}\right] 
    \nonumber\\
    %%%%
    &&
    +\frac{ d_R^{abcd}d_F^{abcd} }{N_R} n_f \left[256 \zeta_2-\frac{256 \zeta_3}{3}-\frac{1280 \zeta_5}{3}\right] \, , 
\eeeq
where $R=A,F$ represents the adjoint (gluon) or fundamental (quark) representation, respectively, as defined in Eq.(\ref{eq:casimirs}). The 5-loop cusp anomalous dimensions have been estimated with uncertainty \cite{Herzog:2018kwj}
\beeq
\Gamma_4 &=& 50000 \pm 40000 \quad (\text{quark}) \, ,	
 \\
 \qquad
\Gamma_4 &=& 30000 \pm 60000 \quad (\text{gluon}) \, .
\eeeq

\subsection{Strong Coupling Constant and Beta Function}

The ingredients are perturbatively expanded in the strong coupling constant, and are evolved to a common scale for the resummation. The strong coupling constant also evolves under its RG equation, 
\beeq
\beta(a_s) = \mu^2 \frac{d}{d\mu^2} a_s (\mu) = - \sum_{i \ge 0} \beta_i a_s^{i+2} \, . 
\eeeq
%%%%
The N$^4$LL resummation requires the perturbative coefficients $\beta_i$ to the 5-loop order, calculated in  \cite{Gross:1973id,Politzer:1973fx,Caswell:1974gg,Jones:1974mm,Egorian:1978zx,Tarasov:1980au,Larin:1993tp,vanRitbergen:1997va,Czakon:2004bu,Baikov:2016tgj,Herzog:2017ohr}, 
%%%%
\beeq
\beta_0&=&\frac{11 C_A}{3}-\frac{2 n_f}{3}\,,\\
\beta_1&=&\frac{34 C_A^2}{3}-\frac{10 C_A n_f}{3}-2 C_F n_f\,,\\
\beta_2&=&\frac{325 n_f^2}{54}-\frac{5033 n_f}{18}+\frac{2857}{2}\,,\\
\beta_3&=&\frac{1093 n_f^3}{729}+n_f^2 \left(\frac{6472 \zeta_3}{81}+\frac{50065}{162}\right)+n_f \left(-\frac{6508 \zeta_3}{27}-\frac{1078361}{162}\right)\nonumber\\
&&+3564 \zeta_3 +\frac{149753}{6}\,,\\
\beta_4
&=&n_f^4 \left(\frac{1205}{2916}-\frac{152 \zeta_3}{81}\right)
+n_f^3 \left(-\frac{48722 \zeta_3}{243}+\frac{460 \zeta_5}{9}+\frac{809 \pi ^4}{1215}-\frac{630559}{5832}\right)\nonumber\\
&&+n_f^2 \left(\frac{698531 \zeta_3}{81}-\frac{381760 \zeta_5}{81}-\frac{5263 \pi ^4}{405}+\frac{25960913}{1944}\right)
\nonumber\\
&&+n_f \left(-\frac{4811164 \zeta_3}{81}\right. \left.+\frac{1358995 \zeta_5}{27}+\frac{6787 \pi ^4}{108}-\frac{336460813}{1944}\right)\nonumber\\
&&+\frac{621885 \zeta_3}{2}  -288090 \zeta_5 +\frac{8157455}{16} -\frac{9801 \pi ^4}{20}\,, 
\eeeq
where we have used the color factors in $n_f$-flavour QCD, and the full color dependence can be found in the original calculations.

\section{Prediction at N$^4$LO}

\label{sec:PredictionN4LO}

The N$^4$LL resummation is sufficient to predict the leading power large-$x$ contributions at N$^4$LO, except for the boundary $\delta(1-x)$ terms which require in particular the four-loop jet function constant terms currently not available. 
The results presented in this section are contained in the ancillary file with the arXiv submission. 

\subsection{$\gamma^* \to q \bar q$}

\beeq
C_{\rm SIA}^{(4)}[q] &=&    
C_F n_f^3 \bigg(\frac{8 \mathL_4}{27}-\frac{232 \mathL_3}{81}+\mathL_2 \bigg(\frac{940}{81}-\frac{32 \zeta_2}{9}\bigg)+\mathL_1 \bigg(\frac{464 \zeta_2}{27}-\frac{17716}{729}\bigg)
\nonumber\\
&&
+\mathL_0 \bigg(\frac{16 \zeta_2^2}{5}-\frac{1864 \zeta_2}{81}+\frac{752 \zeta_3}{243}+\frac{124903}{6561}\bigg)\bigg)                                                                                                                                                                                                                                                                                                       \nn \\   &+& C_A^2 C_F n_f \bigg(\frac{242 \mathL_4}{9}+\mathL_3 \bigg(\frac{352 \zeta_2}{9}-\frac{9502}{27}\bigg)+\mathL_2 \bigg(\frac{176 \zeta_2^2}{5}-\frac{5096 \zeta_2}{9}-352 \zeta_3+\frac{17189}{9}\bigg)
\nonumber\\
&&
+\mathL_1 \bigg(-\frac{3944 \zeta_2^2}{15}+32 \zeta_2 \zeta_3+\frac{68836 \zeta_2}{27}+\frac{5752 \zeta_3}{3}-\frac{2080 \zeta_5}{9}-\frac{2452247}{486}\bigg)
\nonumber\\
&&
+\mathL_0 \bigg(\frac{3536 \zeta_2^3}{189}+\frac{3398 \zeta_2^2}{15}+760 \zeta_2 \zeta_3-\frac{2896561 \zeta_2}{729}+\frac{3056 \zeta_3^2}{9}-\frac{662540 \zeta_3}{243}
\nonumber\\
&&
-\frac{1288 \zeta_5}{9}+\frac{15772433}{2916}\bigg)\bigg)                                                                                                                                                                                                                                                                                                       \nn \\   &+& C_A C_F^2 n_f \bigg(-\frac{704 \mathL_5}{27}+\mathL_4 \bigg(\frac{8120}{27}-\frac{80 \zeta_2}{3}\bigg)+\mathL_3 \bigg(\frac{11984 \zeta_2}{27}+\frac{1216 \zeta_3}{9}-\frac{339134}{243}\bigg)
\nonumber\\
&&
+\mathL_2 \bigg(\frac{2332 \zeta_2^2}{15}-\frac{52772 \zeta_2}{27}-\frac{10600 \zeta_3}{9}+\frac{713162}{243}\bigg)+\mathL_1 \bigg(-\frac{41122 \zeta_2^2}{135}-\frac{1120 \zeta_2 \zeta_3}{3}
\nonumber\\
&&
+\frac{67102 \zeta_2}{81}+\frac{158740 \zeta_3}{81}-\frac{208 \zeta_5}{3}-\frac{326570}{729}\bigg)+\mathL_0 \bigg(\frac{8216 \zeta_2^3}{45}+\frac{68837 \zeta_2^2}{405}+240 \zeta_2 \zeta_3
\nonumber\\
&&
+\frac{703702 \zeta_2}{729}+\frac{752 \zeta_3^2}{3}-\frac{191764 \zeta_3}{81}-\frac{1000 \zeta_5}{3}-\frac{859349}{486}\bigg)\bigg)                                                                                                                                                                                                                                                                                                       \nn \\   &+& n_f \frac{{d_F^{abcd}d_F^{abcd}}}{N_F}  \mathL_1 \bigg(256 \zeta_2-\frac{256 \zeta_3}{3}-\frac{1280 \zeta_5}{3}\bigg)                                                                                                                                                                                                                                                                                                      \nn \\   &+& \frac{{d_A^{abcd}d_F^{abcd}} }{N_F} \mathL_1 \bigg(-\frac{7936 \zeta_2^3}{35}-128 \zeta_2-384 \zeta_3^2+\frac{128 \zeta_3}{3}+\frac{3520 \zeta_5}{3}\bigg)                                                                                                                                                                                                                                                                                                      \nn \\   &+& C_A C_F n_f^2 \bigg(-\frac{44 \mathL_4}{9}+\mathL_3 \bigg(\frac{1540}{27}-\frac{32 \zeta_2}{9}\bigg)+\mathL_2 \bigg(\frac{688 \zeta_2}{9}+16 \zeta_3-\frac{7403}{27}\bigg)
\nonumber\\
&&
+\mathL_1 \bigg(\frac{64 \zeta_2^2}{5}-\frac{9848 \zeta_2}{27}-\frac{688 \zeta_3}{9}+\frac{315755}{486}\bigg)+\mathL_0 \bigg(-\frac{2464 \zeta_2^2}{45}-\frac{128 \zeta_2 \zeta_3}{3}
\nonumber\\
&&
+\frac{408683 \zeta_2}{729}+\frac{41788 \zeta_3}{243}+\frac{32 \zeta_5}{3}-\frac{1414898}{2187}\bigg)\bigg)                                                                                                                                                                                                                                                                                                       \nn \\   &+& C_F^2 n_f^2 \bigg(\frac{64 \mathL_5}{27}-\frac{640 \mathL_4}{27}+\mathL_3 \bigg(\frac{24238}{243}-\frac{736 \zeta_2}{27}\bigg)+\mathL_2 \bigg(\frac{3800 \zeta_2}{27}+\frac{304 \zeta_3}{9}
\nonumber\\
&&
-\frac{52678}{243}\bigg)+\mathL_1 \bigg(\frac{736 \zeta_2^2}{27}-\frac{6148 \zeta_2}{81}-\frac{19304 \zeta_3}{81}+\frac{239633}{1458}\bigg)
\nonumber\\
&&
+\mathL_0 \bigg(-\frac{2848 \zeta_2^2}{135}-\frac{1952 \zeta_2 \zeta_3}{27}-\frac{104 \zeta_2}{729}+\frac{29908 \zeta_3}{81}+\frac{608 \zeta_5}{9}-\frac{45077}{324}\bigg)\bigg)                                                                                                                                                                                                                                                                                                       
\nonumber\\
&+& 
C_F {N_{F,V}} \frac{d_F^{abc}d_F^{abc} }{N_F}
\bigg(\mathL_1 \bigg( -\frac{128 \zeta_2^2}{5}+640 \zeta_2+\frac{896 \zeta_3}{3}-\frac{5120 \zeta_5}{3}+256 \bigg)
\nonumber\\
&&
+\mathL_0 \bigg(\frac{96 \zeta_2^2}{5}-480 \zeta_2-224 \zeta_3+1280 \zeta_5-192  \bigg) \bigg) 
%%%%                                                                                                                                                                                                                                                                                                      
\nn \\   &+& C_F^3 n_f \bigg(\frac{56 \mathL_6}{9}-\frac{164 \mathL_5}{3}+\mathL_4 \bigg(\frac{4630}{27}-\frac{560 \zeta_2}{9}\bigg)+\mathL_3 \bigg(\frac{2032 \zeta_2}{9}+\frac{1888 \zeta_3}{9}-\frac{280}{3}\bigg)
\nonumber\\
&&
+\mathL_2 \bigg(-\frac{728 \zeta_2^2}{5}-344 \zeta_2-936 \zeta_3-\frac{145}{9}\bigg)+\mathL_1 \bigg(\frac{10924 \zeta_2^2}{135}-\frac{6016 \zeta_2 \zeta_3}{9}-\frac{3520 \zeta_2}{27}
\nonumber\\
&&
+\frac{78328 \zeta_3}{27}+\frac{6368 \zeta_5}{9}+\frac{1775}{18}\bigg)+\mathL_0 \bigg(-\frac{93232 \zeta_2^3}{315}-\frac{3202 \zeta_2^2}{27}+\frac{7600 \zeta_2 \zeta_3}{9}
\nonumber\\
&&
+\frac{27625 \zeta_2}{27}-\frac{64 \zeta_3^2}{3}-1518 \zeta_3-\frac{2816 \zeta_5}{3}-\frac{39775}{108}\bigg)\bigg)                                                                                                                                                                                                                                                                                                       \nn \\   &+& C_A^3 C_F \bigg(-\frac{1331 \mathL_4}{27}+\mathL_3 \bigg(\frac{55627}{81}-\frac{968 \zeta_2}{9}\bigg)+\mathL_2 \bigg(-\frac{968 \zeta_2^2}{5}+\frac{4012 \zeta_2}{3}
\nonumber\\
&&
+1452 \zeta_3-\frac{649589}{162}\bigg)+\mathL_1 \bigg(-\frac{20032 \zeta_2^3}{105}+\frac{17996 \zeta_2^2}{15}+528 \zeta_2 \zeta_3-\frac{156238 \zeta_2}{27}
\nonumber\\
&&
-16 \zeta_3^2-\frac{78296 \zeta_3}{9}+\frac{19360 \zeta_5}{9}+\frac{16865531}{1458}\bigg)+\mathL_0 \bigg(-\frac{19448 \zeta_2^3}{189}-\frac{7631 \zeta_2^2}{45}
\nonumber\\
&&
-\frac{8668 \zeta_2 \zeta_3}{3}+\frac{6106508 \zeta_2}{729}-\frac{16808 \zeta_3^2}{9}+\frac{2193533 \zeta_3}{243}+\frac{4180 \zeta_5}{9}-\frac{673606301}{52488}\bigg)\bigg)                                                                                                                                                                                                                                                                                                       \nn \\   &+& C_A C_F^3 \bigg(-\frac{308 \mathL_6}{9}+\mathL_5 \bigg(\frac{998}{3}-48 \zeta_2\bigg)+\mathL_4 \bigg(\frac{4160 \zeta_2}{9}+400 \zeta_3-\frac{27835}{27}\bigg)
\nonumber\\
&&
+\mathL_3 \bigg(\frac{248 \zeta_2^2}{5}-\frac{9856 \zeta_2}{9}-\frac{8800 \zeta_3}{9}-\frac{1534}{3}\bigg)+\mathL_2 \bigg(\frac{1094 \zeta_2^2}{5}-768 \zeta_2 \zeta_3+1822 \zeta_2
\nonumber\\
&&
+\frac{2716 \zeta_3}{3}-720 \zeta_5+\frac{13990}{3}\bigg)+\mathL_1 \bigg(-\frac{2176 \zeta_2^3}{315}+\frac{3347 \zeta_2^2}{135}+\frac{55360 \zeta_2 \zeta_3}{9}-\frac{3539 \zeta_2}{27}
\nonumber\\
&&
+\frac{5984 \zeta_3^2}{3}-\frac{333526 \zeta_3}{27}-\frac{77216 \zeta_5}{9}+\frac{2829}{2}\bigg)+\mathL_0 \bigg(\frac{448888 \zeta_2^3}{315}-\frac{648 \zeta_2^2 \zeta_3}{5}+\frac{377731 \zeta_2^2}{270}
\nonumber\\
&&
-\frac{35012 \zeta_2 \zeta_3}{9}-1632 \zeta_2 \zeta_5-\frac{8859 \zeta_2}{2}-\frac{12008 \zeta_3^2}{3}+\frac{27860 \zeta_3}{3}+\frac{29240 \zeta_5}{3}-\frac{137084}{27}\bigg)\bigg)                                                                                                                                                                                                                                                                                                       \nn \\   &+& C_A^2 C_F^2 \bigg(\frac{1936 \mathL_5}{27}+\mathL_4 \bigg(\frac{440 \zeta_2}{3}-\frac{24040}{27}\bigg)+\mathL_3 \bigg(\frac{864 \zeta_2^2}{5}-\frac{47104 \zeta_2}{27}-\frac{13024 \zeta_3}{9}
\nonumber\\
&&
+\frac{2154563}{486}\bigg)+\mathL_2 \bigg(-\frac{2450 \zeta_2^2}{3}-832 \zeta_2 \zeta_3+\frac{57956 \zeta_2}{9}+\frac{24544 \zeta_3}{3}-1392 \zeta_5
\nonumber\\
&&
-\frac{2254339}{243}\bigg)+\mathL_1 \bigg(\frac{67352 \zeta_2^3}{315}+\frac{39899 \zeta_2^2}{135}+\frac{2720 \zeta_2 \zeta_3}{3}-\frac{40750 \zeta_2}{27}+\frac{3808 \zeta_3^2}{3}
\nonumber\\
&&
-\frac{238400 \zeta_3}{81}+\frac{6472 \zeta_5}{3}-\frac{21987073}{5832}\bigg)+\mathL_0 \bigg(-\frac{288836 \zeta_2^3}{315}+\frac{584 \zeta_2^2 \zeta_3}{15}-\frac{458953 \zeta_2^2}{810}
\nonumber\\
&&
+\frac{49280 \zeta_2 \zeta_3}{27}-928 \zeta_2 \zeta_5-\frac{3798428 \zeta_2}{729}-\frac{1136 \zeta_3^2}{3}-\frac{6500 \zeta_3}{3}+\frac{20248 \zeta_5}{9}+\frac{54699043}{3888}\bigg)\bigg)                                                                                                                                                                                                                                                                                                       \nn \\   &+& 
C_F^4 \bigg(\frac{16 \mathL_7}{3}-28 \mathL_6+\mathL_5 (-32 \zeta_2-18)+\mathL_4 \bigg(240 \zeta_2+\frac{400 \zeta_3}{3}+210\bigg)
\nonumber\\
&&
+\mathL_3 \bigg(-\frac{1488 \zeta_2^2}{5}-168 \zeta_2-1072 \zeta_3+\frac{113}{2}\bigg)+\mathL_2 \bigg(324 \zeta_2^2-448 \zeta_2 \zeta_3-1134 \zeta_2
\nonumber\\
&&
+1392 \zeta_3+4128 \zeta_5-\frac{1299}{2}\bigg)+\mathL_1 \bigg(-\frac{142816 \zeta_2^3}{315}-162 \zeta_2^2+1504 \zeta_2 \zeta_3+\frac{265 \zeta_2}{3}
\nonumber\\
&&
-\frac{256 \zeta_3^2}{3}-1374 \zeta_3+64 \zeta_5-\frac{19885}{24}\bigg)+\mathL_0 \bigg(\frac{992 \zeta_2^3}{21}-\frac{1712 \zeta_2^2 \zeta_3}{5}+\frac{3279 \zeta_2^2}{5}
\nonumber\\
&&
-1880 \zeta_2 \zeta_3+192 \zeta_2 \zeta_5+\frac{2603 \zeta_2}{2}+736 \zeta_3^2+1064 \zeta_3-6744 \zeta_5+3840 \zeta_7+\frac{15605}{16}\bigg)\bigg)                                                                                                                                                                                                                                                                                                       \nn \\   &+& {\gamma_J^{(3)}[q]} \mathL_0   
 \, + \, \delta(1-x) \text{ terms. }   \nn                                                                                                                                                                                                                                                                                                                                                                                                                                                                                                                                                            
\eeeq
This result has not been presented in the literature, and has been checked with the moment-space resummation.

\subsection{$H \to gg$}

\beeq
C_{\rm SIA}^{(4)}[g] &=&  
n_f^4 \bigg(\frac{64 \mathL_3}{81}-\frac{320 \mathL_2}{81}+\mathL_1 \bigg(\frac{1600}{243}-\frac{128 \zeta_2}{27}\bigg)+\mathL_0 \bigg(\frac{640 \zeta_2}{81}-\frac{8000}{2187}\bigg)\bigg)                                                                                                                                                                                                                                                                                                       \nn \\   &+& C_A n_f^3 \bigg(\frac{128 \mathL_4}{27}-\frac{3328 \mathL_3}{81}+\mathL_2 \bigg(\frac{12416}{81}-\frac{1120 \zeta_2}{27}\bigg)+\mathL_1 \bigg(\frac{15488 \zeta_2}{81}+\frac{64 \zeta_3}{81}
\nonumber\\
&&
-\frac{656630}{2187}\bigg)+\mathL_0 \bigg(\frac{272 \zeta_2^2}{27}-\frac{64912 \zeta_2}{243}+\frac{4384 \zeta_3}{243}+\frac{648670}{2187}\bigg)\bigg)                                                                                                                                                                                                                                                                                                       \nn \\   &+& C_F n_f^3 \bigg(\frac{104 \mathL_2}{9}+\mathL_1 \bigg(\frac{512 \zeta_3}{9}-\frac{2548}{27}\bigg)+\mathL_0 \bigg(-\frac{64 \zeta_2^2}{135}-\frac{992 \zeta_2}{27}-\frac{4480 \zeta_3}{27}+\frac{56428}{243}\bigg)\bigg)                                                                                                                                                                                                                                                                                                       \nn \\   &+& C_F^2 n_f^2 \bigg(8 \mathL_1+\mathL_0 \bigg(\frac{3088 \zeta_3}{9}-\frac{1280 \zeta_5}{3}-\frac{271}{27}\bigg)\bigg)                                                                                                                                                                                                                                                                                                       \nn \\   &+& n_f \frac{{d_A^{abcd}d_F^{abcd}} }{N_A}  \mathL_1 \bigg(256 \zeta_2-\frac{256 \zeta_3}{3}-\frac{1280 \zeta_5}{3}\bigg)                                                                                                                                                                                                                                                                                                     \nn \\   &+& \frac{{d_A^{abcd}d_A^{abcd}}}{N_A}   \mathL_1 \bigg(-\frac{7936 \zeta_2^3}{35}-128 \zeta_2-384 \zeta_3^2+\frac{128 \zeta_3}{3}+\frac{3520 \zeta_5}{3}\bigg)                                                                                                                                                                                                                                                                                                     \nn \\   &+& C_A^2 n_f^2 \bigg(\frac{304 \mathL_5}{27}-\frac{10336 \mathL_4}{81}+\mathL_3 \bigg(\frac{171832}{243}-128 \zeta_2\bigg)
+\mathL_2 \bigg(\frac{24560 \zeta_2}{27}
\nonumber\\
&&
+\frac{896 \zeta_3}{9}-\frac{574540}{243}\bigg)+\mathL_1 \bigg(\frac{2456 \zeta_2^2}{45}-\frac{670304 \zeta_2}{243}-\frac{40288 \zeta_3}{81}+\frac{1181833}{243}\bigg)
\nonumber\\
&&
+\mathL_0 \bigg(-\frac{54304 \zeta_2^2}{405}-\frac{7760 \zeta_2 \zeta_3}{27}+\frac{2258489 \zeta_2}{729}+\frac{231920 \zeta_3}{243}+\frac{6832 \zeta_5}{27}-\frac{1201861}{243}\bigg)\bigg)                                                                                                                                                                                                                                                                                                       \nn \\   &+& C_A C_F n_f^2 \bigg(\frac{344 \mathL_3}{9}+\mathL_2 \bigg(256 \zeta_3-\frac{4544}{9}\bigg)+\mathL_1 \bigg(-\frac{512 \zeta_2^2}{45}-\frac{1936 \zeta_2}{9}-\frac{39424 \zeta_3}{27}
\nonumber\\
&&
+\frac{185680}{81}\bigg)+\mathL_0 \bigg(\frac{3488 \zeta_2^2}{135}-\frac{352 \zeta_2 \zeta_3}{3}+\frac{15688 \zeta_2}{27}+\frac{184280 \zeta_3}{81}+\frac{2368 \zeta_5}{9}-\frac{871279}{243}\bigg)\bigg)                                                                                                                                                                                                                                                                                                       \nn \\   &+& C_A^2 C_F n_f \bigg(\frac{100 \mathL_4}{3}+\mathL_3 \bigg(320 \zeta_3-\frac{5480}{9}\bigg)+\mathL_2 \bigg(-\frac{192 \zeta_2^2}{5}-152 \zeta_2-\frac{4832 \zeta_3}{3}+\frac{26288}{9}\bigg)
\nonumber\\
&&
+\mathL_1 \bigg(\frac{2816 \zeta_2^2}{45}+320 \zeta_2 \zeta_3+\frac{7516 \zeta_2}{9}+\frac{189520 \zeta_3}{27}+960 \zeta_5-\frac{920977}{81}\bigg)
+\mathL_0 \bigg(-\frac{128 \zeta_2^3}{5}
\nonumber\\
&&
-\frac{4814 \zeta_2^2}{27}+\frac{4592 \zeta_2 \zeta_3}{9}-\frac{50188 \zeta_2}{27}+1024 \zeta_3^2-\frac{748652 \zeta_3}{81}-\frac{13024 \zeta_5}{9}+\frac{2055173}{162}\bigg)\bigg)                                                                                                                                                                                                                                                                                                       \nn \\   &+& C_A^3 n_f \bigg(\frac{112 \mathL_6}{9}-\frac{4304 \mathL_5}{27}+\mathL_4 \bigg(\frac{90968}{81}-\frac{1520 \zeta_2}{9}\bigg)
+\mathL_3 \bigg(\frac{13952 \zeta_2}{9}+\frac{4000 \zeta_3}{9}
\nonumber\\
&&
-\frac{1235948}{243}\bigg)+\mathL_2 \bigg(\frac{1984 \zeta_2^2}{15}-\frac{185000 \zeta_2}{27}-\frac{33328 \zeta_3}{9}+\frac{3609488}{243}\bigg)
\nonumber\\
&&
+\mathL_1 \bigg(-\frac{11464 \zeta_2^2}{27}-\frac{18368 \zeta_2 \zeta_3}{9}+\frac{3689488 \zeta_2}{243}+\frac{905840 \zeta_3}{81}+\frac{17200 \zeta_5}{9}-\frac{23031829}{729}\bigg)
\nonumber\\
&&
+\mathL_0 \bigg(-\frac{50936 \zeta_2^3}{945}+\frac{133912 \zeta_2^2}{405}+\frac{102944 \zeta_2 \zeta_3}{27}-\frac{10699429 \zeta_2}{729}+\frac{992 \zeta_3^2}{3}-\frac{3952528 \zeta_3}{243}
\nonumber\\
&&
-\frac{70544 \zeta_5}{27}+\frac{63376367}{2187}\bigg)\bigg)                                                                                                                                                                                                                                                                                                       \nn \\   &+& C_A C_F^2 n_f \bigg(-8 \mathL_2+\mathL_1 \bigg(1184 \zeta_3-1920 \zeta_5+\frac{1258}{3}\bigg)
\nonumber\\
&&
+\mathL_0 \bigg(2 \zeta_2-\frac{13024 \zeta_3}{9}+\frac{7040 \zeta_5}{3}-\frac{13382}{27}\bigg)\bigg)                                                                                                                                                                                                                                                                                                       \nn \\   &+& C_A^4 \bigg(\frac{16 \mathL_7}{3}-\frac{616 \mathL_6}{9}+\mathL_5 \bigg(\frac{15628}{27}-80 \zeta_2\bigg)+\mathL_4 \bigg(\frac{8360 \zeta_2}{9}+\frac{1600 \zeta_3}{3}-\frac{250288}{81}\bigg)
\nonumber\\
&&
+\mathL_3 \bigg(-\frac{376 \zeta_2^2}{5}-\frac{43424 \zeta_2}{9}-\frac{37840 \zeta_3}{9}+\frac{2966284}{243}\bigg)+\mathL_2 \bigg(-\frac{7744 \zeta_2^2}{15}-2048 \zeta_2 \zeta_3
\nonumber\\
&&
+\frac{455476 \zeta_2}{27}+\frac{173360 \zeta_3}{9}+2016 \zeta_5-\frac{7795390}{243}\bigg)+\mathL_1 \bigg(-\frac{15304 \zeta_2^3}{35}+\frac{74042 \zeta_2^2}{135}
\nonumber\\
&&
+\frac{85184 \zeta_2 \zeta_3}{9}-\frac{6700964 \zeta_2}{243}+\frac{9488 \zeta_3^2}{3}-\frac{4111304 \zeta_3}{81}-\frac{47080 \zeta_5}{9}+\frac{142480865}{2187}\bigg)
\nonumber\\
&&
+\mathL_0 \bigg(\frac{413204 \zeta_2^3}{945}-\frac{6496 \zeta_2^2 \zeta_3}{15}+\frac{90716 \zeta_2^2}{135}-11972 \zeta_2 \zeta_3-2368 \zeta_2 \zeta_5+\frac{17253320 \zeta_2}{729}
\nonumber\\
&&
-\frac{22352 \zeta_3^2}{3}+\frac{13539112 \zeta_3}{243}+\frac{204940 \zeta_5}{27}+3840 \zeta_7-\frac{242642513}{4374}\bigg)\bigg)                                                                                                                                                                                                                                                                                                                    
\nonumber\\
&+&  {\gamma_J^{(3)}[g]} \mathL_0 \, + \, \delta(1-x) \text{ terms. }                                                                                                                                                                                                                                                                                                                                                                                                                                                                                                                                                              
\eeeq

\subsection{$H \to b\bar b$}

\beeq
C_{\rm SIA}^{(4)}[b] &=&
 C_F n_f^3 \bigg(\frac{8 \mathL_4}{27}-\frac{232 \mathL_3}{81}+\mathL_2 \bigg(\frac{940}{81}-\frac{32 \zeta_2}{9}\bigg)+\mathL_1 \bigg(\frac{464 \zeta_2}{27}-\frac{17716}{729}\bigg)
\nonumber\\
&&
+\mathL_0 \bigg(\frac{16 \zeta_2^2}{5}-\frac{1864 \zeta_2}{81}+\frac{752 \zeta_3}{243}+\frac{124903}{6561}\bigg)\bigg)                                                                                                                                                                                                                                                                                                       \nn \\   &+& C_A^2 C_F n_f \bigg(\frac{242 \mathL_4}{9}+\mathL_3 \bigg(\frac{352 \zeta_2}{9}-\frac{9502}{27}\bigg)+\mathL_2 \bigg(\frac{176 \zeta_2^2}{5}-\frac{5096 \zeta_2}{9}-352 \zeta_3+\frac{17189}{9}\bigg)
\nonumber\\
&&
+\mathL_1 \bigg(-\frac{3944 \zeta_2^2}{15}+32 \zeta_2 \zeta_3+\frac{68836 \zeta_2}{27}+\frac{5752 \zeta_3}{3}-\frac{2080 \zeta_5}{9}-\frac{2452247}{486}\bigg)
\nonumber\\
&&
+\mathL_0 \bigg(\frac{3536 \zeta_2^3}{189}+\frac{3398 \zeta_2^2}{15}+760 \zeta_2 \zeta_3-\frac{2896561 \zeta_2}{729}+\frac{3056 \zeta_3^2}{9}-\frac{662540 \zeta_3}{243}
\nonumber\\
&&
-\frac{1288 \zeta_5}{9}+\frac{15772433}{2916}\bigg)\bigg)                                                                                                                                                                                                                                                                                                       \nn \\   &+& C_A C_F^2 n_f \bigg(-\frac{704 \mathL_5}{27}+\mathL_4 \bigg(\frac{8120}{27}-\frac{80 \zeta_2}{3}\bigg)+\mathL_3 \bigg(\frac{11984 \zeta_2}{27}+\frac{1216 \zeta_3}{9}-\frac{358142}{243}\bigg)
\nonumber\\
&&
+\mathL_2 \bigg(\frac{2332 \zeta_2^2}{15}-\frac{57524 \zeta_2}{27}-\frac{11032 \zeta_3}{9}+\frac{964334}{243}\bigg)+\mathL_1 \bigg(-\frac{10730 \zeta_2^2}{27}-\frac{1120 \zeta_2 \zeta_3}{3}
\nonumber\\
&&
+\frac{308842 \zeta_2}{81}+\frac{227356 \zeta_3}{81}-\frac{208 \zeta_5}{3}-\frac{4986920}{729}\bigg)+\mathL_0 \bigg(\frac{8216 \zeta_2^3}{45}+\frac{235697 \zeta_2^2}{405}
\nonumber\\
&&
+656 \zeta_2 \zeta_3-\frac{2326994 \zeta_2}{729}+\frac{752 \zeta_3^2}{3}-\frac{393652 \zeta_3}{81}-\frac{1000 \zeta_5}{3}+\frac{1263017}{243}\bigg)\bigg)                                                                                                                                                                                                                                                                                                       \nn \\   &+& n_f \frac{{d_F^{abcd}d_F^{abcd}}}{N_F}     \mathL_1  \bigg(256 \zeta_2-\frac{256 \zeta_3}{3}-\frac{1280 \zeta_5}{3}\bigg)                                                                                                                                                                                                                                                                                                   \nn \\   &+& \frac{{d_A^{abcd}d_F^{abcd}}}{N_F}     \mathL_1 \bigg(-\frac{7936 \zeta_2^3}{35}-128 \zeta_2-384 \zeta_3^2+\frac{128 \zeta_3}{3}+\frac{3520 \zeta_5}{3}\bigg)                                                                                                                                                                                                                                                                                                   \nn \\   &+& C_A C_F n_f^2 \bigg(-\frac{44 \mathL_4}{9}+\mathL_3 \bigg(\frac{1540}{27}-\frac{32 \zeta_2}{9}\bigg)+\mathL_2 \bigg(\frac{688 \zeta_2}{9}+16 \zeta_3-\frac{7403}{27}\bigg)
\nonumber\\
&&
+\mathL_1 \bigg(\frac{64 \zeta_2^2}{5}-\frac{9848 \zeta_2}{27}-\frac{688 \zeta_3}{9}+\frac{315755}{486}\bigg)+\mathL_0 \bigg(-\frac{2464 \zeta_2^2}{45}-\frac{128 \zeta_2 \zeta_3}{3}
\nonumber\\
&&
+\frac{408683 \zeta_2}{729}+\frac{41788 \zeta_3}{243}+\frac{32 \zeta_5}{3}-\frac{1414898}{2187}\bigg)\bigg)                                                                                                                                                                                                                                                                                                       \nn \\   &+& C_F^2 n_f^2 \bigg(\frac{64 \mathL_5}{27}-\frac{640 \mathL_4}{27}+\mathL_3 \bigg(\frac{25966}{243}-\frac{736 \zeta_2}{27}\bigg)+\mathL_2 \bigg(\frac{4088 \zeta_2}{27}+\frac{304 \zeta_3}{9}-\frac{71776}{243}\bigg)
\nonumber\\
&&
+\mathL_1 \bigg(\frac{736 \zeta_2^2}{27}-\frac{23572 \zeta_2}{81}-\frac{18728 \zeta_3}{81}+\frac{806603}{1458}\bigg)+\mathL_0 \bigg(-\frac{5728 \zeta_2^2}{135}-\frac{1952 \zeta_2 \zeta_3}{27}
\nonumber\\
&&
+\frac{233932 \zeta_2}{729}+\frac{28324 \zeta_3}{81}+\frac{608 \zeta_5}{9}-\frac{60185}{108}\bigg)\bigg)                                                                                                                                                                                                                                                                                                       \nn \\   &+& C_F^3 n_f \bigg(\frac{56 \mathL_6}{9}-\frac{164 \mathL_5}{3}+\mathL_4 \bigg(\frac{6070}{27}-\frac{560 \zeta_2}{9}\bigg)+\mathL_3 \bigg(\frac{2608 \zeta_2}{9}+\frac{1888 \zeta_3}{9}-\frac{5660}{9}\bigg)
\nonumber\\
&&
+\mathL_2 \bigg(-\frac{728 \zeta_2^2}{5}-\frac{2336 \zeta_2}{3}-936 \zeta_3+\frac{12299}{9}\bigg)+\mathL_1 \bigg(\frac{41164 \zeta_2^2}{135}-\frac{6016 \zeta_2 \zeta_3}{9}+\frac{9500 \zeta_2}{27}
\nonumber\\
&&
+\frac{86968 \zeta_3}{27}+\frac{6368 \zeta_5}{9}-\frac{27791}{18}\bigg)+\mathL_0 \bigg(-\frac{93232 \zeta_2^3}{315}-\frac{8458 \zeta_2^2}{27}+\frac{7024 \zeta_2 \zeta_3}{9}+\frac{10505 \zeta_2}{27}
\nonumber\\
&&-\frac{64 \zeta_3^2}{3}-\frac{17818 \zeta_3}{9}-\frac{2816 \zeta_5}{3}+\frac{104099}{108}\bigg)\bigg)                                                                                                                                                                                                                                                                                                       \nn \\   &+& C_A^3 C_F \bigg(-\frac{1331 \mathL_4}{27}+\mathL_3 \bigg(\frac{55627}{81}-\frac{968 \zeta_2}{9}\bigg)+\mathL_2 \bigg(-\frac{968 \zeta_2^2}{5}+\frac{4012 \zeta_2}{3}+1452 \zeta_3
\nonumber\\
&&
-\frac{649589}{162}\bigg)+\mathL_1 \bigg(-\frac{20032 \zeta_2^3}{105}+\frac{17996 \zeta_2^2}{15}+528 \zeta_2 \zeta_3-\frac{156238 \zeta_2}{27}-16 \zeta_3^2-\frac{78296 \zeta_3}{9}
\nonumber\\
&&
+\frac{19360 \zeta_5}{9}+\frac{16865531}{1458}\bigg)+\mathL_0 \bigg(-\frac{19448 \zeta_2^3}{189}-\frac{7631 \zeta_2^2}{45}-\frac{8668 \zeta_2 \zeta_3}{3}+\frac{6106508 \zeta_2}{729}
\nonumber\\
&&
-\frac{16808 \zeta_3^2}{9}+\frac{2193533 \zeta_3}{243}+\frac{4180 \zeta_5}{9}-\frac{673606301}{52488}\bigg)\bigg)                                                                                                                                                                                                                                                                                                       \nn \\   &+& C_A C_F^3 \bigg(-\frac{308 \mathL_6}{9}+\mathL_5 \bigg(\frac{998}{3}-48 \zeta_2\bigg)+\mathL_4 \bigg(\frac{4160 \zeta_2}{9}+400 \zeta_3-\frac{35755}{27}\bigg)
\nonumber\\
&&+\mathL_3 \bigg(\frac{248 \zeta_2^2}{5}-\frac{16192 \zeta_2}{9}-\frac{11392 \zeta_3}{9}+\frac{26720}{9}\bigg)+\mathL_2 \bigg(\frac{1094 \zeta_2^2}{5}-768 \zeta_2 \zeta_3+\frac{13718 \zeta_2}{3}
\nonumber\\
&&
+\frac{13300 \zeta_3}{3}-720 \zeta_5-\frac{12529}{3}\bigg)+\mathL_1 \bigg(-\frac{2176 \zeta_2^3}{315}-\frac{173341 \zeta_2^2}{135}+\frac{57088 \zeta_2 \zeta_3}{9}-\frac{51839 \zeta_2}{27}
\nonumber\\
&&
+\frac{5984 \zeta_3^2}{3}-\frac{312466 \zeta_3}{27}-\frac{68576 \zeta_5}{9}+\frac{53395}{18}\bigg)+\mathL_0 \bigg(\frac{448888 \zeta_2^3}{315}-\frac{648 \zeta_2^2 \zeta_3}{5}+\frac{483427 \zeta_2^2}{270}
\nonumber\\
&&
-\frac{38036 \zeta_2 \zeta_3}{9}-1632 \zeta_2 \zeta_5-\frac{50249 \zeta_2}{54}-\frac{11144 \zeta_3^2}{3}+\frac{50998 \zeta_3}{9}+\frac{22760 \zeta_5}{3}-\frac{53948}{27}\bigg)\bigg)                                                                                                                                                                                                                                                                                                       \nn \\   &+& C_A^2 C_F^2 \bigg(\frac{1936 \mathL_5}{27}+\mathL_4 \bigg(\frac{440 \zeta_2}{3}-\frac{24040}{27}\bigg)+\mathL_3 \bigg(\frac{864 \zeta_2^2}{5}-\frac{47104 \zeta_2}{27}-\frac{13024 \zeta_3}{9}
\nonumber\\
&&
+\frac{2259107}{486}\bigg)+\mathL_2 \bigg(-\frac{2450 \zeta_2^2}{3}-832 \zeta_2 \zeta_3+\frac{63764 \zeta_2}{9}+\frac{25336 \zeta_3}{3}-1392 \zeta_5
\nonumber\\
&&
-\frac{6034493}{486}\bigg)+\mathL_1 \bigg(\frac{67352 \zeta_2^3}{315}+\frac{132239 \zeta_2^2}{135}+\frac{2864 \zeta_2 \zeta_3}{3}-\frac{310264 \zeta_2}{27}+\frac{3808 \zeta_3^2}{3}
\nonumber\\
&&
-\frac{1000304 \zeta_3}{81}+\frac{7912 \zeta_5}{3}+\frac{112625501}{5832}\bigg)+\mathL_0 \bigg(-\frac{288836 \zeta_2^3}{315}+\frac{584 \zeta_2^2 \zeta_3}{15}-\frac{1321171 \zeta_2^2}{810}
\nonumber\\
&&
-\frac{22324 \zeta_2 \zeta_3}{27}-928 \zeta_2 \zeta_5+\frac{5378278 \zeta_2}{729}-\frac{5456 \zeta_3^2}{3}+\frac{56149 \zeta_3}{3}-\frac{8048 \zeta_5}{9}-\frac{44913479}{3888}\bigg)\bigg)                                                                                                                                                                                                                                                                                                       \nn \\   &+& C_F^4 \bigg(\frac{16 \mathL_7}{3}-28 \mathL_6+\mathL_5 (78-32 \zeta_2)+\mathL_4 \bigg(240 \zeta_2+\frac{400 \zeta_3}{3}-150\bigg)+\mathL_3 \bigg(-\frac{1488 \zeta_2^2}{5}
\nonumber\\
&&
-376 \zeta_2-1072 \zeta_3+\frac{533}{2}\bigg)+\mathL_2 \bigg(324 \zeta_2^2-448 \zeta_2 \zeta_3+198 \zeta_2+1584 \zeta_3+4128 \zeta_5-420\bigg)
\nonumber\\
&&
+\mathL_1 \bigg(-\frac{142816 \zeta_2^3}{315}-\frac{1402 \zeta_2^2}{5}+928 \zeta_2 \zeta_3-\frac{1097 \zeta_2}{3}-\frac{256 \zeta_3^2}{3}-5726 \zeta_3-1856 \zeta_5
\nonumber\\
&&
+\frac{55217}{24}\bigg)+\mathL_0 \bigg(\frac{992 \zeta_2^3}{21}-\frac{1712 \zeta_2^2 \zeta_3}{5}+\frac{1587 \zeta_2^2}{5}-2008 \zeta_2 \zeta_3+192 \zeta_2 \zeta_5+\frac{473 \zeta_2}{2}
\nonumber\\
&&
+736 \zeta_3^2+3734 \zeta_3-120 \zeta_5+3840 \zeta_7-\frac{25765}{16}\bigg)\bigg)     
\nonumber\\
&+& {\gamma_J^{(3)}[q]} \mathL_0\, + \, \delta(1-x) \text{ terms. }                                                                                                                                                                                                                                                                                                     
\eeeq

This result reproduces all the N$^3$LL terms given in \cite{Blumlein:2006pj}, and provides in addition the $\mathL_1$ and $\mathL_0$ terms in the N$^4$LL resummation. Note that many of the terms for $H \to b \bar b$ agree with those for  $\gamma^* \to q \bar q$, with the difference coming from the form factors.

\bibliographystyle{JHEP}
\bibliography{ThresholdResummationSIA}{}

\end{document}